\documentclass[a4,epsf,showpacs]{revtex4-1}

\usepackage{amsmath}
\usepackage{amssymb}
\usepackage{graphicx}
\usepackage[usenames]{color}

\begin{document}

\newcommand{\blue}[1]{\textcolor{black}{#1}}

\newcommand{\beq}{\begin{equation}}
\newcommand{\eeq}{\end{equation}}
\newcommand{\beqa}{\begin{eqnarray}}
\newcommand{\eeqa}{\end{eqnarray}}
\newcommand{\bmat}{\begin{displaymath}}
\newcommand{\emat}{\end{displaymath}}

\newcommand{\eq}[1]{Eq.~(\ref{#1})}

\newcommand{\lan}{\langle}
\newcommand{\ran}{\rangle}

\newcommand{\tav}[1]{\left\lan #1 \right\ran}

\title{Replica theory of the rigidity of structural glasses}

\author{Hajime Yoshino}

\affiliation{Department of Earth and Space Science, Faculty of Science,
 Osaka University, Toyonaka 560-0043, Japan}


\begin{abstract}
We present a first principle scheme to compute the rigidity, i.~e. the shear-modulus of structural glasses at finite temperatures using the cloned liquid  theory, 
which combines the replica theory and the liquid theory. 
With the aid of the replica method which enables disentanglement of
thermal fluctuations in liquids into intra-state and inter-state fluctuations, we extract the rigidity of metastable amorphous solid states in the supercooled liquid and glass phases.
The result can be understood intuitively without replicas.
As a test case, we apply the scheme to the supercooled and glassy state of a binary mixture of soft-spheres.
The result compares well with the shear-modulus obtained by a previous molecular dynamic simulation. 
The rigidity of metastable states is significantly reduced with respect
 to the instantaneous rigidity, namely the Born term, due to
non-affine responses caused by displacements of particles inside cages at all temperatures down to $T=0$. 
It becomes nearly independent of temperature below the Kauzmann temperature $T_{\rm K}$.
At higher temperatures in the supercooled liquid state, the non-affine correction to the rigidity becomes stronger suggesting melting of the metastable solid state.
Inter-state part of the static response implies jerky,
intermittent stress-strain curves with static analogue of yielding at mesoscopic scales.
\end{abstract}


\maketitle

\clearpage

\section{Introduction}
\label{sec-introduction}

Rigidity is a distinct character of solids including crystals and glasses \cite{Anderson}. 
It is quantified unambiguously by the shear-modulus
\cite{landau-lifshitz} which
represents strength of materials against shear deformation, which changes 
the {\it shape} of the containers but not their volume.
The shear-modulus is finite in solids but zero in gasses and liquids 
so that, much like the order parameters, it 
distinguishes solids from other states of matters,
in sharp contrast to the bulk modulus which is finite even in gasses.
We denote the shear-modulus simply as rigidity in the present paper.
It is the rigidity of a Goldstone mode 
which reflects the breaking of the translational symmetry
much as the rigidity of the spin-wave in ferromagnets
which reflects the breaking of the rotational symmetry.
The significance of the shear-modulus can be hardly appreciated at the
level of the macroscopic elastic theory which treats all elastic constants on the same footing.
We must go back to the microscopic scales to study the mechanism of the rigidity of solids. 
Unfortunately the rigidity of solids is usually taken for granted and regarded merely as a fitting or adjustable parameter.
In the present paper we dig into this basic problem and attempt to develop a theoretical approach to compute the
shear-modulus of structural glasses at finite temperatures starting from microscopic Hamiltonians. 
Microscopic computation of the rigidity of structural glasses
would be  useful also for practical applications of amorphous materials.

A class of amorphous solid or glasses can be obtained through supercooling liquids. 
First one cools a liquid down to the supercooled liquid state, the metastable liquid phase
below the melting transition temperature $T_{\rm m}$. 
A supercooled liquid behaves as a visco-elastic material. 
On one hand the system behaves as a solid with a finite effective rigidity 
for a long time before it flows.
Within this time scale, which is called as the $\beta$-regime, the configuration of the particles
does not evolve much but the particles fluctuate mostly within
narrow spaces around themselves called as {\it cage} created by the
surrounding other particles. On the other hand the system behaves as a
fluid with high viscosity at longer time scales called as the $\alpha$-regime. 
In this regime the configuration of the cages themselves are reorganized. 
The two qualitatively different relaxation
mechanisms manifest themselves as two step relaxations, i.~e. $\alpha$
and $\beta$-relaxations,  in various observables such as the
intermediate scattering function, dielectric susceptibility and e.t.c. \cite{Angell-Ngai-McKenna-McMillan-Martin}.
Upon lowering the temperature, the viscosity or the structural
relaxation time $\tau_{\alpha}$ increases enormously. Eventually $\tau_{\alpha}$ exceeds
the laboratory time scale at some glass transition temperature $T_{\rm g}$ and the system
falls out-of equilibrium. As the result we are left with a piece of
glass with a finite rigidity, which behaves as a solid in all practical means. This is the glass
transition observed in practice. Thus the rigidity of glasses is completed by enormous increase of the viscosity
in sharp contrast to crystals where the rigidity is established abruptly by 1st order phase transitions with no appreciable precursors. 
The rich visco-elastic or rheological properties, which are absent in crystals, give distinct flavor to supercooled liquids and glasses
and make them useful in practical applications.
Apart from the interests to the outstanding question that whether an ideal, thermodynamic glass transition
exists or not, development of first principle, microscopic theories to describe
such an out-of equilibrium amorphous state of matters is strongly desired.

Among various theoretical proposals\cite{berthier-biroli}, the Random First Order Transition
(RFOT) theory \cite{Kirkpatrick-Thirumalai-1987,Kirkpatrick-Wolynes-1987a,RFOT,Cavagna-review,RFOT-review} provides a useful working ground to study
the supercooled liquids and amorphous solids in a unified manner. 
The basic scenario of the RFOT theory goes as follows at the mean field level.
By lowing the temperature down to the so called dynamical transition
temperature $T_{\rm d}$, exponentially larger number of metastable
amorphous solid states emerge
so that the structural entropy or the complexity becomes finite. The
ideal, equilibrium glass transition takes place at the putative Kauzmann
temperature $T_{\rm K}$ due to the entropy crisis mechanism \cite{Kauzmann}. Within the RFOT
theory, $T_{\rm d}$ is equivalent to the critical temperature $T_{\rm
c}$ predicted by the mode coupling theory (MCT) where $\tau_{\alpha}$
diverges \cite{MCT,MCT2}. 
As is well known, the dynamical transition itself happens only at the
mean-field level and  $T_{\rm d}$ would remain in real systems at most
as a crossover temperature. The metastable solid states can have only finite life times in finite dimensional systems.
True divergence of the viscosity, if any, may occur
approaching $T_{\rm K}$ as envisioned by Adam, Gibbs and
DiMarzio\cite{gibbs-dimarzio-1958,adm-gibbs-1965}. An important concern of the RFOT theory
is to seek for a theoretical foundation of the scenario \cite{RFOT,eastwood-wolynes-2002,bouchaud-biroli-2004,biroli-bouchaud-cavagna-grigera-verrocchio-2008,RFOT-review}.

At the mean-field level, the RFOT theory is backed up by the cloned liquid method \cite{mezard-parisi-1999, coluzzi-mezard-parisi-verrocchio-1999,parisi-zamponi,Berthier-Jacquin-Zamponi-2011} and the mode-couping theory (MCT) \cite{MCT,MCT2}. 
\blue{Consistency between the two approaches is currently 
investigated intensively \cite{ikeda-miyazaki-2010}.}
The cloned liquid method provides a first principle scheme to compute the equilibrium and {\it quasi}-equilibrium properties of the supercooled liquids and amorphous solids starting from
microscopic Hamiltonians with the help of the liquid theory
\cite{hansen-mcdonald} combined with the replica method \cite{monasson-1995,Franz-Parisi-1995}. 
In the nutshell the cloned liquid theory, in its simplest formulations, views the metastable
amorphous solids through an effective Einstein model
in which each particle is subjected to a virtual Hookian spring with one end fixed
at the mean position of each particle.  The configuration of the latter
is just that of a liquid. The virtual Hookian spring is meant to mimic
the {\it cage effect}: the spring constant is inversely proportional to
the size of the cage $A$, which is the order parameter of the theory. 
Physically, existence of amorphous states with finite cage size $A$ 
means breaking of the translational symmetry.

In the present paper we study rigidity of structural glasses against shear
at finite temperatures using the fluctuation formula of the
rigidity \cite{squire}.  We evaluate it using the cloned liquid theory.
This would amount to a formulation of an effective Debye model starting
from the effective Einstein model, which is a necessary step to go
beyond the mean-field theory. Long wave length elastic deformations are
the essential low energy excitations in solids and play important roles
much as spin-waves in ferromagnets. For example localized plastic events  which are elementary
steps of the flow in glasses \cite{Goldstein,Argon,shoving-model} are
known to accompany smooth elastic deformations around them,
which naturally implies elastic free-energy barriers and
interactions with each other through the long-ranged elastic couplings \cite{Eshelby,Bulatov-Argon,maloney-lemaitre-2004a,maloney-lemaitre-2006}. 

We show that the cloned liquid method
enables decomposition of a generic response function into two parts: the intra-state
responses corresponding to the $\beta$-relaxation
and inter-state responses, which correspond to the
$\alpha$-relaxation. Based on this scheme we compute the
intra-state rigidity $\hat{\mu}$, which presumably represents the
effective rigidity of supercooled liquids and the rigidity of glasses.
On the other hand, the inter-state fluctuation reduces the rigidity down to $0$.
In general, it is reasonable to expect that the $\beta$-relaxation only weakly changes
between ideal equilibrium and out-of equilibrium situations while the
$\alpha$-relaxation strongly changes \cite{Corberi-Cugliandolo-Yoshino-2010}. 
Thus the quasi-static approximation for the intra-state responses would be valid for 
both the equilibrated supercooled liquids and out-of equilibrium glasses. 

We find that the intra-state rigidity $\hat{\mu}$, i.~e. the rigidity of
the metastable glassy states is significantly reduced with respect to
the Born term \cite{Born-Huang} which describes the instantaneous, affine response  
by a {\it non}-affine correction term which represents 
stress relaxations caused by relaxation of particles inside cages,
i.~e. the $\beta$-relaxation, at all temperatures down to the zero temperature limit.
Our result implies a characteristic temperature dependence of the intra-state rigidity.
On one hand,  it is nearly constant below the Kauzmann temperature $T_{\rm K}$.
On the other hand, it strongly depends on the temperature above $T_{\rm K}$.
To test our scheme, we applied the method to a binary mixture of soft spheres. 
We found the theoretical prediction compares well with the previous 
result by Barrat et. al. (1988) \cite{barrat-roux-hansen-klein-1988} 
obtained by a molecular dynamic simulation. 

An interesting general problem is how glasses melt.
It has been proposed a long time ago by Born  \cite{Born} 
that melting of solids may be signaled by vanishing of the rigidity.
Although this rigidity crisis scenario 
obviously does not apply to the equilibrium liquid-solid transitions 
which are 1st order phase transitions, whether it is relevant for 
the melting of superheated metastable crystals approaching the spinodal temperature $T_{\rm s}$
from below is an intriguing question
\cite{Tallon-1989,wang,sorkin,Jin-2001}. 
Interestingly enough, there is an intimate analogy \cite{KZ2011}
between the melting of metastable amorphous solids at the 
dynamical transition temperature $T_{\rm d}$ (or the MCT critical
temperature $T_{\rm c}$) and the melting
of superheated metastable crystals at $T_{\rm s}$.
\blue{At the mean-field
level $T_{\rm d}$ is regarded precisely as the temperature 
above which the metastable
amorphous solids states become absent in equilibrium. 
Moreover it has been pointed out that the
underlying mechanism of the dynamical critical phenomena found by MCT is 
the qualitative change of the free-energy landscape at around 
$T_{\rm d}$ \cite{grigera-2002,Cavagna-review}.}
Then an interesting question is whether the rigidity crisis
scenario also applies to the melting of the amorphous solids.

For clarity let us note that there are important classes of amorphous
 solids other than the glasses obtained by supercooling simple liquids.
In a class of systems including gels, polymer glasses and rubbers, formation of disordered networks of molecules and
colloids are important. Another important class is the granular matters
which are athermal, i.~e. the temperature play no essential roles. 
How the rigidity emerge in these amorphous systems have also attracted a lot of
 interests and some microscopic theoretical approaches have been developed \cite{goldpart-goldenfeld-1989,zippelius-group,castillo-goldbart,Wyart-2005,hecke-2010}.

The organization of the paper is as follows.
In the next section, we discuss the background of the present work
with some short reviews on related works, basic concepts 
and prepare basic tools needed in the analysis of the rigidity.
In sec. \ref{sec-rigidity}, we develop our scheme to compute the
rigidity of structural glasses and apply it to the case of a binary mixture of
soft-spheres. Finally in sec. \ref{sec-discussions} and \ref{sec-conclusions} we summarize the
preset work and discuss some related problems. In the appendices we 
present some technical details. The brief account of the present work has been reported in
\cite{YM2010}. 

\section{The background}
\label{sec-background}

In this section first we discuss some basic aspects of response to externally induced
shear deformations in sec. \ref{subsec-linear-response-to-shear}.
\blue{Then in sec. \ref{subsec-melting-orderparameter}
we discuss a generic mean-field picture which suggests
an intimate connection between the order parameter and rigidity in solids.}
Finally  we review the cloned liquid method in sec. \ref{subsec-cloned-liquid}. 

\subsection{Linear response to shear}
\label{subsec-linear-response-to-shear}

\subsubsection{Response to shear : a paradox and a lesson}
\label{subsubsec-paradox}

A material is said to be in a solid state if its rigidity (shear-modulus) $\mu$ is positive. 
Let us consider $N$ particles put in rectangular container which can be deformed by simple shear.
Under a simple shear of shear-strain $\gamma$, the volume $V$ (and thus the number density
$\rho=N/V$ of the particles) remains unchanged and only the {\it shape} of the container changes.
(See. Fig.~\ref{fig-shear-coordinate}) 

Let us consider how much change occurs in the free-energy $F$ of the system by the simple shear.
Assuming that the free-energy $F=N f$ is an analytic function of the
shear-strain $\gamma$, we can expand the free-energy par particle $f(\gamma)=F(\gamma)/N$ as,
\beq
f(\gamma)=f(0)+\sigma \gamma + \frac{\gamma^{2}}{2} \mu + \ldots
\label{eq-free-ene-expansion}
\eeq
where we find the stress $\sigma$ and the rigidity $\mu$ as,
\beq
\sigma=\frac{d f}{d \gamma} \qquad \mu=\frac{d^{2} f}{d \gamma^{2}}.
\label{eq-stress-shearmodulus}
\eeq

However, there is a plain fact that thermodynamic free-energy
(par particle) $f$ should {\it not} depend on the {\it shape} of the container
(remember that the density remains unchanged under the simple shear deformation), 
\beq
\lim_{N \to \infty} F({\gamma})/N = {\rm const}.
\eeq
Thus we conclude that {\it rigidity defined in the thermodynamic sense
must always be  zero, whether the system is a liquid or a solid} \cite{Penrose} ! Apparently 
this goes against to our basic intuition that in solids $\mu > 0$.

The paradox described above suggests that the thermodynamic limit $N \to
\infty$ and small shear-strain limit $\gamma \to 0$ do {\it not} commute in
solids: linear response theory which is built in the limit $\lim_{N \to
\infty} \lim_{\gamma \to 0}$ {\it fails} to predict what actually happens in the
thermodynamics $\lim_{\gamma \to 0} \lim_{N \to \infty}$. 
The breakdown of the commutation of the two limits is a consequence of the breaking of the transnational symmetry in solids. 
Physically the break down of the linear response theory  means that {\it  as 
a system becomes a solid, not only the elasticity but plasticity must emerge simultaneously}.
The plasticity, which means non-linear responses like yielding or stress
drops, recovers the translational symmetry at the
macroscopic level required by the thermodynamics.
In this respect, the idealized elastic bodies which appear in macroscopic 
continuum descriptions are purely hypothetical and thermodynamically
unsound objects. In the present paper we actually limit ourselves to the
linear response theory but we shall always keep in our mind the fact that the linear response theory must fail.

Finally let us note that we strictly consider {\it shear-strain control} protocols 
instead of {\it shear stress control} protocols in the present paper. 
In the latter case, static formulations are impossible and one essentially
studies the rheology where one finds that  ``everything flows under shear''. 
Physically the fact that even
perfect crystals flow  \cite{sauset-biroli-kurchan} is intimately 
related to the {\it static} paradox discussed above and the plasticity
is the mechanism of the flow of solids under shear.

\subsubsection{Static fluctuation formula}
\label{subsubsec-static-fluctuation-formula}

\begin{figure}[b]
\includegraphics[width=0.4\textwidth]{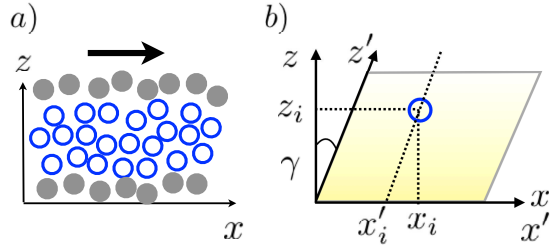}
\caption{
Schematic picture of the system under static shear. 
a) The mobile particles (open circles)
are bounded by ``wall particles'' (filled circle).
b) Laboratory frame $(x,y,z)$ and sheared frame $(x',y',z')$.
}
\label{fig-shear-coordinate}
\end{figure}

Let us consider a generic system of $N$
particles ($i=1,2,\ldots,N$) of mass $m$ at position ${\bf
r}_{i}=(x_{i},y_{i},z_{i})$ in the laboratory frame. 
More generally we denote 
components of a position vector ${\bf r}_{i}$ in 
a $d$-dimensional space as $x_{i}^{\mu}$ with $\mu=1,2,\ldots,d$.
We assume the Hamiltonian is given by,
\beq
{\cal H}=K+U \qquad K=\sum_{i=1}^{N}\frac{{\bf p}^{2}_{i}}{2m} \qquad U=\sum_{i < j} v(r_{ij}) \ ,
\label{eq-hamiltonian}
\eeq
where ${\bf r}_{ij} \equiv {\bf r}_{i}-{\bf r}_{j}$ and $r_{ij} \equiv |{\bf r}_{ij}|$
is the distance between the $i$-th and $j$-th particles.
The  term $K$ is the kinetic energy with ${\bf p}_{i}$ being 
the momentum of the $i$-th particle. For the potential energy $U$ 
we assume the simple two body interactions just for the sake of simplicity.

The free-energy of the system at temperature $T$ can be written
in terms of the partition function $Z$ of the canonical ensemble as,
\beq
-\beta F(T,{\cal V}) = \log Z  \qquad Z =\frac{1}{N!}\int_{\cal V} \prod_{i=1}^{N}
\frac{d^{d}r_{i}}{\Lambda^{d}} \exp \left( -\beta \sum_{i<j}v(r_{ij})\right)
\label{eq-def-free-energy}
\eeq
where $\beta=1/k_{\rm B}T$ is the inverse temperature, $\Lambda$ is the thermal de Brogile length $\Lambda=h/\sqrt{2\pi m k_{\rm B}T}$ with $k_{\rm B}$ being the Boltzmann's constant.  
In \eq{eq-def-free-energy},
the symbol ${\cal V}$ is meant to represent symbolically the integration volume, i. e. the container of the system including not only its volume $V$ but also its {\it shape}, which will be important in the following.

In order to study the rigidity against simple shear deformation,
we consider a container with two boundary walls which are normal 
to the $z$-axis and separated from each other by distance $L_{z}$ as shown in Fig.\ref{fig-shear-coordinate}.
To impose a shear-strain $\gamma$ on the system, we simply displace 
the top wall by an amount $\gamma L_{z}$ into $x$-direction.
The shear-deformation changes the boundary of the integration 
volume in \eq{eq-def-free-energy}. Let us denote the deformed boundary 
symbolically as ${\cal V}(\gamma)$. However, note that the volume of the system 
$V$ and thus the number density $\rho=N/V$ remain constant under this simple shear deformation.

Now we wish to find an expansion of the free-energy in power series of
the shear-strain $\gamma$ as \eq{eq-free-ene-expansion}. 
To this end, it is convenient to change the integration variables
to those of a {\it sheared frame} with $x'$,$y'$ and $z'$
which are related to the original laboratory frame as, $(x,y,z)=(x'+z' \gamma, y',z')$
(See Fig.~\ref{fig-shear-coordinate}). 
Note that the boundary of the original integration variables ${\cal V}(\gamma)$
evolves with the shear-strain $\gamma$ but that for the new variables
remains the same as the original one ${\cal V}(0)={\cal V}$.
But now we have to express the Hamiltonian, which is written in terms of the original coordinates, by the new ones. This can be done by simply expanding the Hamiltonian in power series of the shear-strain $\gamma$ assuming that it is small. In practice it is useful to notice,
\beq
\frac{d}{d\gamma}= \sum_{i < j} z_{ij}\frac{\partial}{\partial x_{ij}},
\eeq
where $x_{ij} \equiv x_{i}-x_{j}$ and $z_{ij} \equiv z_{i}-z_{j}$.
As the result, one easily finds a microscopic expression of the shear-stress defined in \eq{eq-stress-shearmodulus},
\beq
\sigma= \left. \frac{1}{N}\frac{d U}{d\gamma} \right |_{\gamma=0} =
\frac{1}{N} \sum_{i < j}  \sigma({\bf r}_{ij}) \qquad \sigma({\bf r}_{ij})=  
\left. \frac{d v(r_{ij})}{d\gamma} \right |_{\gamma=0}=\left. r v^{(1)}(r) \right |_{r=r_{ij}} \hat{x}_{ij}\hat{z}_{ij}. 
\label{eq-def-stress}
\eeq
where $\hat{x}_{ij} \equiv (x_{i}-x_{j})/r_{ij}$ and 
$\hat{z}_{ij} \equiv (z_{i}-z_{j})/r_{ij}$, and $v^{(n)}\equiv d^{n}v(r)/dr^{n}$.
Similarly the explicit expression of the 
rigidity defined in \eq{eq-stress-shearmodulus} is found as \cite{squire},
\beq
\mu= \langle b \rangle - N \beta [\langle \sigma^{2} \rangle  - \langle \sigma \rangle^{2}],
\label{eq-def-shearmodulus}
\eeq
with 
\beq
 b=\left. \frac{1}{N}\frac{d^{2} U}{d \gamma^{2}} \right |_{\gamma=0}=\frac{1}{N}\sum_{i <
j} b({\bf r}_{ij}) \qquad b({\bf r}_{ij})=
\left.  \frac{d^{2} v(r_{ij})}{d\gamma^{2}}\right |_{\gamma=0}=
\hat{z}_{ij}^{2} \left [
 r^{2}v^{(2)}(r)
\hat{x}_{ij}^{2}
+  r v^{(1)}(r)
(1- \hat{x}_{ij}^{2}) \right]_{r=r_{ij}}.
\label{eq-def-born}
\eeq
In \eq{eq-def-shearmodulus}  $\langle \ldots \rangle$ denotes the  thermal average evaluated with zero shear-strain $\gamma=0$,
\beq
\langle \ldots \rangle = Z^{-1} \int \prod_{i=1}^{N}
\frac{d^{d}r_{i}}{\Lambda^{d}} \exp \left( -\beta \sum_{i<j}v(r_{ij})\right)\ldots. 
\label{eq-thermal-average}
\eeq

The formula \eq{eq-def-shearmodulus} is the static fluctuation formula
of the rigidity \cite{squire}. The 1st term $b$ on the r. h. s of
\eq{eq-def-shearmodulus} is the so called Born term \cite{Born-Huang}.
It represents the {\it affine} response of the system against shear, which is finite even in simple liquids. 
The 2nd term is the {\it non-affine} correction term due to stress
relaxations which play crucial roles as we discuss in the
present paper.  Let us call it as the fluctuation term in the following. 
The difference of the nature of the affine and non-affine terms will
become more clear shortly by discussing the dynamic response against
shear. It may be instructive to note that quite analogous fluctuation formula is known for the
rigidity of spin-waves in ferromagnets or the helicity modulus of superconductors \cite{Chaikin-Lubenski} where
the non-affine part is represented by correlation functions of spin or super currents.

In liquids, the rigidity $\mu$ vanishes. It means 
an {\it exact} cancellation of the Born term and the fluctuation terms.
The exact cancellation may look somewhat surprising but it is
a direct consequence of the translational symmetry in liquids.
The vanishing of the rigidity in liquids can also be seen formally
by considering Mayer expansions, i.~e. expansions in power
	series of the number density $\rho=N/V$, of the free-energy of
	liquids. Such an expansion should converge in liquids 
	just like high temperature expansions are convergent in paramagnets.
At each order in the expansion, one finds finite sized clusters. 
An infinitesimal shear-deformation, which is just
an infinitesimal change on the boundary condition, should not change
the contribution from the finite sized cluster to the free-energy per unit volume in the thermodynamic limit $V \to \infty$. Thus the rigidity is
zero at {\it each} order of the density in the thermodynamic limit.
The vanishing of the rigidity at the level of linear response means
the two limits $V \to \infty$ and $\gamma \to 0$ {\it do} commute in
liquids (see sec. \ref{subsubsec-paradox}).
Similarly one can consider high temperature expansions in spin systems
by which one arrives at the same conclusion that the spin-wave stiffness must be zero in paramagnets.

\blue{Finally let us note that it is possible to obtain the zero temperature 
limit of the fluctuation formula of the rigidity \eq{eq-def-shearmodulus}
within a harmonic approximation as we discuss in sec. \ref{sec-IS-shearmodulus}.
It is very instructive to note that the vanishment of the rigidity in the unjamming transition of
a class of systems with contact forces at zero temperature 
is due to an exact cancellation of the Born term and the non-affine correction term
 \cite{hecke-2010,Zaccone-Romano-2011}.}
 
\subsubsection{Dynamic fluctuation formula}

Now let us briefly discuss more general, dynamic fluctuation formula for
the linear response to shear-strains which describes the elasticity and viscosity in a unified manner.
Although our focus is put on the elasticity rather than the viscosity in the present paper, 
the overview will become useful in discussions. 
Such a unified view point has been developed also in Ref. \cite{Williams-Evans-2009,Williams-Evans-2010,Williams-2011}. 

Linear response against shear deformations can be seen experimentally 
by measurements of the linear-viscosity (See for example \cite{dyre-group,Weitz97,mckenna-group1,mckenna-group2}).  In sec. \ref{subsubsec-paradox}
we emphasized that for arbitrary small but finite strength of
perturbation $\gamma$, {\it static} linear response break down in the $N
\to \infty$ limit in solids. Similarly {\it dynamic} linear response theory will
break down at large enough time scales by overwhelming non-linear contributions.
However, it is still possible to delay the dominance of the non-linear responses and bring it out of a given observation time (or frequency) window by choosing sufficiently small shear-strain $\gamma$ \cite{Weitz97}. Of course one has to choose smaller $\gamma$ for larger time window.

Within the equilibrium linear response theory the shear-stress $\sigma(t)$ at time $t$ can be related to small changes of the shear-strain $ \gamma(t')$ 
in the past $t \geq t' \geq -\infty$ as  \cite{Williams-Evans-2009},
\beq
\delta \sigma(t)=b \gamma(t)-\beta \int^{t}_{-\infty} dt' 
\frac{\partial C_{\sigma}(t,t')}{\partial t'} \gamma(t').
\label{eq-linear-response-1}
\eeq
Here $b$ is the born term defined in \eq{eq-def-born}.
The correlation function $C_{\sigma}(t,t')$ is the shear-stress auto-correlation function defined as,
\beq
C_{\sigma}(t,t')\equiv N \langle \sigma(t)\sigma(t')\rangle
\label{eq-stress-stress}
\eeq
We show in Appendix \ref{appendix-ft} a simple derivation of \eq{eq-linear-response-1}.
The 1st and 2nd term on the r.h.s. of \eq{eq-linear-response-1} 
represents respectively the instantaneous rigidity, which describes the
affine part of response to shear, and the non-affine correction term due
to shear-stress relaxations.

By doing an integration by parts in the r.~h.~s. of
\eq{eq-linear-response-1} we can find an alternative expression for the linear response,
\beq
\delta \sigma(t)=(b - N \beta \langle \sigma^{2} \rangle )\gamma(t) + N \beta \langle \sigma \rangle^{2} \gamma(-\infty)+\beta \int^{t}_{-\infty} dt' 
C_{\sigma}(t,t')\dot{\gamma}(t')
\label{eq-linear-response-2}
\eeq
where $\dot{\gamma}(t)\equiv d  \gamma(t')/dt'$ is the shear-strain rate.

More specifically, let us consider two typical experimental protocols,
\begin{itemize}
\item Step like shear : $\gamma(t)=\gamma\theta(t)$ 

The stress relaxation after then step like perturbation can be found to be,
\beq
\mu(t) \equiv \delta \sigma(t)/\gamma= b-\beta (C_{\sigma}(t,t)-C_{\sigma}(t,0)).
\label{eq-mu-t}
\eeq
The initial value is nothing but the Born term 
\beq
\mu(0)=b,
\label{eq-mu-0}
\eeq
and it relaxes down to the static rigidity $\mu$ defined in \eq{eq-def-shearmodulus} in the large time limit,
\beq
\mu=\lim_{t\to \infty} \mu(t)
= b-N\beta ( \langle \sigma^{2}  \rangle-\langle \sigma \rangle^{2} ).
\label{eq-shearmodulus-again}
\eeq
Thus the remnant stress under external strain becomes finite in solids and vanishes in liquids at the level of the linear response.

Furthermore, in the case of liquids the rigidity vanishes $\mu=0$. This implies (see \eq{eq-shearmodulus-again}),
\beq
b=N\beta \langle \sigma^{2} \rangle.
\label{eq-b-sigma-variance}
\eeq
Here we used $\langle \sigma \rangle=0$ which must always hold in liquids. 
It is worth to mention that the relation \eq{eq-b-sigma-variance} means the born term (instantaneous
      rigidity) and variance of the thermal fluctuation of the stress are
intimately related in liquids in equilibrium. As a result, the stress relaxation \eq{eq-mu-t} becomes directly proportional to the shear-stress auto-correlation function,
\beq
\mu(t)=\beta C_{\sigma}(t,0).
\label{eq-mu-t-stress-stress}
\eeq

\item Constant shear-rate : $\dot{\gamma}(t)=\dot{\gamma}$ 

Let us consider here liquids for which the rigidity is zero $\mu=0$.
Then from \eq{eq-linear-response-2} we find,
\beq
\delta \sigma/\dot{\gamma}=\eta
\qquad \eta \equiv \beta \int_{0}^{\infty} d\tau C_{\sigma}(\tau)
\label{eq-green-kubo}
\eeq
where $\eta$ is the shear-viscosity. The last equation is nothing but
the Green-Kubo formula for the shear-viscosity \cite{kubo}.

\end{itemize}

Usually we tend to think that rigidity and viscosity are very different properties. In this respect it is very interesting 
to note here that both the non-affine correction term of the rigidity, which appears in the 2nd term on the r. h. s of  \eq{eq-shearmodulus-again},
and the viscosity given by \eq{eq-green-kubo} are related to the spontaneous thermal fluctuation of the shear stress. 
Quite remarkably recent numerical studies of glasses have shown that the non-affine correction term is significant even in the zero temperature limit  \cite{Tanguy2002,maloney-lemaitre-2004b}.
\blue{(The zero temperature limit of the fluctuation formula of the rigidity \eq{eq-def-shearmodulus} is given by \eq{eq-mu-IS-AQS}). }
This implies rigidity and viscosity are related in glasses.

\subsubsection{Visco-elasticity}
\label{subsubsec-two-step-relaxation}

\begin{figure}[t]
\includegraphics[width=0.8\textwidth]{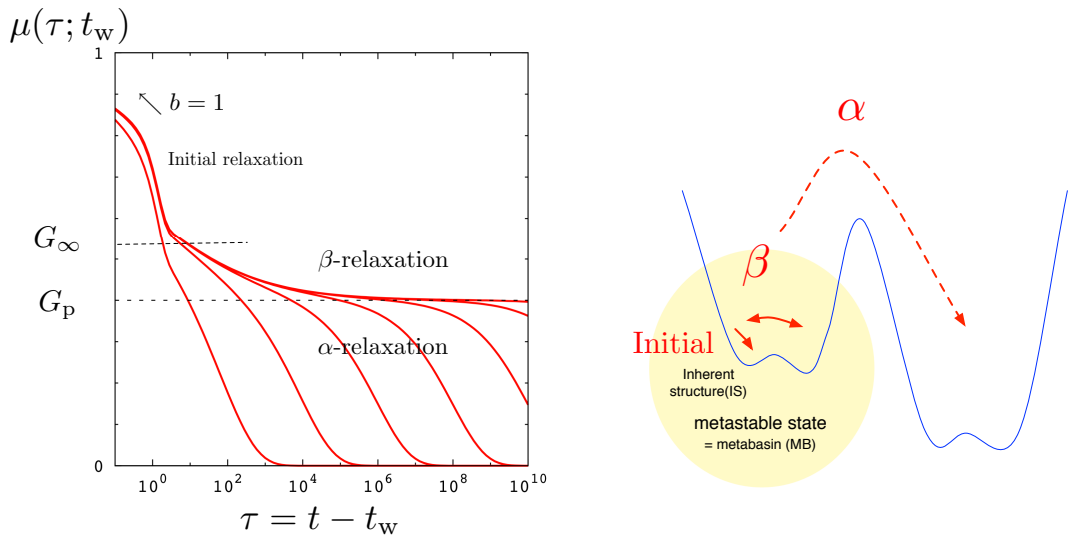}
\caption{Schematic, idealized picture of stress relaxation of
 supercooled liquid/glass under aging (left) 
and an intuitive free-energy landscape picture (right).
The stress relaxation function $\mu(\tau;t_{\rm w})$ is presented in a unit such that $\mu(0)=b=1$.
Here $b$ is the Born term \cite{Born-Huang} which represents the instantaneous rigidity of system.
The time $\tau$ is presented in a unit such that the microscopic time scale $\tau_{0}$, corresponding to typical vibration time of the particles, becomes $1$.
The waiting time $t_{\rm w}$ increases from the left to the right curves.
If the relaxation time $\tau_{\alpha}$ is finite, the curves converge to
a limiting (equilibrium) curve 
at large enough waiting time $t_{\rm w} \gg \tau_{\alpha}$. 
We consider that the stress relaxation consists of three regimes.
Within the initial very short time regime whose duration is just of the order of
the unit time scale $\tau_{0}$, the stress relaxes rapidly starting from
 the instantaneous rigidity $b$ down to some value, which is often
 called (somewhat misleadingly) $G_{\infty}$.
Most likely this is due to \blue{fast modes of} purely elastic relaxation in the vicinity 
of inherent structures (IS) \cite{Stillinger-Weber-1982,Stillinger-1995} and that $G_{\infty}$
roughly corresponds to rigidity of inherent structures $\mu_{\rm IS}$. 
In the subsequent $\beta$-relaxation regime, the stress value remains
within a plateau-like region in the range $G_{\rm p}  < \mu(\tau)  < G_{\infty}$.
We argue that this relaxation takes place within metastable states, 
each of which is a metabasin (MB) \cite{Doliwa-Heuer-2003,Heuer-2008} i.~e. a union of ISs with large enough mutual overlaps rather
than individual ISs. Furthermore we argue that $G_{\rm p}$ corresponds to rigidity of
 metastable states  $\hat{\mu}$ which we analyze in detail in the present paper.
The reduction of $\hat{\mu}$ with respect to the rigidity of inherent structures $\mu_{\rm IS}$ is
attributed to thermal excitations within metastable states, such as mutually independent, localized plastic deformations.
Thus the gap between $G_{\rm p}$ and $G_{\infty}$ is expected to become smaller at lower temperatures.
Finally the stress relaxes down to $0$ by the $\alpha$-relaxation by
 which  the system explores larger phase space.
While the $\alpha$ relaxation strongly depends on the waiting times $t_{\rm w}$, the initial and  $\beta$ relaxations depend on 
 $t_{\rm w}$ only weakly (quasi-equilibrium). 
}
\label{fig-two-step-func-model}
\end{figure}

Supercooled liquids are visco-elastic materials because of the two-step relaxation: 
the shear-stress auto-correlation function $C_{\sigma}(\tau)$ or equivalently the stress-relaxation $\mu(\tau)$ (see
\eq{eq-mu-t-stress-stress}) exhibit plateau like behaviours as shown schematically in Fig.~\ref{fig-two-step-func-model}.

The stress relaxation function $\mu(\tau)$ starts from the instantaneous rigidity, i.~e. the born term $b$ at time $\tau=0$ (see \eq{eq-mu-0}). 
Very at the beginning it exhibits a rapid decay as indicated in Fig.~\ref{fig-two-step-func-model},
which we call as 'initial relaxation'. The duration of the initial relaxation is expected to be of the order of the vibration time $\tau_{0}=\sqrt{m a^{2}/\epsilon}$ 
with $a$ being the microscopic unit length scale and $\epsilon$ being the unit energy (or collision time $\rho^{-1/d}\sqrt{m/k_{\rm B}T}$ ). 
\blue{The rapid initial relaxation can be easily seen in numerical simulations \cite{sciortino-group,furukawa,Okamura-Yoshino-Yukawa}. 
Whether it lies within the time window of experiments would depend on
the specific systems under study \cite{dyre-group,Weitz97,mckenna-group1,mckenna-group2}.}

Then the relaxation slows down significantly exhibiting a plateau like behavior
\cite{sciortino-group,furukawa,mckenna-group2,Okamura-Yoshino-Yukawa},
which we call as '$\beta$-relaxation'.
Let us call the height of the {\it onset} of the plateau like region as 
$G_{\infty}$ in the present paper.
If the plateau is ideally extended for a long enough time scale, 
the system would appear as a solid with a finite rigidity  (see \eq{eq-shearmodulus-again}), which is the main concern of this paper.

\blue{The temporal solid state is characterized by the cage
structure: the particles are collectively confined by the surrounding
particles within a narrow space called as {\it cage}.
Within the framework of the RFOT, including both the MCT \cite{MCT,MCT2}
and cloned liquid approaches \cite{mezard-parisi-1999, parisi-zamponi}, the size of the cage $A$ is viewed as the order parameter which
indicates the presence of amorphous solid state (see sec. \ref{sec-cage-size}).}

Finally the $\alpha$ relaxation starts at longer time scales due to reorganization of the cages by
which the residual stress $\mu(\tau)$ eventually decays down to $0$ so that the viscosity $\eta$  \eq{eq-green-kubo} can be defined.
As the temperature is lowered the relaxation time of the supercooled liquid $\tau_{\alpha}$ 
increases rapidly and eventually exceeds the experimental time scale. As the result the system falls out of equilibrium in experimental time scales. 

Still visco-elastic properties at lower temperatures can be studied systematically by observing aging effects \cite{Struik,Corberi-Cugliandolo-Yoshino-2010}
which allow us to explore the glassy phase below $T_{\rm K}$.
For example, a typical stress-relaxation protocol \cite{mckenna-group1,mckenna-group2} goes as follows: 
1) the system is prepared in the equilibrium 
liquid state at a high enough temperature above
the glass transition temperature $T_{\rm g}$ , 2) the temperature of the bath is 
quenched to a target temperature $T$ below $T_{\rm g}$, 3) the system is left
in a quiescent state for a waiting time $t_{\rm w}$ after which 4) a finite shear-strain
of amplitude $\gamma$ is applied,
\beq
\delta \gamma(t)=\gamma \theta(t-t_{\rm w}).
\eeq
Again the stress relaxation starts from the instantaneous rigidity,
the born term $b$ and exhibits the two step relaxations (+ initial relaxation) as shown in Fig.~\ref{fig-two-step-func-model}. 
\blue{But now the stress relaxation
depends not only on the time $\tau$ elapsed after switching on the shear-strain
but also on the waiting time $t_{\rm w}$ thus we denote 
it as $\mu(\tau;t_{\rm w})$.}
Most notable feature is that $\alpha$-relaxation becomes strongly dependent of the waiting time $t_{\rm w}$
while the $\beta$-relaxation only weakly depends on the waiting time $t_{\rm w}$.
\blue{Then it would be meaningful to define a limiting relaxation curve for the $\beta$-regime,
\beq
\mu_{\rm eq} (\tau) \equiv \lim_{t_{\rm w} \to \infty}\mu(\tau;t_{\rm w}).
\label{eq-mu-eq}
\eeq
}
As suggested by the studies of aging effects, in particular the dynamical mean-field theory of spin-glasses \cite{Cuku},
it is natural to expect that the fluctuation formula \eq{eq-mu-t-stress-stress} still holds in the $\beta$-regime
and that a slightly modified fluctuation formula applies for the $\alpha$-regime (See Appendix \ref{appendix-ft}).

The system will behave as a solid with finite rigidity 
as long as the time scale lies within the plateau-like region.
However there is an ambiguity to define such a rigidity because the plateau is not completely flat at finite time scales.
As noted above we call the value at the onset of the plateau-like region as $G_{\infty}$ in the present paper. 
Physically it is more interesting to examine the value of the plateau in the asymptotic limit (See Fig.~\ref{fig-two-step-func-model}),
\blue{
\beq
G_{\rm p} \equiv \lim_{\tau \to \infty}\mu_{\rm eq}(\tau)
\label{eq-Gp}
\eeq
Note that as long as the system is ergodic, which is certainly the case
at high enough temperatures in the liquid phase, 
the limiting curve \eq{eq-mu-eq} covers not only the $\beta$-relaxation but also the $\alpha$-relaxation by which the shear stress relaxes down to $0$.
Thus for this $G_{\rm p}$ to become non-zero, there must be ergodicity 
breaking, i.~e. the ideal glass transition. 
If the putative ergodicity breaking take place at some temperature, say
$T_{\rm NE}$, then the asymptotic value of the plateau at $T_{\rm NE}$
can also be obtained following the equilibrium liquid state at $T > T_{\rm NE}$ as,
\beq
G_{\rm p}(T_{\rm NE})=\lim_{\tau \to \infty} \lim_{T \to T_{\rm NE}^{+}}\mu_{\rm eq}(\tau)
\eeq
It is not known yet whether such a true ergodicity breaking (ideal glass transition) really exist in real systems. }
However even if it does not exist in the rigorous sense, 
we may take a pragmatic view and look for some crossover height $G_{\rm p}$  between more $\beta$-{\it like} regime and more $\alpha$-{\it like} regime.

\subsubsection{Free-energy landscape picture}
\label{sec-landscape}

Let us discuss the visco-elasticity of supercooled liquids within 
the phenomenological free-energy landscape picture shown schematically in Fig.~\ref{fig-two-step-func-model}
\cite{Goldstein,Stillinger-Weber-1982,Stillinger-1995,Doliwa-Heuer-2003,Heuer-2008}. 
\blue{The following arguments are based on simple intuitions on some basic 
energetics so that they are admittedly naive in many respects and thus they should be taken with a grain of salt.}
\blue{Here we focus on deeply supercooled, low temperature regime. In sec. \ref{section-landscape-highT}
we briefly discuss the free-energy landscape at higher temperatures.}

The most basic ingredients in the free-energy landscape at low temperatures 
are local energy minima 
which are called as inherent structures (IS) \cite{Stillinger-Weber-1982,Stillinger-1995}.
In the phase space, each inherent structure is surrounded by a local region called as basin whose
energy landscape can be described well by a purely elastic Hamiltonian
(See \eq{eq-hamiltonian-IS} below).

A set of inherent structures and their basins  may be grouped together
as a {\it metabasin} (MB) \cite{Doliwa-Heuer-2003,Heuer-2008}.
From the point of view of the mean-field theory which we discuss in
sec. \ref{subsec-cloned-liquid}, a metabasin rather than an inherent structure should be identified as a {\it metastable state}
which are {\it pure states} in the mean-field theory \cite{biroli-monasson}.
Within the mean-field theory, metastable states are characterized by the
cage size $A$ (See \eq{eq-cage-size-static}). Two different inherent structures, say IS$_{1}$ and IS$_{2}$ 
are associated with a common metastable state if the average mean-squared distance is smaller than the cage size,
\beq
\frac{1}{2Nd} \sum_{i=1}^{N} ({\bf r}_{i}^{{\rm IS}_{1}}-{\bf r}_{i}^{{\rm IS}_{2}})^{2}  < A
\label{eq-distance-two-ISs}
\eeq
Here ${\bf r}_{i}^{{\rm IS}_{1}}$ and  ${\bf r}_{i}^{{\rm IS}_{2}}$ are configurations of the particles at the two ISs. 
\blue{For example consider two sets of particle configurations such that
one configuration is created from the other by a localized plastic deformation 
involving just a few particles. Differences between the two configurations 
on the rest of the system will be 
just smooth relative elastic deformations whose amplitude decay to $0$ at long distances 
from the region where the localized plastic deformation took place \cite{Goldstein}.
Such a pair of configurations, both of which are energy minima,  can be easily generated explicitly for example by athermal quasi-static shear (AQS) processes \cite{maloney-lemaitre-2004a,maloney-lemaitre-2004b,Tanguy2006,maloney-lemaitre-2006,lemaitre-maloney-2006,karmakar-lemaitre-lerner-procaccia-2010,procaccia-2011}. In the thermodynamic limit $N \to \infty$ the 
mean-squared distance between such a pair of states 
which appears on the left hand side of \eq{eq-distance-two-ISs} will vanish. 
Although they are distinct inherent structures which are not accessible 
to each other by continuous elastic deformations, so that each one is out of 
the basin of the other, 
they should be regarded as belonging to the same metastable state since
their thermodynamic properties must be the same.
This is an extreme example but the argument can be repeated 
for the cases involving two, three, ... localized plastic deformations
which are independent from each other.}

Within a given metastable state or metabasin, there must be an
inherent structure which
has the lowest energy, i.~e. the ground state within the metabasin. All other inherent structures are excited states in the metabasin.
As the temperature increases, not only the ground state but also 
the excited states become more populated. \blue{Moreover different 
metastable states may merge together by raising the temperature.
As the result the entropy of the metabsin increases with the temperature.}

We consider that the stress relaxation of deeply supercooled liquids 
may be described within the free-energy landscape picture 
as follows (see Fig.~\ref{fig-two-step-func-model}).
First, we consider that the initial rapid relaxation is mainly due to  
the dynamics in the basins of individual ISs, which are dominated by fast
modes of the purely {\it elastic} dynamics which we discuss in sec. \ref{sec-IS-shearmodulus}. 
\blue{The subsequent slower process in the plateau-region, i.e. the $\beta$-relaxation
would involve possibly two different  mechanisms. One is due to slower  modes of the
elastic dynamics around inherent structures. The other is due to thermally activated 
transitions between different inherent structures in a common metabasin (metastable state).
The latter is  necessarily inelastic ({\it plastic})
deformations as we discuss in sec. \ref{subsubsec-Rigidity-of-metabasin-or-intra-state-rigidity}. 
}
Finally the $\alpha$-relaxation may be attributed to 
thermally activated transitions between different metabasins (metastable states).

The above phenomenological picture implies thermal averaging may be factorized as,
\beq
\langle \ldots \rangle = [\![\langle \ldots \rangle_{\rm MS} ]\!] 
\qquad  \langle \ldots \rangle_{\rm MS} \sim  [ \langle \ldots
\rangle_{\rm IS}  ]_{\rm MB} 
\label{eq-av-factorization}
\eeq
Here $\langle \ldots \rangle_{\rm MS}$ and  $[\![\ldots ]\!]$ represent respectively thermal averaging 
within a metastable state and over different metastable
states. Similarly $\langle \ldots \rangle_{\rm IS}$ and $[ \ldots ]_{\rm MB}$
represent respectively thermal averaging  in the vicinity of an inherent
structure (IS) and over different ISs within a common metabasin (MB) (or metastable state). 
Then thermal fluctuations of a physical quantity, say $O$, could be decomposed into
{\it intra-state} and {\it inter-state} fluctuations as,
\beq
\langle  O^{2} \rangle-  \langle  O \rangle^{2}
=[\![\langle  O^{2} \rangle_{\rm MS}]\!] -  [\![\langle  O \rangle_{\rm MS}]\!]^{2}
=\underbrace{[\![\langle  O^{2} \rangle_{\rm MS} - \langle  O \rangle^{2}_{\rm MS}]\!]}_{\mbox{intra-state fluctuation (initial$+ \beta$)}}
+  \underbrace{[\![\langle  O \rangle_{\rm MS}^{2}]\!] - [\![\langle  O \rangle_{\rm MS}]\!]^{2}}_{\mbox{inter-state fluctuation ($\alpha$)}}.
\label{eq-decompose-fluctuation} 
\eeq
Quite remarkably the cloned liquid method enables disentanglement of the intra-state and
inter-state fluctuations and decompose thermal fluctuations into the two
parts as \eq{eq-decompose-fluctuation} as we discuss later in sec \ref{subsubsec-disentangle}.

Moreover, the intra-state fluctuation itself may be decomposed into
thermal fluctuations around inherent structures
and those between different inherent structures in a common metabasin, 
\beq
\langle  O^{2} \rangle_{\rm MS} - \langle  O
\rangle^{2}_{\rm MS}=
[\langle  O^{2} \rangle_{\rm IS}]_{\rm MB} - [\langle  O
\rangle_{\rm IS}]^{2}_{\rm MB}
=\underbrace{[\langle  O^{2} \rangle_{\rm IS}
-
\langle  O \rangle^{2}_{\rm IS}]_{\rm MB}}_{\mbox{fluctuation around IS (initial)}}
+\underbrace{[\langle  O \rangle^{2}_{\rm IS}]_{\rm MB}
- [\langle  O \rangle_{\rm IS}]^{2}_{\rm MB}}_{\mbox{inter-IS fluctuation ($\beta$)}}
\label{eq-decompose-fluctuation-within-a-metabasin}
\eeq
\blue{Let us note that $\beta$-relaxation may not be entirely due to the thermal activation
between different inherent structures but partly due to slower modes of the elastic fluctuations
around inherent structures as discussed above. How the two different relaxations mechanism are mixed in the dynamics is not clear at the moment.}

\subsubsection{Hierarchy of rigidities}
\label{subsubsec-hierarchy-rigidity}

The hierarchy of the thermal fluctuations naturally implies {\it
hierarchy of rigidities}.
The rigidity as defined by the static fluctuation formula \eq{eq-def-shearmodulus} (or \eq{eq-shearmodulus-again})) 
contains non-affine corrections represented by the correlation function of the shear-stress fluctuations.
By decomposing the latter into several pieces following the above prescription we find,
\begin{eqnarray}
\mu= \underbrace{ [\![ \underbrace{ [ b -N \beta  \left( \langle \sigma^{2} \rangle_{\rm IS}
-   \langle \sigma \rangle^{2}_{\rm IS}\right)]_{\rm MB}  }_{\mu_{\rm IS}}
-N \beta  \left ( [ \langle \sigma \rangle^{2}_{\rm IS}]_{\rm MB}  
-   [  \langle \sigma \rangle_{\rm IS}]_{\rm MB}^{2} \right ) \!]] }_{\hat{\mu}}
-N \beta \left ( [\![ \langle
\sigma \rangle^{2}_{\rm MS}]\!]   -   [\![  \langle \sigma \rangle_{\rm MS}]\!]^{2}\right) 
\label{eq-decompose-mu}
\end{eqnarray}
where we introduced the rigidity of inherent structures $\mu_{\rm IS}$,
\beq
\mu_{\rm IS} \equiv   b -N \beta  \left(  \langle \sigma^{2} \rangle_{\rm IS}
-   \langle \sigma \rangle^{2}_{\rm IS}\right)
\label{eq-mu-IS-def}
\eeq
and the rigidity of metabasin or {\it intra-state} rigidity \blue{$\hat{\mu}$},
\beq
\hat{\mu} \equiv b - N\beta   \left( \langle
\sigma^{2} \rangle_{\rm MS} - \langle \sigma \rangle_{\rm MS} ^{2}\right)
\label{eq-plateau-modulus}
\eeq
with $b$ being the Born term \eq{eq-def-born}.

We speculate that the rigidity of inherent structures $\mu_{\rm IS}$ and that of
metastable states (metabasins) $\hat{\mu}$
corresponds respectively to the upper and lower bound of the plateau region ($\beta$-relaxation) 
of the shear-stress relaxation  (see Fig.~\ref{fig-two-step-func-model}),
\begin{eqnarray}
G_{\infty}  & \simeq & \mu_{\rm IS}=N \beta \langle \sigma \rangle_{\rm IS}^{2}  \label{eq-mu-p-fast-mu-IS}\\
G_{\rm p}  & \simeq & \hat{\mu}=N \beta   \langle \sigma\rangle_{\rm MS}^{2}  \label{eq-mu-p-slow-mu-state}
\end{eqnarray}
Since the initial and $\beta$-relaxations 
weakly depend on whether the system is in equilibrium or out-of equilibrium, we
expect that the quasi-static evaluations of  $G_{\infty}$ and $G_{\rm p}$ 
are valid for the supercooled liquid in equilibrium as well as glasses out-of equilibrium. 

The last equations of \eq{eq-mu-p-fast-mu-IS} and
\eq{eq-mu-p-slow-mu-state} relates the rigidity at different
levels to the variance of the shear-stress fluctuations at the corresponding levels.
It is easy to see that they must hold in liquids. First note that the total rigidity
must vanish $\mu=0$ in liquids. Then note that the stress must
vanish after fully taking the thermal average  
\blue{$\langle \sigma\rangle=[\![\langle \sigma \rangle_{\rm MS} ]\!] = 0$}. Using these two in 
the expression \eq{eq-decompose-mu} one immediately finds the last equations
of \eq{eq-mu-p-fast-mu-IS} and \eq{eq-mu-p-slow-mu-state}.
Note also an analogous relation 
\eq{eq-b-sigma-variance} which reads
\beq
b=\blue{\mu(t=0)=N\beta \langle \sigma^{2} \rangle}.
\eeq
It relates the born term $b$ (instantaneous rigidity) to the variance of the {\it total}
thermal fluctuation of the stress $\langle \sigma^{2} \rangle$.

\subsubsection{Rigidity of inherent structures}
\label{sec-IS-shearmodulus}

Here let us discuss more explicitly
the rigidity $\mu_{\rm IS}$ of inherent structures (IS).
An IS \cite{Stillinger-Weber-1982} is a configuration of the particles in an
energy minimum around which the particles make purely elastic
fluctuations.
In the close vicinity of an IS with the particle configuration
$\{({\bf r}_{\rm IS})_{i}\}$ $(i=1,2,\ldots,N)$, the potential 
part of the Hamiltonian \eq{eq-hamiltonian} is approximated as,
\beq
U=U(\{(x_{i}^{\mu})_{\rm IS}\})+U_{\rm el}(\{(u_{i}^{\mu})_{\rm IS}\})
+ \mbox{unharmonic corrections}.
\label{eq-hamiltonian-IS}
\eeq
Here $U_{\rm el}(\{(u_{i}^{\mu})_{\rm IS}\})$ is the harmonic part of the energy,
\beq
U_{\rm el}(\{(u_{i}^{\mu})_{\rm IS}\})\equiv
\frac{1}{2}\sum_{i,j}\sum_{\mu \nu}
H_{ij}^{\mu \nu} u_{i}^{\mu}u_{j}^{\nu}
\qquad u_{i}^{\mu} \equiv x_{i}^{\mu}-(x_{\rm IS})_{i}^{\mu}
\eeq
with the matrix $H$ being the Hessian matrix,
\beq
H_{ij}^{\mu \nu}
=\delta_{\mu \nu}
\delta_{ij}\sum_{k} 
v^{\mu \mu}(r_{i,k})
-v^{\mu \nu}(r_{i,j})
\qquad v^{\mu \nu}(r_{i,j})\equiv \left. \frac{\partial^{2}v(r_{ij})}{\partial x^{\mu}_{i}\partial x^{\mu}_{j}}\right |_{\rm IS}
\label{eq-hessian}
\eeq
where $\left . \ldots \right |_{\rm IS}$ means to take derivatives at the IS, namely  $x^{\mu}_{i}=(x_{\rm IS})^{\mu}_{i}$.
Then the intra-IS rigidity \eq{eq-mu-IS-def}
can be expressed fully microscopically \cite{Lutsko1989,maloney-lemaitre-2004b,lemaitre-maloney-2006} as,
\beq
\mu_{\rm IS}=b -  \frac{1}{N}\sum_{i=1}^{N} {\bf \Xi}_{i} \cdot ({\cal H}^{-1} {\bf \Xi})_{i},
\label{eq-mu-IS-AQS}
\eeq
where ${\cal H}^{-1}$ is the inverse of the Hessian matrix \eq{eq-hessian}
and the ${\bf \Xi}$ is defined as,
\beq
{\bf \Xi}_{i}=\left. \nabla_{i} \sigma \right |_{\rm IS}
\label{eq-def-xi}
\eeq
with $\sigma$ being the shear-stress \eq{eq-def-stress}
and $\nabla_{i}=(\partial/\partial x_{i},\partial/\partial y_{i},\partial/\partial z_{i},\ldots)$.
More precisely, the born term \eq{eq-def-born}
at the IS can be written as,
\beq
b=\frac{1}{N}\sum_{ij}H^{xx}_{ij} (z_{\rm IS})_{i} (z_{\rm IS})_{j},
\label{eq-born-IS}
\eeq
and the ${\bf \Xi}$ at the IS can be written as,
\beq
\Xi^{\mu}_{i}= \sum_{k}H^{\mu x}_{i k}(z_{\rm IS})_{k} \qquad (\mu=x,y,z).
\label{eq-xi-IS}
\eeq
Within the harmonic approximation to the potential \eq{eq-hamiltonian-IS}, the fluctuation formula of the rigidity at finite temperatures
\eq{eq-def-shearmodulus} reduces exactly to \eq{eq-mu-IS-AQS} \cite{Lutsko1989}.
\blue{Non-harmonic corrections which can be treated perturbatively amount at most to
$O(T)$ terms which vanish in the limit $T \to 0$.}


It is instructive to rewrite the fluctuation formula 
\eq{eq-mu-IS-AQS} in terms of the eigen modes of the Hessian matrix as \cite{maloney-lemaitre-2004b,maloney-lemaitre-2006,lemaitre-maloney-2006},
\beq
\mu_{\rm IS}=b -  \frac{1}{N}\sum_{\lambda} 
\frac{ \Xi_{\lambda}^{2}}{\lambda}.
\label{eq-mu-IS-AQS-mode}
\eeq
Here $\lambda$'s are the eigen values of the Hessian matrix which satisfy
the equations,
\beq
\sum_{(j,\nu)}H_{ij}^{\mu \nu} U_{(j,\nu),\lambda}= \lambda U_{(i,\mu),\lambda}
\label{eq-eigenvalue-equation}
\eeq
with $U_{(i,\mu),\lambda}$ being the normalized eigen vectors.
The eigen values $\lambda$ of the Hessian matrix at the inherent
structures (energy minima) are not negative by definition. In \eq{eq-mu-IS-AQS-mode}
we also introduced the projection of the $\Xi$ field onto the eigen modes,
\beq
\Xi_{\lambda}=\sum_{(i,\mu)}(U^{\dag})_{\lambda,(i,\mu)} \Xi^{\mu}_{i}.
\eeq

The shear-stress relaxation $\mu(t)$ can be studied easily within the
harmonic approximation.
For simplicity we assume the equations of motion with viscous damping,
\beq
m \frac{d^{2} u^{\mu}_{i}}{dt^{2}}+
m\eta \frac{d u^{\mu}_{i}}{dt}=-\sum_{(j,\nu)}H^{\mu \nu}_{ij}u_{j}^{\nu}
\label{eq-motion-elastic}
\eeq
with $m$ and $\eta$ being the mass of a particle and shear-viscosity.
The coupled equations of motions can be transformed to independent equations
of motions for the eigen modes $u_{\lambda}\equiv\sum_{(i,\mu)}(U^{\dag})_{\lambda,(i,\mu)} u^{\mu}_{i}$,
\beq
\frac{d^{2} u_{\lambda}}{dt^{2}}+
\eta \frac{d u_{\lambda}}{dt}=-\omega^{2}(\lambda) u_{\lambda}
\label{eq-motion-elastic-decomposition}
\eeq
where  we introduced characteristic frequencies $\omega$ defined as,
\beq
\omega(\lambda) \equiv \sqrt{\lambda/m}
\eeq
We further simplify the problem considering 
the over-damped limit neglecting the inertial term.
Then the shear-stress relaxation $\mu(t)$ after the step like perturbation
is obtained as,
\beq
\mu(t)=b-\int_{0}^{\infty} d\omega D(\omega)
\frac{\Xi^{2}_{\omega}}{m \omega^{2}}
[1-e^{-(\omega^{2}/\eta) t}].
\label{eq-mu-t-decomposition}
\eeq
where we introduced the distribution function $D(\omega)$ of the characteristic
frequencies $\omega$,
\beq
D(\omega) \equiv \frac{1}{N}\sum_{\lambda} \delta(\omega-\sqrt{\lambda/m}).
\eeq
Note that in the large time limit $t \to \infty$, $\mu(t)$ converges
to the static rigidity $\mu_{\rm IS}$ given by  \eq{eq-mu-IS-AQS-mode} as it should.

The spectrum of eigen values of the Hessian matrix of the ISs is known to 
constitute a continuous band extending down to $\omega=0$ so that
the stress relaxation $\mu(t)$ cannot be a simple exponential relaxation
even in the harmonic approximation.
However convergence of $\mu(t)$ to the asymptotic value $\mu_{\rm IS}$
would be not too slow if the integrand of \eq{eq-mu-t-decomposition}
is not singular in the $\omega \to 0$ limit.
Indeed the numerical analysis of inherent structures of a model structural
glass system in \cite{lemaitre-maloney-2006} 
(See Fig.~5 of the latter reference) suggests this is the case
such that the a large fraction of the whole reduction of the
rigidity from $b$ to $\mu_{\rm IS}$ would be attained
within the order of the microscopic time scales. 
This observation is consistent with our view
that the initial rapid decay of stress-relaxation  (see
Fig.~\ref{fig-two-step-func-model}) is most likely purely elastic
dynamics
in the close vicinity of ISs.


\subsubsection{Rigidity of metabasin}
\label{subsubsec-Rigidity-of-metabasin-or-intra-state-rigidity}

\begin{figure}[h]
\includegraphics[width=0.4\textwidth]{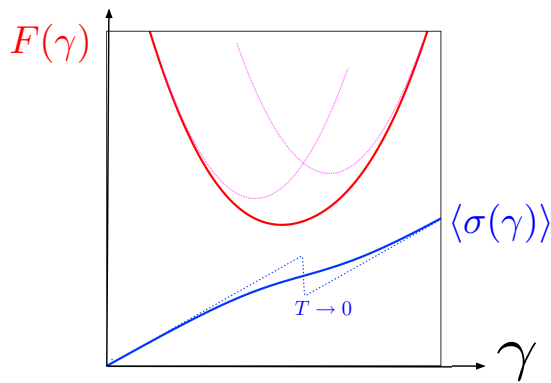} 
\caption{Schematic free-energy landscape of metabasin under external strain and corresponding static stress-strain curve. 
The free-energy $F(\gamma)$ (bold line) and stress $\langle \sigma(\gamma) \rangle =dF(\gamma)/d\gamma$ (bold line ) is plotted against the $\gamma$-axis.
In this example two inherent structures (IS) with rigidity $\mu_{\rm IS}$, represented by the parabolas,  belong to the metabasin. At zero temperature, the level crossing between
the two result in a discontinuous stress drop in the stress strain curve.
At finite temperatures, the singularity at the level-crossing is removed such that the {\it renormalized} rigidity of the metabasin $\hat{\mu}$ can be defined.}
\label{fig-stress-strain-renormalization}
\end{figure}

In sharp contrast to the purely elastic deformations discussed above,
the transitions between different ISs, even within a common MB (See Fig.~\ref{fig-two-step-func-model}), necessarily 
involve some inelastic ({\it plastic}) deformations \cite{Goldstein}.
It is natural to expect that they are thermally excited plastic
deformations, which are mostly independent from each other.
We consider such processes, \blue{together with slower modes of the elastic
dynamics},  contribute to the slower relaxation
around the plateau, i.~e. the $\beta$-relaxation \cite{Stillinger-1995}.

\blue{Let us come back to the example of the pair of IS  which we discussed
at the beginning of sec. \ref{sec-landscape}.} They are very closely related to
each other such that the transition between the two involves only a localized plastic deformation involving 
a few particles. Such transitions between two ISs are already beyond the scope of purely harmonic theories. In such a process,  typically an eigen value of the Hessian matrix becomes negative at the transition \cite{maloney-lemaitre-2004b}.
This means that elasticity as defined in the most strict sense
breaks down {\it immediately} moving just a little bit away from the close vicinity of an IS.

However it is reasonable to expect that some sort of 
{\it renormalized} harmonic theory which takes into account the effects of some weak plasticity should exist at macroscopic level. 
In this respect it is instructive to recall, for example, the fact that the rigidity
of spin waves of ferromagnets can still be defined in the presence of some small densities of 
topological defects such as pairs of vorticies (dislocations).
In the latter case, thermally excited topological defects renormalizes the the rigidity down to a smaller value as the temperature is increased to higher temperatures 
and eventually enables melting of otherwise elastic system \cite{Kosterlitz-Thouelss,Chaikin-Lubenski}.

Our picture for the rigidity of metabasin can be summarized as the following. 
Different ISs within a common metastable state or MB have large enough mutual overlaps in the sense of \eq{eq-distance-two-ISs}.
They are likely to be related to each other by localized plastic
deformations
\cite{Goldstein,maloney-lemaitre-2004a,maloney-lemaitre-2004b,Leonforte,Tanguy2006,maloney-lemaitre-2006,lemaitre-maloney-2006}. 
Among all ISs within a common MB there is an IS which has the lowest
energy, i.~.e the ground state within the MB. All other ISs in the same
MB are excited states, which can acquire larger statistical weights at higher temperatures.
In a sense the dynamics within a MB is ``reversible'' \cite{Goldstein,Doliwa-Heuer-2003,Heuer-2008} because a transition to an excited
state would be often followed by a return to the ground state. From \eq{eq-mu-IS-def} and \eq{eq-plateau-modulus} we find the
intra-state rigidity $\hat{\mu}$ is reduced with respect to the rigidity
of IS $\mu_{\rm IS}$ due to the stress fluctuations between different ISs in the common MB,
\beq
\hat{\mu}=\mu_{\rm IS} - N \beta \left(  \langle \langle \sigma
\rangle^{2}_{\rm IS}\rangle_{\rm MB}
-   \langle \langle \sigma \rangle_{\rm IS} \rangle^{2}_{\rm MB}\right).
\eeq
In our interpretation, the non-affine correction which appears on the
r.~h.~s. is {\it plastic} in sharp contrast to the {\it elastic} non-affine
correction in the basins of inherent structures discussed in sec. \ref{sec-IS-shearmodulus}.
In the low temperature limit, however, the thermal excitations will be
suppressed so that it is natural to expect that the two rigidities become the same,
\beq
\lim_{T \to 0}\hat{\mu}=\lim_{T \to 0}\mu_{\rm IS}.
\label{eq-T-0-mu-mu}
\eeq

In order to illustrate our picture in an alternative way, we show an intuitive free-energy landscape picture 
of a metabasin plotted along the axis of external strain $\gamma$ in Fig.~\ref{fig-stress-strain-renormalization} together with the corresponding {\it static} stress-strain curve. 
Local energy landscape of the basins around each inherent structure should be described well as a parabola with curvature $\mu_{\rm IS}$.
The positions of the bottoms of the parabolas of different ISs  would be slightly displaced from each other. 
At a given $\gamma$, the ISs can ranked according to their energy: ground state, 1st excited state, e.t.c. Along the $\gamma$-axis there can be
level crossings among the ISs, i.~e. plastic deformations.  At zero temperature each level crossing would be reflected on the stress-strain curve as a stress-drop.
However at finite temperatures, the {\it free-}energy landscape and the corresponding stress-strain curve should be rounded. Then it is natural to expect that the {\it free}-energy landscape of the metabasin
as a whole may be described well as a large parabola with curvature
$\hat{\mu}$ which is reduced with respect to $\mu_{\rm IS}$ at finite temperatures. 


\subsubsection{Free-energy landscape at higher temperature}
\label{section-landscape-highT}

Let us finally turn to discuss on the free-energy landscape at higher
temperatures which will provide us useful insights on how glasses melt.
Within the
RFOT \cite{Kirkpatrick-Thirumalai-1987,Kirkpatrick-Wolynes-1987a,RFOT,Cavagna-review,RFOT-review},
the critical behaviours predicted by the mode
coupling theory (MCT)\cite{MCT,MCT2} are considered as
reflections of melting or emergence of amorphous metastable solid 
states.
If this is the case there must be an intimate analogy 
between the MCT criticality and the spinodal criticality of superheated crystals.
In sec. \ref{subsec-melting-orderparameter} we will discuss implication of
this point of view on the rigidity problem.

It is important to note however that, at variance to the spinodal
criticality of superheated crystals,
the MCT criticality is relevant not only in the low temperature side
{\it but also in the high temperature side} above $T_{\rm c}$.
Actually MCT itself found the criticality in the higher temperature side.
This point is very important in supercooled liquids
which exhibit the two-step relaxations as the one shown in
Fig.~\ref{fig-two-step-func-model} not only at temperatures lower than
$T_{\rm c}$ but also at {\it higher} temperatures around $T_{\rm c}$.

The key notion to describe the free-energy landscape of supercooled liquids at $T \to T_{\rm c}^{+}$ 
, corresponding the inherent structures for the lower temperature side, are {\it saddles}.
This has been substantiated by extensive studies of a class of 
solvable mean-field spin-glass models
and realistic model glass forming systems \cite{Kurchan-Laloux,grigera-2002,Cavagna-review}.
The basic result goes as follows. The dynamics of a super-cooled liquid 
at $T \to T_{\rm c}^{+}$ can be viewed as
a point in the phase space going through a narrow channel along which the energy
can be decreased. As the time goes on, the number of eigen modes 
of the temporal Hessian matrix with negative eigen values becomes less and less.
In the large time limit the system ends up in a marginally stable state:
all the eigen values of the temporal Hessian matrix turns out to be
positive but the lowest one is touching zero.  According to this picture, the main 
mechanism of the slow dynamics as $T \to T_{\rm c}^{+}$ is a sort of critical dynamics
distinct from the activated dynamics which becomes relevant at lower temperatures.

The essential difference between saddles and inherent structures are due to
the slow modes with smaller eigen values. The saddles which are relevant for the long time
dynamics as $T \to T_{\rm c}^{+}$ contain a large number of positive eigen values
and a small number of slightly negative eigen values.
Then it is tempting to speculate that a somewhat 
similar elastic description of the stress relaxation as expressed by \eq{eq-mu-t-decomposition}
would be also possible for $T> T_{\rm c}$ for short enough time scales.

Conversely if a supercooled liquid initially at a temperature below $T_{\rm c}$ is heated up to
a temperature above $T_{\rm c}$, the temporal Hessian matrix will acquire a fraction of 
negative eigen-values. Even with one negative eigen value, 
the system will yield by arbitrary weak shear perturbations 
leading to collapse of a metastable amorphous solid state. 
This suggests that the static rigidity of a (temporal) metastable amorphous state 
would vanish approaching $T_{\rm c}$ from below.

\subsection{Connection between the order parameter and rigidity : mean-field points of view}
\label{subsec-melting-orderparameter}

As we discussed in sec \ref{subsubsec-paradox}
emergence of rigidity $\mu >0$ in solids means break down of the commutation of
the two limits, namely the thermodynamic $N \to \infty$
and small shear-strain limit $\gamma \to 0$. 
Thus the rigidity detects the liquid-solid phase transition as sensitively as the
order parameters. Here we show that several mean-field theories 
generically suggest an explicit connection between the rigidity and the order parameter.

In the introduction of the present paper we mentioned the Born's
rigidity crisis scenario \cite{Born} which says that melting of solids
may be signaled by vanishing of the rigidity $\mu$. 
The problem is {\it how} it vanishes: continuously or discontinuously?
Usually Born's conjecture is interpreted as suggesting continuous vanishment.
This is an interesting and much debated question in the case of superheated
crystals \cite{Tallon-1989,wang,sorkin,Jin-2001} approaching their metastability limit at $T_{\rm s}$ \cite{binder-review}. 
The same question is relevant also for supercooled liquids 
and glasses since,
as emphasized in a recent work \cite{KZ2011}, 
the dynamical transition at $T_{\rm d}$ (or the MCT critical temperature
$T_{\rm c}$) can be regarded as an analogue of the spinodal behaviour
of superheated crystals at $T_{\rm s}$.
We show below that some mean-field theories indeed suggest a common, universal behaviour 
of the rigidity approaching the metastability limits of solids 
including both the crystalline and amorphous systems.

In the following we first develop a simple mean-field description of the rigidity of a superheated crystalline system (ferromagnet).
Then we recall the predictions by the MCT on the fate of the rigidity of amorphous solids. 

\subsubsection{Rigidity in a GL theory for superheating}
\label{subsubsec-rigidity-superheated-ferro}

\begin{figure}[t]
\includegraphics[width=0.4\textwidth]{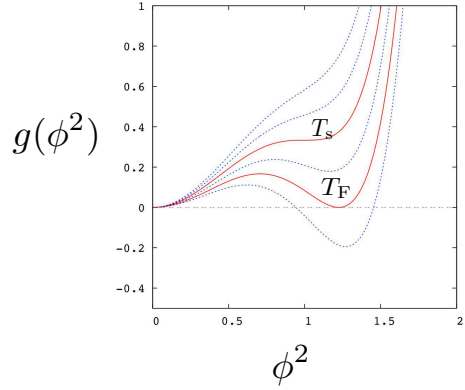}
\caption{A model Landau potential. The potential is defined in \eq{eq-GL-model}.
The temperatures are $T=5/8,6/8,\ldots,10/8$ from the bottom to the top curves.
The curves at the spinodal temperature $T_{\rm s}=1$
and the 1st order phase transition $T_{\rm F}=3/4$ are indicated.
The two minima on the left and right corresponds respectively to the paramagnetic and ferromagnetic state.
The equivalent effective potential of structural glasses and related
spin-glass systems is the so called Franz-Parisi
 potential \cite{Franz-Parisi-1997} for which the horizontal axis should
 be interpreted as the axis for the Edwards-Anderson order parameter $q_{\rm EA}$ or the cage size $A$. In the latter case the two minima on the left and right
corresponds respectively to the paramagnetic state $q=0$ (liquid $A=\infty$) and
 spin-glass state $q > 0$ (glass $A < \infty$). The two temperatures $T_{\rm s}$ and $T_{\rm
 F}$ corresponds respectively to the dynamical transition temperature
 $T_{\rm d}$ (or the MCT transition temperature $T_{\rm c}$) and the
 Kauzmann transition temperature $T_{\rm K}$.
}
\label{fig-spinodal}
\end{figure}

Let us first discuss the fate of the rigidity of superheated crystalline
systems approaching their metastability limits. 
For simplicity we consider Ginzburg-Landau (GL) theory for a 
{\it vectorial} spin system which can realize superheated, metastable
ferromagnetic phase.
By slightly generalizing the usual scalar GL theory
for superheating \cite{binder-review}, 
we consider an $O(2)$ symmetric $\phi^{6}$ theory described by a
free-energy functional,
\beq
F\left[\vec{\phi}\right]=\int d^{d}r \left[
\frac{c}{2} \left(\nabla \vec{\phi}\right)^{2}+g\left(|\vec{\phi}|^{2}\right)\right] \qquad g(y)=T y-y^{2}+\frac{y^{3}}{3}
\label{eq-GL-model}
\eeq
where $\vec{\phi}({\bf r})=(\phi_{1}({\bf r}),\phi_{2}({\bf r}))$ is a
continuous vectorial field in a $d$-dimensional space.
The profile of the Landau potential $g(\phi^{2})$ is shown in Fig.~\ref{fig-spinodal}.

Let us first examine the equilibrium states.
The equilibrium states are uniform in space and parametrized by
the amplitude of the order parameter  $M=|\phi|=\sqrt{\phi_{1}^{2}+\phi_{2}^{2}}$.
At temperatures below the spinodal temperature $T_{\rm s}=1$, there is a metastable ferromagnetic state
with non-zero amplitude of the order parameter $M >0$. Equilibrium 1st order phase transition from the paramagnetic
state $M=0$ to the ferromagnetic state $M >0$ takes place at a temperature
$T_{\rm F}=3/4$ lower than the spinodal temperature $T_{\rm s}=1$.

Upon heating the metastable ferromagnet above $T_{\rm F}$ up to $T_{\rm s}$,
the amplitude of the order parameter $M$ exhibits
a discontinuous jump from a finite value $M > 0$ (metastable
ferromagnetic state) to zero $M=0$ (paramagnetic state ) signaled by a
preceding square-root singularity,
\beq
M(T)-M(T_{\rm s}) \propto \sqrt{T_{\rm s}-T}.
\label{eq-order-parameter-ferro}
\eeq
The mixed character of 1st and 2nd order phase transitions is the generic
feature of the spinodal criticality \cite{binder-review}: 
the order parameter vanishes discontinuously much as usual 1st order equilibrium 
melting transitions but the jump accompanies a preceding
square-root behaviour as $T \to T^{-}_{\rm s}$.

Now let us consider spatial fluctuation of the order parameter. To this
end it is convenient to re-parametrize the order parameter using 'angular variable' $\theta({\bf r})$ as,
\beq
\vec{\phi}({\bf r})=|\vec{\phi}|({\bf r})(\cos \theta({\bf r}), \sin
\theta({\bf r})).
\eeq
Then the free-energy functional \eq{eq-GL-model} can be rewritten as,
\beq
F\left[|\vec{\phi}\|,\theta\right]=
\int d^{d}r  
\left[
\frac{\mu}{2} (\nabla \theta)^{2}
+\frac{c}{2}\left(\nabla |\vec{\phi}|\right)^{2}
+g\left(|\vec{\phi}|^{2}\right)
\right]  
\label{eq-GL-model-2}
\eeq
with
\beq
\mu=c|\vec{\phi}|^{2}.
\label{eq-rigidity-spin-wave-ferro}
\eeq
Note that the last two terms of the integrand in \eq{eq-GL-model-2}
appear also in the usual
scalar GL theory. On the other hand, the 1st term of the integrand can be
naturally interpreted as the 'spin-wave' term. 

An important observation is that the rigidity $\mu$ of spin-waves is 
simply proportional to the square of the order parameter at this
mean-field level \cite{Chaikin-Lubenski}.
In particular, assuming that the amplitude of
the order parameter is uniform in space and takes the equilibrium value $M(T)$ we find,
\beq
\mu(T)-\mu(T_{\rm s}) \propto c\sqrt{T_{\rm s}-T}.
\label{eq-mu-ferro-spinodal}
\eeq
using \eq{eq-order-parameter-ferro} in \eq{eq-rigidity-spin-wave-ferro}.
Thus at the pure mean-field level the rigidity of the spin-wave $\mu$ 
also exhibits a discontinuous behavior preceded by a square-root 
singularity as $T \to T_{\rm s}^{-}$. 

Note that the situation is quite different from the case of the usual second order ferromagnetic
transition (e.~g. $\phi^{4}$ theory) \cite{Chaikin-Lubenski}. 
In the latter case 
the amplitude of the order parameter exhibit a similar square-root behaviour
$M(T) \propto \sqrt{T_{\rm c}-T}$ approaching the critical
temperature $T_{\rm c}$ from below {\it but with no discontinuity at}
$T_{\rm c}$. Then using the latter in \eq{eq-rigidity-spin-wave-ferro} one finds
that the spin-wave stiffness scales as $\mu(T) \propto c(T_{\rm c}-T)$.

\subsubsection{Mode coupling theory}
\label{subsubsec-MCT}

The mode coupling theory (MCT) \cite{MCT,MCT2} predicts that the density auto-correlation 
function $C_{\rho}(\tau)$, such as the self-intermediate scattering function, 
exhibits a two-step relaxation with a plateau similar to the one shown in 
Fig.~\ref{fig-two-step-func-model} in supercooled liquids

and that the ergodocity breaking takes place at the MCT critical temperature $T_{\rm c}$:
the $\alpha$-relaxation time $\tau_{\alpha}$ diverges as $T \to T^{+}_{\rm c}$.
Physically this means emergence of amorphous solid states with the cage structure.
The amplitude of the plateau of the density auto-correlation
function $C_{\rho}(\tau)$ is related to the cage size $A$ which is the order parameter
within the RFOT  (see \eq{eq-cage-size-dynamic}). The MCT predicts that 
the cage size $A$ exhibits a discontinuous jump from $0$ to a finite value 
as $T \to T_{\rm c}^{+}$ followed by a square-root behaviour for $T < T_{\rm c}$,
\beq
A(T_{\rm c})-A(T) \propto \sqrt{T_{\rm c}-T}.
\label{eq-A-sqrt-singularity}
\eeq
Note that the behaviour is very similar to that of the order parameter
of the superheated crystalline system \eq{eq-order-parameter-ferro}.

The cage size $A$ can also be defined in a static matter (see \eq{eq-cage-size-static}).
In the cloned liquid approach \cite{mezard-parisi-1999,coluzzi-mezard-parisi-verrocchio-1999,parisi-zamponi}, it is obtained by extremizing an effective
free-energy which is a function of $A$ (see sec.\ref{subsubsec-molecular-liquid}). The effective free-energy
corresponds to the so called Franz-Parisi potential \cite{Franz-Parisi-1997} which
has essentially the same qualitative features as the Landau potential shown in Fig.~\ref{fig-spinodal}. 
The cloned liquid approach \cite{parisi-zamponi} 
predicts that the minimum of the potential corresponding to the metastable
amorphous solid state $A < \infty$ disappears approaching the dynamical transition temperature
$T_{\rm d}$ from below, i.~e. as  $T \to T_{\rm d}^{-}$.
The dynamical transition temperature $T_{\rm d}$ is considered to corresponds to the MCT critical temperature 
$T_{\rm c}$. The resultant behaviour of the cage size $A$ is the same as that
predicted by the MCT \eq{eq-A-sqrt-singularity}.
\blue{Quantitative comparisons of the results by the dynamic and static
approaches are currently undertaken intensively \cite{ikeda-miyazaki-2010}.}

Within the convectional MCT, the behavior of the the stress-autocorrrelation function
$C_{\sigma}(\tau)$ is obtained by {\it projecting} it onto the density auto-correlation function $C_{\rho}(\tau)$
\cite{nagele-bergenholtz, sciortino-group}. The former is assumed to be
simply proportional to the square of the latter $C_{\sigma}(\tau) \propto C_{\rho}(\tau)^{2}$ 
which follows from the so called factorization approximation \cite{nagele-bergenholtz}. 
As the result the MCT predicts a similar plateau behaviour for the stress relaxation function $\mu(\tau)$
as for the density auto-correlation function $C_{\rho}(\tau)$ : 
the asymptotic plateau value of the the stress-relaxation curve 
$\mu(\tau)=\beta C_{\sigma}(\tau)$ (see \eq{eq-mu-t-stress-stress})
exhibits a discontinuous jump from $0$ to a finite value
$G_{\rm p}(T_{\rm c})$ as $T \to T_{\rm c}^{+}$ followed by a 
square-root behaviour for $T < T_{\rm c}$,
\beq
G_{\rm p}(T_{\rm c})-G_{\rm p}(T) \propto \sqrt{T_{\rm c}-T}.
\label{eq-gp-sqrt-singularity}
\eeq
Note that this is again very similar to the behaviour of the rigidity 
of the superheated crystalline system \eq{eq-mu-ferro-spinodal}.

\subsubsection{Discussion}

The discussions presented above suggest common features of the rigidity
of crystalline and amorphous solids at the mean-field
level. First of all, the rigidity does look like the order parameter: 
it is just the square of the order parameter.
As the result, it vanishes discontinuously preceded by a square-root
behaviour approaching the spinodal like melting temperature from below.
Although the preceding square-root singularity somehow signals the melting,
the discontinuous jump of the rigidity disagrees with the usual interpretations
of the Born's original rigidity crisis scenario \cite{Born,Tallon-1989,wang,sorkin}. 

The phenomenological GL approach like the one presented in sec \ref{subsubsec-rigidity-superheated-ferro} 
is simple and instructive by itself. The local mean-field (Landau) potential $g(\phi^{2})$ (see Fig. \ref{fig-spinodal}) 
has an analogous one for glasses, namely the Franz-Parisi potential \cite{Franz-Parisi-1997} as we mentioned above.
Indeed we demonstrate in Appendix \ref{appendix-spin-wave-vectorSG} that 
this approach can be naturally extended to the case of vectorial spin-glass system using some replica field theoretic formalism, which implies again the same generic behaviour at $T_{\rm d}$: discontinuous spin-wave stiffness preceded by a square-root singularity. 

However in the main part of the present paper (sec \ref{sec-rigidity}), we will analyze the static rigidity of glasses starting from a microscopic level using the fluctuation formula
of the shear-modulus (see sec \ref{subsubsec-static-fluctuation-formula}) rather than working with the phenomenological GL like approach. 
As the result we will obtain the rigidity of the metastable amorphous solid states,
i.~e. the intra-state rigidity $\hat{\mu}$ \eq{eq-plateau-modulus}
in power series of the cage size $A$ as \eq{eq-mu-A} which reads,
\beq
\hat{\mu}(A)=c_{0}-c_{1} A + O(A^{2}).
\eeq
with $c_{0}$ and $c_{1}$ are positive constants which we evaluated explicitly.
Thus we do find an intimate connection between the rigidity and the order parameter.
Based on the result, we will come back to the question on the
fate the rigidity in sec. \ref{subsec-discussion-melting}.

Finally, we should remind ourselves after all that the spinodal like criticality only exists at the pure
mean-field level and that to what extent the signatures of such a criticality remains in realistic systems is an open question \cite{binder-review}.


\subsection{Cloned liquid theory}
\label{subsec-cloned-liquid}

\subsubsection{Cage size: order parameter}
\label{sec-cage-size}

The basic picture of the cloned liquid
theory\cite{mezard-parisi-1999,coluzzi-mezard-parisi-verrocchio-1999,parisi-zamponi}
is to view the amorphous
solid state such that each particle is frozen in a disordered
configuration which looks exactly like a liquid but allowed to fluctuate
around the mean position within a finite range called as the cage size $A$. 
The cage size $A$ is the order parameter of the theory which is
determined self-consistently by extremizing a free-energy function $G(A)$.
The order parameter distinguishes between the liquid state $A=\infty$
and the solid state $A < \infty$. Such an idea has 
been proposed originally in the context of density functional theory for
amorphous solids \cite{Hall-Wolynes-1987,Singh1985a,Kirkpatrick-Wolynes-1987b,Kirkpatrick-Thirumalai-1989}.

An important feature of the order parameter $A$ is that it can be viewed both dynamically and statically \cite{Kirkpatrick-Wolynes-1987b,Kirkpatrick-Thirumalai-1989} 
much as the Edwards-Anderson order parameter of spin-glasses \cite{Edwards-Anderson}.
First suppose that we monitor the dynamics of the system with the Hamiltonian
\eq{eq-hamiltonian} keeping track of the positions of the particles ${\bf r}_{i}(t)$ $(i=1,2,\ldots,N)$ 
as a function of time $t$. Then measuring the mean-squared displacement (MSD)
of the particles, we may detect existence a metastable
state if the MSD tends to saturate to a finite value in a large time limit, 
\begin{eqnarray}
A_{\rm d} \equiv \lim_{t \to \infty}\frac{1}{2Nd}\sum_{i=1}^{N}\langle ({\bf r}_{i}(t)
 -{\bf r}_{i}(0))^{2}\rangle_{\rm ini}.
\label{eq-cage-size-dynamic}
\end{eqnarray}
Here $\langle \ldots \rangle_{\rm ini}$ is the average over diffident
initial conditions. Indeed the mode coupling theory (MCT) \cite{MCT,MCT2} has detected a transition from the liquid phase
$A_{\rm d} = \infty$ to a metastable solid phase $A_{\rm d}
< \infty$ at the MCT critical temperature $T_{\rm c}$. 
Alternatively the same cage size can also be obtained in a static way 
using two {\it replicas}, say $a$ and $b$, which have no mutual interactions but
obey exactly the same Hamiltonian \eq{eq-hamiltonian},
\begin{eqnarray}
A_{\rm d}=A \equiv \frac{1}{2Nd} \sum_{i=1}^{N}\langle ({\bf r}_{i}^{a}-{\bf r}_{i}^{b})^{2} \rangle
\label{eq-cage-size-static}
\end{eqnarray}
where ${\bf r}^{a}_{i}(t)$ and ${\bf r}^{b}_{i}(t)$ are particle
configuration of the replica $a$ and $b$ respectively.

\subsubsection{Mean-field scenario}
\label{subsubsec-mean-field-picture}

At the mean field level, the RFOT theory
\cite{Kirkpatrick-Thirumalai-1987,Kirkpatrick-Wolynes-1987a,RFOT,RFOT-review} assumes the following idealized features.
First we assume that there are metastable states $\alpha=1,2,\ldots$ with free-energy $N f_{\alpha}$.
The metastable states are characterized by the cage size $A$ discussed above: 
1) if we prepare the system in any one of the metastable states, the
system remains there such that $A_{\rm d}$ as defined in \eq{eq-cage-size-dynamic} is finite
and 2) if we prepare two replicas in that metastable state, they remain
there  such that $A_{\rm s}$ as defined in \eq{eq-cage-size-static} is finite
which takes the same value as $A_{\rm d}$. Then the total free-energy of the supercooled liquid state
and the ideal glass state can be written formally as,
\beq
-\beta F(T)=\log Z \qquad Z=\sum_{\alpha} e^{-\beta N f_{\alpha}}=
\int df e^{-N(\beta f-\Sigma(f,T))}.
\label{eq-TAP-free-energy}
\eeq
In the last equation the complexity (structural entropy)  $\Sigma(f,t)$ is introduced as,
\beq
N\Sigma(f,t) \equiv \log \sum_{\alpha} \delta (f-f_{\alpha}(T)).
\label{eq-complexity-def}
\eeq
We assume that there are exponentially large number of metastable states in the supercooled and glass phase such that the complexity is finite.
Apparently this description is the same as the free-energy landscape picture based on the notion of inherent structure \cite{Stillinger-1995,Stillinger-Weber-1982}.
However, as we noted in sec. \ref{sec-landscape}, the metastable
state in the context of the mean-field theory corresponds to {\it metabasins} rather than individual inherent structures. 

At the dynamical transition $T_{\rm d}$, the complexity $\Sigma >0$ emerges. Usually the complexity  $\Sigma(f,T)$ is a function of $f$
which is convex upward and vanishes continuously approaching some 
$f_{\rm min}(T)$ from above. (See Fig.~\ref{fig-m-T-diagram-complexity} b))
The total free-energy of the system below $T_{\rm d}$ is given by the saddle point value
of the free-energy $f$ which dominate the integral in the 2nd equation of \eq{eq-TAP-free-energy}. Similarly, the structural entropy is given by the value of the complexity at the saddle point. Upon lowering the temperature, the free-energy and the complexity at the saddle point
decreases. Finally the structural entropy vanishes 
at the Kauzmann transition temperature $T_{\rm K}$, i.~.e the entropy crisis.
To summarize, apparently different groups of metastable states become thermodynamically relevant at different
temperatures in the supercooled liquid state $T > T_{\rm K}$ while
the equilibrium statistical weight condenses onto a certain amorphous ``ground state'' at $T < T_{\rm K}$.

The scenario described above is known to hold precisely in a class of mean-field
spin-glass models \cite{REM,Kirkpatrick-Thirumalai-1987,Kirkpatrick-Wolynes-1987a,crisanti-sommers-1992,Cavagna-review-sg} in which the states $\alpha=1,2,\ldots$ correspond to
the solutions of the equation of states called as the Thouless-Anderson-Palmer equations.
For the structural glass, the cloned liquid approach substantiated the scenario starting from
microscopic Hamiltonian as we sketch below.

\subsubsection{Cloning}
\label{subsubsec-cloning}

Now let us recall the idea of {\it cloning}
\cite{monasson-1995,Franz-Parisi-1997,mezard-parisi-1999}. We consider
$m$ replicas (copies) $a=1,2,\ldots,m$ of the original system which stay
in the {\it same} metastable state (thus the name 'cloning'). The free-energy of the cloned system can be written formally as,
\beq
-\beta m F_{m}(T)=\log Z_{m}
\qquad
Z_{m}=\sum_{\alpha} e^{-m\beta N f_{\alpha}(T)}=\int df e^{-N(m\beta f-\Sigma(f,T))}.
\label{eq-TAP-clone}
\eeq
Note that there is only one summation over the states $\alpha$ instead
of $m$ summations which is because we are assuming here that different
replicas are staying in the same metastable state. We will recall
shortly later (sec. \ref{subsubsec-molecular-liquid}) how to realize such a state in practice.

\begin{figure}[h]
\includegraphics[width=0.4\textwidth]{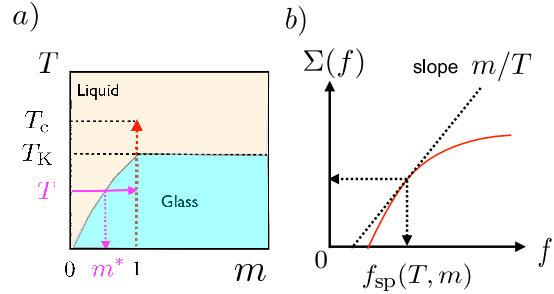}
\caption{Cloned liquid method: 
a) schematic $m-T$ phase diagram b) complexity
}
\label{fig-m-T-diagram-complexity}
\end{figure}

Now we may regard $m$ as a variational real number.  
Then we can notice that the saddle point $f_{\rm sp}(T,m)$ which dominate the integral in the 2nd
equation of \eq{eq-TAP-clone} can be moved around by varying $m$ {\it without}
changing the temperature $T$ itself (See Fig.~\ref{fig-m-T-diagram-complexity} b)).
As we increase $m$, the saddle point $f_{\rm sp}(T,m)$ becomes smaller 
and at some critical $m^{*}(T)$, the complexity at the saddle point \eq{eq-complexity}
vanishes, just like the physical ($m=1$) Kauzmann transition. This observation
suggests that even at very low temperatures below the putative Kauzmann
transition temperature $T_{\rm K}$,  we may choose sufficiently small
$m$ such that the cloned system as a whole still remain effectively in a
liquid phase
where the structural entropy is still finite (See Fig.~\ref{fig-m-T-diagram-complexity} a)).
In the latter regime, we may compute the free-energy $F_{m}(T)$ of the
whole system regarding it simply as a liquid using the conventional
liquid theory \cite{hansen-mcdonald} as we recall shortly later.
Then the value of the complexity can be evaluated as  \cite{monasson-1995},
\beq
\Sigma(f_{\rm sp}(T,m),T)=\frac{1}{N}\frac{m^{2}}{T}\frac{\partial
F_{m}(T)}{\partial m} 
\label{eq-complexity}
\eeq
The whole profile of the complexity curve $\Sigma(f,T)$ at a fixed temperature $T$
can be ``scanned'' (See Fig.~\ref{fig-m-T-diagram-complexity} b)) by varying $m$ by which the position of the saddle
point $f_{\rm sp}(T,m)$,
\beq
f_{\rm sp}(T,m)=\frac{1}{N} \frac{\partial (m  F_{m}(T))}{\partial m} 
\eeq
can be varied at will.
Now we can locate the critical point $m^{*}(T)$ where the structural entropy vanishes 
\beq
\Sigma(f_{\rm sp}(T,m^{*}),T)=0
\label{eq-m-star}
\eeq
for each temperature $T$. Further increase of $m$ does not change the
saddle point value of the free-energy. This implies the value of the
free-energy in the glass phase $T < T_{\rm K}$ is given by,
\beq
\lim_{m \to 1^{-}} F_{m}(T)= F_{m^{*}}(T).
\label{eq-m-1-star}
\eeq
The Kauzmann temperature $T_{\rm K}$ is located by (See Fig.~\ref{fig-m-T-diagram-complexity} b)),
\beq
m^{*}(T_{\rm K})=1
\label{eq-Tk}
\eeq  
At higher temperatures $T> T_{\rm K}$, $m^{*}=1$.

Thus the system in the glass phase $T < T_{\rm K}$  can be viewed as a
liquid at an effective temperature,
\beq
T^{*}=\frac{T}{m^{*}(T)}
\label{eq-Teff}
\eeq
Typically it is found that $m^{*}(T) \sim T/T_{\rm K}$ so that 
$T^{*}=T/m^{*}(T) \sim T_{\rm K}$ \cite{mezard-parisi-1999,coluzzi-mezard-parisi-verrocchio-1999}.
This implies configuration of the system below $T_{\rm K}$ is almost
like that of the liquid at $T_{\rm K}$ independently of the temperature $T$.
On the other hand, in the supercooled liquid state $T > T_{K}$,
$m^{*}(T)=1$ so that $T^{*}=T$. 

\subsubsection{Disentangle intra-state and inter-state responses}
\label{subsubsec-disentangle}

Let us make here a general remark that the cloning method allows disentanglement of
generic linear response functions into intra-state and inter-state responses
much as anticipated by the phenomenological free-energy landscape picture discussed in
sec. \ref{sec-landscape} (See \eq{eq-decompose-fluctuation}). 
In the present paper we fully makes use of this feature to analyze the rigidity of structural glasses.
This is a strong advantage of the use of the replica method because the
two responses are physically quite different in glassy systems: $\beta$ and $\alpha$ relaxations as we
discussed in sec. \ref{subsubsec-two-step-relaxation}. 
In Appendix \ref{appendix-p-spin} we demonstrate the method works
exactly in a typical mean-field spin-glass model.

Suppose that we have a physical observable of our interest, say $O$,
which can be coupled to a conjugate external field $h$ so that the Hamiltonian of the system can be written as,
\beq
H=H_{0}-h O,
\eeq
where $H_{0}$ is the original Hamiltonian of the system. Suppose that there are many metastable states $\alpha=1,2,\ldots$ such that the free-energy of the system can be written 
as \eq{eq-TAP-free-energy}, which reads,
\beq
-\beta F(T,h)=\log Z \qquad Z=\sum_{\alpha} e^{-N \beta f_{\alpha}(T,h)}.
\eeq
The static susceptibility,
\beq
\chi\equiv \left. \frac{1}{N} \frac{\partial \langle O \rangle}{\partial h} \right |_{h=0}
=\left. \frac{1}{N} \frac{\partial^{2} F(T,h)}{\partial h^{2}}  \right |_{h=0}
=\frac{\beta}{N} \left( \langle O^{2}\rangle - \langle O\rangle^{2} \right),
\label{eq-chi-generic}
\eeq
can be formally decomposed as,
\beq
\chi=\hat{\chi}+\tilde{\chi}
\label{eq-chi-decomposition}
\eeq
with
\beq
\hat{\chi} \equiv 
[\![ \chi_{\alpha} ]\!] \qquad 
\tilde{\chi} \equiv 
\frac{\beta}{N} \left([\![O^{2}_{\alpha}]\!]-[\![O_{\alpha}]\!]^{2} \right)
\label{eq-def-intra-inter-state-susceptibility}
\eeq
where we introduced averaging over different metastable states,
\beq
[\![\ldots ]\!] \equiv \sum_{\alpha}P_{\alpha}(h)\ldots \qquad P_{\alpha}(h)\equiv \frac{e^{-N \beta f_{\alpha}(T,h)}}{Z}
\eeq
and defined thermal averages within metastable states,
\beq
O_{\alpha} \equiv \langle O \rangle_{\alpha}=N \left. \frac{\partial f_{\alpha}(h)}{\partial h}  \right |_{h=0}
\qquad 
\chi_{\alpha}=\frac{\beta}{N} \left(
\langle O^{2} \rangle_{\alpha}
-\langle O \rangle^{2}_{\alpha}
 \right)
=\left. \frac{\partial^{2} f_{\alpha}(h)}{\partial h^{2}} \right |_{h=0}
\label{eq-intrastate-averages}
\eeq
Here $\langle \ldots \rangle_{\alpha}$ stands for a thermal average with
respect to thermal fluctuations within the metastable state $\alpha$.
By inspecting the 1st and 2nd term in the decomposition of the
susceptibility \eq{eq-chi-decomposition}, 
namely $\hat{\chi}$ and $\tilde{\chi}$
which are defined in \eq{eq-def-intra-inter-state-susceptibility},
one can realize that they can be naturally interpreted as
intra-state and inter-state susceptibilities respectively. 

Now let us turn to consider linear response of a {\it cloned} system. The Hamiltonian of the replicated system with replicas ($a=1,2,\ldots,m$) can
be written as,
\beq
H_{m}=\sum_{a=1}^{m} \left[ (H_{0})^{a}-h^{a} O^{a} \right].
\eeq
Here we have put probing field $h^{a}$ on each replica
($a=1,2,\ldots,m$). The free-energy of the cloned system reads,
\beq
-\beta m F_{m}(T,\{h^{a}\})= \log Z_{m}(T,\{h^{a}\}) \qquad Z_{m}(T,\{h^{a}\})=\sum_{\alpha}e^{-N \beta \sum_{a=1}^{m}f_{\alpha}(T,h^{a})}
\label{eq-F-clone-field}
\eeq
Note that different replicas stay in the same metastable state $\alpha$ but subjected to different probing fields $h^{a}$ ($a=1,2,\ldots,m$). Then we find a generic susceptibility of the cloned system,
\beq
\chi_{ab} \equiv \left. \frac{m}{N}\frac{\partial^{2}
F_{m}(T,\{h^{a}\})}{\partial h^{a}\partial h^{b}} \right |_{\{h^{a}=0\}}
=\frac{\beta}{N} \left( \langle O^{a}O^{b}\rangle 
- \langle O^{a}\rangle \langle O^{b}\rangle \right)
=\hat{\chi}_{m}\delta_{ab}+\tilde{\chi}_{m}
\label{eq-chi-decomposition-clone}
\eeq
with
\beq
\hat{\chi}_{m} \equiv 
[\![ \chi_{\alpha} ]\!]_{T,m} \qquad 
\tilde{\chi}_{m} \equiv 
\frac{\beta}{N} \left([\![O^{2}_{\alpha}]\!]_{T,m}-[\![O_{\alpha}]\!]_{T,m}^{2} \right)
\eeq
where
\beq
[\![\ldots ]\!]_{T,m} \equiv \sum_{\alpha}P_{\alpha}(h,T,m)\ldots \qquad P_{\alpha}(h,T,m)\equiv \frac{e^{-N m\beta f_{\alpha}(T,h)}}{Z_{m}} \qquad
Z_{m}=\sum_{\alpha}e^{-N m\beta f_{\alpha}(T,h)}
\eeq

Here $\hat{\chi}$ and $\tilde{\chi}$ are nothing but the intra-state and inter-state
susceptibilities defined in
\eq{eq-def-intra-inter-state-susceptibility} but at the effective 
temperature $T/m$.
By taking trace over the replica index we find, 
\beq
\sum_{b=1}^{m}\chi_{ab}=\hat{\chi}_{m}+m\tilde{\chi}_{m}.
\eeq
In the limit $m \to 1$ this becomes nothing but the total susceptibility
of the original system (see \eq{eq-chi-decomposition}),
\beq
\chi=\lim_{m \to 1^{-}} \sum_{b=1}^{m}\chi_{ab}=\hat{\chi}_{m^{*}(T)}+m^{*}(T)\tilde{\chi}_{m^{*}(T)}.
\label{eq-total-response}
\eeq
Here we are assuming that we evaluate the terms on the
r.~h.~s. in the liquid state (See \eq{eq-m-1-star}).

Let us list below some observations and remarks,
\begin{itemize}

\item In the cloned ``liquid'' system, the total susceptibility
      \eq{eq-total-response} is just that of a (molecular) liquid
      state at temperature $T^{*}=T/m^{*}$.

\item By comparing \eq{eq-chi-decomposition-clone} and
      \eq{eq-chi-decomposition} we notice that replica trick allows us
      to disentangle the intra-state ($\beta$-relaxation) and
      inter-state response ($\alpha$-relaxation) hided in the (cloned)
      liquid:  to find the intra-state susceptibility $\hat{\chi}$ one
      simply needs to look at the coefficient of the terms with
      $\delta_{ab}$ in the susceptibility matrix $\chi_{ab}$.

\item Note that the decomposition can be done
      not only in the glass phase where $m=m^{*}(T)< 1$ but also at higher
      temperatures where $m=m^{*}(T)=1$ as long as the metastable states
      (with finite cage size $A$) exist. This is made possible by introducing
      $m$ clones and taking $m \to 1$ limit afterwards (see Appendix \ref{appendix-p-spin}).

\item  In the standard interpretation of the RSB solution for the glass phase \cite{Parisi-1983,Cavagna-review-sg}, which is based on solvable mean-field spin-glass models (
       see Appendix \ref{appendix-p-spin}), the factor $m^{*}(T) ( \leq 1))$ which appears in front of $\tilde{\chi}$ in \eq{eq-total-response} is understood as the probability 
       that two independent replicas stay at different metastable states
       (here the replicas are not the 'cloned' ones).
       In a more physical term, it roughly corresponds to the
       probability that a given system stay at an low lying 
       'excited' metastable
       state rather than 'ground' metastable state in the glass
       phase. The parameter $m^{*}(T)$ is decreasing function of the
       temperature $T$ and typically behaves as $m \sim T/T_{\rm K}$ \cite{REM}.

\end{itemize}

The idea presented above is generic but needs to be implemented more explicitly.
In sec. \ref{sec-recipe-disentangle}, we present a recipe to
do realize the decomposition using the cage expansion.

\subsubsection{Molecular liquid}
\label{subsubsec-molecular-liquid}

Now we recall a microscopic implementation of the idea of the cloned liquid \cite{mezard-parisi-1999}.
The idea is to consider a molecular liquid state: a liquid with an assembly of 
hypothetical ``molecules'' $i=1,2,\ldots,N$ with the center of mass (CM)
at ${\bf R}_{i}$, which itself consists of original particles of
different replicas $a=1,2,\ldots,m$ at ${\bf r}_{i}^{a}$ . The task goes
as follows. First one put an artificial confining potential to a create molecular state: we start from a 'forced' cloned system whose free-energy  \eq{eq-TAP-clone} is given by,
\beqa
-\beta m F_{m}= \log Z_{m} \qquad 
Z_{m} = \frac{1}{(N!)^{m}}
\int \prod_{i=1}^{N}\prod_{a=1}^{m} \frac{d^{d}r^{a}_{i}}{\Lambda^{d}} e^{-\beta H_{m}} 
\eeqa
where the potential part of the Hamiltonian given by,
\beq
H_{m}=\sum_{a=1}^{m}\sum_{i < j} v(r^{a}_{ij}) 
-\frac{1}{4\alpha}\sum_{i=1}^{N}\sum_{a,b=1}^{m}|{\bf r}_{i}^{a}-{\bf r}_{i}^{b}|^{2}.
\eeq
Here the 1st term on the r.h.s is the two-body interaction potential \eq{eq-hamiltonian}.
The 2nd term is introduced to confine the particles in different replicas 
to develop a 'forced' molecular state. 
The cage size $A$ of the resultant system is defined as,
\beq
\frac{m(m-1)}{2}Nd
A \equiv \frac{1}{4}\sum_{i=1}^{N}
\sum_{a,b=1}^{m}\langle ({\bf r}^{a}_{i}-{\bf r}^{b}_{i})^{2}\rangle
=\frac{\partial (\beta m F_{m}(\alpha))}{\partial (1/\alpha)}.
\eeq
After computing the free-energy $F_{m}(\alpha)$, a Legendre transformation,
\beq
\beta m G_{m}(A)= {\rm ext}_{\alpha} \left\{ \beta m F_{m}(\alpha) - \frac{m(m-1)}{2}Nd \frac{A}{\alpha} \right\}
\label{eq-legendre-transform}
\eeq
yields a free-energy function $G(A)$. Here ${\rm extr.}_{\alpha}$ stands
for extrimization with respect to $\alpha$. Subsequent extremization of the latter with respect to the cage size $A$,
\beq
0= \left. \frac{\partial G_{m}(A)}{\partial A} \right
|_{A=A_{\rm extr}}
\label{eq-G-A-ext}
\eeq
yields the cage size $A=A_{\rm extr}$ of the desired cloned liquid state, which remains after switching-off the artificial confining potential $1/\alpha \to 0$. Then the complexity
can be computed using \eq{eq-complexity}. The final step is to find $m^{*}(T)$ at which 
the complexity vanishes (See \eq{eq-m-star}). For clarity we recall the
basic steps of the computation \cite{mezard-parisi-1999,coluzzi-mezard-parisi-verrocchio-1999}
in the following. Especially the parameter $\alpha=\alpha(A)$ determined
below as \eq{eq-variance-molecular-1st} is indispensable in our analysis
of the rigidity of glasses in sec.~\ref{sec-rigidity}.

Anticipating that thermal fluctuations of the molecular
liquid can be factorized into 
two parts: thermal fluctuations inside molecules and the thermal fluctuations
of the CM positions of the molecules,  
we decompose the positions of the particles as,
\beq
{\bf r}^{a}_{i}={\bf R}_{i}+{\bf u}_{i}^{a}
\label{eq-molecular-coordinate}
\eeq
where ${\bf u}_{i}^{a}$ are the fluctuations around the CM position of
the molecules at ${\bf R}_{i}$. Then the integration variables can be changes as
$
\int \prod_{i=1}^{N}\prod_{a=1}^{m}d^{d}r^{a}_{i} 
=\int \prod_{i=1}^{N}d^{d}R_{i} \prod_{a=1}^{m} d^{d}r^{a}_{i} \delta^{d} (R_{i}-\frac{1}{m}\sum_{a=1}^{m}r_{i}^{a})
=\int \prod_{i=1}^{N}d^{d}R_{i} \prod_{a=1}^{m} d^{d}u^{a}_{i} m^{d}\delta^{d} \left(\sum_{a=1}^{m}u_{i}^{a}\right)  \nonumber
$. 
The averages over thermal fluctuations inside molecules is given by,
\beq
\langle \ldots \rangle_{\rm cage}\equiv \int \prod_{i=1}^{N}\prod_{a=1}^{m} d^{d}u_{i}^{a}
P({\bf u_{i}}^{a}) \ldots \qquad 
P(\{{\bf u}^{a}\}) \equiv\left( \frac{(2\pi \alpha)^{m-1}}{m} \right)^{-d/2}
\delta^{d} \left(\sum_{a=1}^{m}{\bf u}^{a}\right)
e^{-\frac{1}{4 \alpha}\sum_{a,b=1}^{m}
|{\bf u}^{a}-{\bf u}^{b}|^{2}}
\label{eq-pdf-molecular}
\eeq
which yields in particular,
\beq
\langle (u_{i}^{a})^{\mu}(u_{j}^{b})^{\nu} \rangle_{\rm cage}=-(1-m\delta_{ab})\frac{\alpha}{m^{2}}\delta_{ij}\delta_{\mu\nu}.
\label{eq-variance-molecular}
\eeq
Assuming first $\alpha$ is small so that the fluctuations $u$'s are small one finds,
\beqa
&& Z_{m}=\frac{1}{N!}
 \sqrt{
\frac{(2\pi \alpha/\Lambda^{2})^{m-1}}{m^{m}}
}^{Nd}m^{dN}
\int \prod_{i=1}^{N}\frac{d^{d}R_{i}}{\Lambda^{d}}
 \prod_{i < j} 
\exp  \left [ 
  -\beta m v_{\rm eff}(R_{i}-R_{j}) + \ldots \right]
\label{eq-Zm}
\eeqa
where we introduced the effective potential,
\beq
v_{\rm eff}(r)=v(r)-(1-m)\frac{\alpha}{m^{2}}\nabla^{2}v(r)+\ldots
\label{eq-veff}
\eeq 
where $\nabla=(\partial/\partial x,\partial/\partial y,\partial/\partial z)$.
This implies averages over thermal fluctuation of the CM positions of
 the molecules can be written as,
\beq
\langle \ldots \rangle_{T/m,[v_{\rm eff}(r)]} \equiv\int \prod_{i} dR_{i} W(\{R_{i}\})_{T/m,[v_{\rm eff}(r)]}
\ldots
\label{eq-averaging-effective}
\eeq
with
\beq
W(\{R_{i} \})_{T,[v(r)]} =\frac{
e^{- \beta \sum_{i < j}v(R_{ij})}}{\int \prod_{i=1}^{N} d^{d}R_{i}
 e^{  -\beta \sum_{i < j} v(R_{ij})}}.
\label{eq-averaging-effective-W}
\eeq 

Now the free-energy of the molecular liquid can be expressed as,
\beq
-\beta m F_{m}(\alpha)= 
N\log \left(
 \sqrt{
\frac{(2\pi \alpha/\Lambda^{2})^{m-1}}{m^{m}}
}^{d}
m^{d} \right) 
-\beta m F_{\rm liq}(T/m)[v_{\rm eff}(r)]
+\ldots
\label{eq-free-energy-1st}
\eeq
\blue{Here the 1st term on the r.h.s is due to thermal fluctuations inside the
molecules. The 2nd term  $F_{\rm liq}(T/m)[v(r)]$ is the free-energy
associated with the configurations of the CM positions of the molecules
which turns out to be just the free-energy 
of a liquid at temperature $T^{*}=T/m$,}
\beq
-\beta m F_{\rm liq}(T/m)[v(r)] \equiv \log Z_{\rm liq}(T/m)[v(r)] \qquad Z_{\rm liq}(T)[v(r)] \equiv \frac{1}{N!}
\int \prod_{i=1}^{N}\frac{d^{d}R_{i}}{\Lambda^{d}}
e^{-\beta  \sum_{i< j}v(R_{i}-R_{j})} 
\label{eq-free-ene-liquid}
\eeq
Assuming $\alpha$ is small we find,
\beq
-\beta m F_{\rm liq}(T/m)[v_{\rm eff}(r)]=
-\beta m F_{\rm liq}(T/m)[v(r)]
+\beta (1-m) \frac{\alpha}{m^{2}}\sum_{i <
j} \langle \nabla^{2} v(R_{ij}) \rangle_{*}+\ldots
\eeq
where we introduced a short hand notation,
\beq
\langle \ldots \rangle_{*}\equiv \langle \ldots \rangle_{T/m,[v(r)]}
\label{eq-pdf-CM}
\eeq

The resultant free-energy of the molecular state 
after the Legendre transformation \eq{eq-legendre-transform}
is obtained up to 1st order in the  expansion of the cage size $A$ as,
\beq
\frac{\beta}{N} G_{m}(A)=-\frac{d}{2m} \log(m)+\frac{d(1-m)}{2m}+\frac{d(1-m)}{2m}\log(2\pi
A/\Lambda^{2})- A \frac{1-m}{m}\frac{1}{N}\sum_{i < j} \beta \langle \nabla^{2}
v(r_{ij}) \rangle_{*}
+\frac{\beta}{N} F_{\rm liq}(T/m)
+O(A^{2})
\label{eq-G-1storder}
\eeq
More precisely the Legendre transformation is  done as follows: we
expand the parameter $\alpha$ in power series of $A$ as  \cite{mezard-parisi-1999},
\beq
\alpha=\alpha_{1}A+\alpha_{2}A^{2}+\alpha_{3}A^{3}\ldots
\label{eq-alpha-expansion}
\eeq
The coefficients $\alpha_{1}$ and $\alpha_{2}$ are fixed by the 1st order cage expansion while $\alpha_{3}$ requires the 2nd order cage expansion.

Since we consider $m \leq 1$, we have to maximize (rather than minimize)
the free-energy $G_{m}(A)$ with respect to $A$. The equation
$0 = \partial G_{m}(A)/\partial A $ \eq{eq-G-A-ext} reads,
\beq
0=\frac{1}{A}-\frac{2\beta}{d} \frac{1}{N} \sum_{ i < j}
\langle \nabla^{2}v(r_{ij})\rangle_{*} +O(A).
\label{eq-sp-A}
\eeq
Then we find the cage size of the metastable glassy state
\blue{at the level of the 1st order cage expansion} as,
\beq
A=\frac{d}{2\beta \frac{1}{N} \sum_{ i < j}
\langle \nabla^{2}v(r_{ij})\rangle_{*}} .
\label{eq-cage-size-1st}
\eeq

Finally the glass transition line $m=m^{*}(T)$ (See Fig.~\ref{fig-m-T-diagram-complexity}) is determined following the prescription discussed in sec. \ref{subsubsec-cloning}.
To this end we use $F_{m}(\alpha(A))$ obtained above in
\eq{eq-complexity} and locate $m=m^{*}(T)$ where the complexity $\Sigma$
vanishes. The Kauzmann transition 
temperature $T_{\rm K}$ is determined by \eq{eq-Tk}.

\subsubsection{The effective Einstein model: thermal fluctuations of molecular liquid}
\label{sec-fluctuation-molecular-liquid}

The analysis by the cage expansion implies the molecular liquid state can be described as follows.
The coordinates of the particles are decomposed as \eq{eq-molecular-coordinate} which reads,
\bmat
{\bf r}^{a}_{i}={\bf R}_{i}+{\bf u}_{i}^{a},
\emat
where ${\bf R}_{i}$ is the CM position of the molecule and ${\bf u}_{i}$ is the deviation of the particle belonging to the $a$-th replica from the CM.
The statistical weights of the configurations of the CM are given by \eq{eq-averaging-effective} which is that of a liquid with renormalized potential $v_{\rm eff}(r)$
given by \eq{eq-veff}. On the other hand, the statistical weights of the configuration of the displacements inside the molecules are given by \eq{eq-pdf-molecular}).
Therefore the thermal average in the molecular liquid state becomes factorized as,
\begin{eqnarray}
\langle \ldots \rangle = \langle \langle \ldots \rangle_{\rm cage}\rangle_{T/m,[v_{\rm eff}(r)]} 
=\int \prod_{i=1}^{N} d^{d}R_{i}W_{T/m,[v_{\rm eff}(r)]}(\{R_{i}\})
\prod_{a=1}^{m} d^{d}u_{i}^{a} P(\{ {\bf} u^{a}_{i}\})
\label{eq-average-clonedliquid}
\end{eqnarray}
where $\langle \ldots \rangle_{\rm cage}$ 
is the average over thermal fluctuations inside the molecules  defined in \eq{eq-pdf-molecular}) 
and $\langle \ldots \rangle_{T/m,[v_{\rm eff}(r)]}$ is the average over thermal fluctuations of the CM positions of the molecules defined in \eq{eq-averaging-effective}.

The key parameter which defines the renormalized potential $v_{\rm eff}(r)$ and the width of the fluctuations inside molecules is the parameter $\alpha$.
Going back to \eq{eq-alpha-expansion} we find the parameter $\alpha=\alpha(A)$ as,
\beq
\frac{\alpha}{m^{2}}=2\frac{A}{m}+O(A^{3})=\frac{d}{\beta^{*} \frac{1}{N}\sum_{i < j}\langle \nabla^{2}v(r_{ij}) \rangle_{*}}+O(A^{3})
\label{eq-variance-molecular-1st}
\eeq
where the cage size $A$ is given by \eq{eq-cage-size-1st}.
This follows from the fact that $\alpha_{1}A=mA$ and $\alpha_{2}A^{2}=mA$  (the latter can be seen using Eq.(38) and (40) of \cite{mezard-parisi-1999}).
The result implies in particular that,
\beq
\langle (u^{a}_{i})^{\mu}(u^{b}_{j})^{\nu}  \rangle_{\rm cage}
= \langle u^{2}  \rangle_{\rm cage} \delta_{ij}\delta_{\nu \mu}(m\delta_{ab}-1)
\qquad 
\langle u^{2}  \rangle_{\rm cage}
\equiv \frac{d}{\frac{\beta^{*}}{N}\sum_{i < j} \langle \nabla^{2}v(r_{ij}) \rangle_{*}}=\frac{k_{\rm B}T^{*}}{\kappa_{\rm eff}}
\label{eq-u-square}
\eeq
where we introduced
\beq
\kappa_{\rm eff} \equiv \frac{\frac{1}{N}\sum_{i < j} \langle \nabla^{2}v(r_{ij}) \rangle_{*}}{d}
\label{eq-hookian-const}
\eeq
\eq{eq-u-square} follows from \eq{eq-pdf-molecular}, \eq{eq-variance-molecular} and \eq{eq-variance-molecular-1st}. 

It is instructive to note that the above result could have been obtained without using replicas but more intuitively. Suppose that we focus on a single particle, say $i$-th particle, and imagine that all other particles $j (\neq i)$ are {\it frozen} in a certain configuration.
Then the surrounding particles create a static effective potential
$U_{i}=\frac{1}{2}\sum_{\j (\neq i)}v(r_{ij})$ for the $i$-th particle:
the Einstein model. Then the thermal average of the fluctuation of the particle 
can be evaluated as,
\beq
\langle {\bf u}_{i}^{2} \rangle \simeq \frac{k_{\rm B}T}{(1/2)\sum_{j (\neq i)} \nabla^{2}v(r_{ij})/d}.
\eeq
Here we assumed isotropic fluctuations inside cages. Thus we find $\kappa_{\rm eff}$ defined in \eq{eq-hookian-const} can be viewed as the Hookian spring constant of the effective Einstein oscillator.

The above discussion suggests that the basic idea of cage expansion is very similar in spirit to consider energy landscape in the vicinity of inherent structures (see sec. \ref{sec-landscape}).
However there is an important difference : the configurations of the center of mass positions of the molecules ${ R_{i} }$ are by no means restricted to energy minima but allowed
to take {\it any} configurations of a liquid thermalized at effective temperature $T_{\rm eff}=T/m^{*}$ (see \eq{eq-averaging-effective}). As we discuss soon, this point is important 
in linear responses to external perturbations at finite temperatures.

Using the statistical weights \eq{eq-average-clonedliquid} thermal averages of physical quantities can be done as follow.
Consider an observable  $O(\{r^{a}_{ij}\})$ which is a function of the distances between particles $r^{a}_{ij}=|{\bf r}^{a}_{i}-{\bf r}^{a}_{j}|$.
Using the molecular coordinate \eq{eq-molecular-coordinate} and
assuming that the fluctuations $u^{a}_{i}$'s are small, it can be expanded as,
\begin{eqnarray}
 O(\{r^{a}_{ij}\}) &=&O(\{R_{ij}\})+\sum_{i < j}\sum_{a}\sum_{\mu}\frac{\partial O(\{r^{a}_{ij}\})}{\partial (r_{ij}^{a})^{\mu}}(u_{i}^{a}-u_{j}^{a})^{\mu}\nonumber\\
&& +\frac{1}{2}\sum_{i < j, k < l}\sum_{a,b} \sum_{\mu,\nu}
\frac{\partial^{2}O(\{r^{a}_{ij}\})}{\partial (r^{a}_{ij})^{\mu}\partial(r^{b}_{kl})^{\nu}}
 (u^{a}_{i}-u^{a}_{j})^{\mu}(u^{b}_{k}-u^{b}_{l})^{\nu} \ldots.
\end{eqnarray}
Then the thermal average can be done as,
\beqa
\langle O(\{r^{a}_{ij}\}) \rangle 
&& = \langle O(\{R_{ij}\}) \rangle_{T/m,[v_{\rm eff}(r)]}
-\frac{1}{2}\langle u^{2} \rangle_{\rm cage}
\sum_{i} \sum_{j_{1} (\neq i)}\sum_{j_{2} (\neq i)}
\sum_{a,b}(1-m\delta_{ab})
 \sum_{\mu}
\left \langle
\frac{\partial^{2}O(\{r^{a}_{ij}\})}{\partial (r^{a}_{ij_{1}})^{\mu}\partial(r^{b}_{ij_{2}})^{\mu}}
\right \rangle_{*} + \ldots
\nonumber \\
&& = \langle O(\{R_{ij}\}) \rangle_{*} 
+ \beta^{*} (1-m) \langle u^{2} \rangle_{\rm cage}
\sum_{i < j } 
\left [
\langle O(\{R_{ij}\}) \nabla^{2} v(R_{ij}) \rangle_{*}
- \langle O(\{R_{ij}\})\rangle_{*} 
\langle \nabla^{2} v(R_{ij}) \rangle_{*}
\right ] + \ldots
\nonumber \\
&& 
-\frac{1}{2}\langle u^{2} \rangle_{\rm cage}
\sum_{i} \sum_{j_{1} (\neq i)}\sum_{j_{2} (\neq i)}
\sum_{a,b}(1-m\delta_{ab})
 \sum_{\mu}
\left \langle
\frac{\partial^{2}O(\{r^{a}_{ij}\})}{\partial (r^{a}_{ij_{1}})^{\mu}\partial(r^{b}_{ij_{2}})^{\mu}}
\right \rangle_{*}  + \ldots
\label{eq-O-cage-expansion}
\eeqa
where $\alpha/m^{2}=2A/m$ as given in \eq{eq-variance-molecular-1st}. 
The evaluation of the averages
$\langle \ldots \rangle_{*}$ 
(see \eq{eq-pdf-CM}) can be done using methods of standard liquid theory \cite{hansen-mcdonald}
at temperature $T/m$. The value of the cage size $A=A(T)$ and the
parameter $m=m^{*}(T)$ must be determined beforehand. 

\subsubsection{Evaluations of intra-state and inter-state thermal
   fluctuations by the cage expansion}
\label{sec-recipe-disentangle}

In sec. \ref{subsubsec-disentangle} we discussed the possibility to disentangle the intra-state and inter-state responses using a cloned system.
Here we present a more precise prescription to do it explicitly using the cage expansion.
In Appendix \ref{appendix-p-spin} we demonstrate that the prescription works exactly in a mean-field spin-glass model.
In the next section we compute the rigidity applying the following 
prescription to evaluate shear-stress correlation function
which appears in the static fluctuation formula of the 
rigidity \eq{eq-def-shearmodulus}. 

Here let us consider the generic susceptibility considered
in sec. \ref{subsubsec-disentangle} (see \eq{eq-chi-decomposition-clone}),
\beq
\chi_{ab}
=\frac{\beta}{N} \left( \langle O_{a}O_{b}\rangle 
- \langle O_{a}\rangle \langle O_{b}\rangle \right).
\eeq
Using the above prescription \eq{eq-O-cage-expansion} it becomes,
\begin{eqnarray}
\chi_{ab} &=& \frac{\beta}{N} \left( \langle
O^{2}(\{R_{ij}\})\rangle_{T/m,[v_{\rm eff}(r)]} -
\langle O(\{R_{ij}\})\rangle_{T/m,[v_{\rm eff}(r)]}^{2} \right)
\nonumber \\
&&  - \frac{\beta}{N}\sum_{ij} \sum_{\mu \nu} \left \langle 
\frac{\partial O}{\partial (x_{i}^{a})^{\mu}}
\frac{\partial O}{\partial (x_{j}^{b})^{\nu}} \right \rangle_{*}
(1-m\delta_{ab}) \langle (u_{i})^{\mu}(u_{j})^{\nu}  \rangle_{\rm cage}
\end{eqnarray}
with $\langle (u_{i})^{\mu}(u_{j})^{\nu}  \rangle_{\rm cage}$ given by \eq{eq-u-square}.
Comparing the above result with 
the last equation of \eq{eq-chi-decomposition-clone} we find the intra-state susceptibility as,
\beq
\hat{\chi}=\frac{\beta^{*}}{N}\sum_{ij} \sum_{\mu \nu} \left \langle 
\frac{\partial O}{\partial (x_{i}^{a})^{\mu}}
\frac{\partial O}{\partial (x_{j}^{b})^{\nu}} \right \rangle_{*}
\langle (u_{i})^{\mu}(u_{j})^{\nu} \rangle_{\rm cage}
=\frac{\beta^{*}}{N}\sum_{i} \sum_{\mu} 
\left  \langle \left( 
\frac{\partial O}{\partial (x_{i}^{a})^{\mu}}\right)^{2} \right   \rangle_{*} 
\langle u^{2} \rangle_{\rm cage}
\label{eq-intra-chi-cage}
\eeq
with $\beta^{*}=m \beta$, and the inter-state susceptibility, 
\beq
\tilde{\chi}=
\frac{\beta}{N} \left( \langle
O^{2}(\{R_{ij}\})\rangle_{T/m,[v_{\rm eff}(r)]} -
\langle O(\{R_{ij}\})\rangle_{T/m,[v_{\rm eff}(r)]}^{2} \right)
-
\frac{\beta}{N}\sum_{i} \sum_{\mu} 
\left  \langle \left( 
\frac{\partial O}{\partial (x_{i}^{a})^{\mu}}\right)^{2} \right   \rangle_{*} 
\langle u^{2} \rangle_{\rm cage}.
\label{eq-inter-chi-cage}
\eeq
Finally the total susceptibility \eq{eq-total-response} becomes,
\beq
\chi= \frac{\beta^{*}}{N} \left( \langle
O^{2}(\{R_{ij}\})\rangle_{T/m,[v_{\rm eff}(r)]} -
\langle O(\{R_{ij}\})\rangle_{T/m,[v_{\rm eff}(r)]}^{2} \right)
\label{eq-total-chi-cage}
\eeq

Although we have only performed the cage expansion up to 1st order in \eq{eq-intra-chi-cage},\eq{eq-inter-chi-cage} and \eq{eq-total-chi-cage}, 
we expect that the basic structure of the result is generic.
Higher order cage expansion will yield more accurate effective potential $v_{\rm eff}(r)$ \cite{parisi-zamponi} 
which will refine the averages over the CM positions of the molecules.

\subsubsection{Discussion: responses inside metastable states}
\label{sec-intra-state-responses}

Let us focus on the responses within metastable states and discuss it more in detail.
Since fluctuations in the close vicinity of an IS is completely 
characterized by the elastic hamiltonian \eq{eq-hamiltonian-IS},
the susceptibility $\chi_{\rm IS}$ associated with an IS can be obtained
just like the 1st equation of \eq{eq-intra-chi-cage} but replacing
the correlator $\langle (u_{i})^{\mu}(u_{j})^{\nu} \rangle_{\rm cage}$ by,
\beq
\langle (u_{i})^{\mu}(u_{j})^{\nu}  \rangle_{\rm IS} = (\beta H^{-1})_{ij}^{\mu \nu}
\eeq
with $H^{-1}$ being the inverse of the Hessian matrix \eq{eq-hessian}. As the result we find,
\beq
\chi_{\rm IS}=\frac{1}{N}\sum_{ij} \sum_{\mu \nu} 
\left. \frac{\partial O}{\partial (x_{i}^{a})^{\mu}} \right |_{\rm IS}
\left. \frac{\partial O}{\partial (x_{j}^{b})^{\nu}} \right |_{\rm IS}
 (H^{-1})_{ij}^{\mu \nu}
\label{eq-chi-IS}
\eeq
where the derivatives are took at the IS. This must be compared with the 
intra-state susceptibility $\hat{\chi}$ defined in
\eq{eq-intra-chi-cage} together with \eq{eq-u-square}, which reads,
\beq
\hat{\chi}=\frac{1}{N}\sum_{i} \sum_{\mu} 
\left  \langle \left( \frac{\partial O}{\partial (x_{i}^{a})^{\mu}}\right)^{2}  \right   \rangle_{*}
\frac{1}{\kappa_{\rm eff}}
\label{eq-intra-chi-cage-again}
\eeq
where we used \eq{eq-hookian-const}.

Comparing the two susceptibilities $\chi_{\rm IS}$ and $\hat{\chi}$ we find that they are apparently similar to some extent. 
In both cases, the fluctuation of the observable $O$ is induced by small fluctuation of particles around some {\it reference state(s)}.
We also notice that the cage expansion approach amounts to replace the 
full Hessian matrix by that of an effective Einstein model we discussed in sec. \ref{sec-fluctuation-molecular-liquid}.
Validity of such an approximation should be examined critically. A better approximation may be obtained for example by doing partial harmonic
re-summation exactly \cite{mezard-parisi-1999}.

However here let us focus on a more important difference between the
two susceptibilities that comes from the difference of their {\it reference state(s)} themselves.
Note that the reference state for $\hat{\chi}$ as defined in \eq{eq-intra-chi-cage-again}
is a {\it thermalized liquid configuration} within a metastable state at a finite temperature $T^{*}=T/m^{*}(T)$,
where all sorts of thermal excitations\blue{,  not only the elastic but also
plastic deformations, are included. On the other hand the reference
state for $\chi_{\rm IS}$ is just an {\it energy minimum}.} 
This implies for instance, 
\beq
\left (\left. \frac{\partial O}{\partial (x_{i}^{a})^{\mu}} \right |_{\rm IS} \right)^{2} <
\left \langle 
\left (
\frac{\partial O}{\partial (x_{i}^{a})^{\mu}}  \right)^{2}
\right \rangle.
\label{eq-dev-o-is-mb}
\eeq
The above observation suggests that in general the intra-state susceptibility $\hat{\chi}$
is {\it larger} than that of the inherent structures $\chi_{\rm IS}$,
\beq
 \hat{\chi} > \chi_{\rm IS}.
\label{eq-hat-chi-chi-IS}
\eeq
In sec \ref{sec-landscape} we argued that a metastable would be identified
as a metabasin (MB), which is a union of inherent structures (IS), rather than a
single IS. Then the difference between the two susceptibilities \eq{eq-hat-chi-chi-IS} 
should be attributed to thermal fluctuations between different inherent structures
within a common metabasin, which are necessarily {\it plastic} deformations. 
In the supercooled liquid state $T > T_{K}$, $m^{*}(T)=1$ so that
$T^{*}=T$. This implies the difference between the two susceptibilities 
increases with temperature at $T > T_{K}$.
On the other hand, in the glass phase $T < T_{K}$ one typically finds $m^{*}(T) \sim T/T_{\rm K}$ so that
$T^{*}=T/m^{*}(T) \sim T_{\rm K}$ \cite{mezard-parisi-1999,coluzzi-mezard-parisi-verrocchio-1999}.
This suggests that $\hat{\chi}$ becomes nearly independent of
temperatures below $T_{\rm K}$ \blue{such that $\tilde{\chi} \sim
\chi_{\rm IS}$.}

We argued in sec.~\ref{subsubsec-two-step-relaxation} that the same intra-state susceptibility can be observed
either in equilibrium or out-of equilibrium based on the observation 
that in general the $\beta$-relaxation shows little aging effects.
In this respect, it is interesting to note that the intra-state
susceptibility $\hat{\chi}$ reflect the underlying equilibrium state.
Especially the above discussion suggests even the Kauzmann temperature
$T_{\rm K}$ itself may be reflected in the temperature dependence of $\hat{\chi}$.

\section{Rigidity of glasses}
\label{sec-rigidity}

\subsection{Shear on the cloned liquid}
\label{subsec-shear-on-the-cloned-liquid}

In sec. \ref{subsubsec-paradox}  and
sec. \ref{subsubsec-static-fluctuation-formula} we discussed some
general aspects of static response to shear. The rigidity is defined as the 2nd derivative of the free-energy with respect to the shear-strain $\gamma$ as in \eq{eq-stress-shearmodulus} whose explicit expression, the fluctuation formula, is given in \eq{eq-def-shearmodulus}. Now let us consider to shear the cloned liquid.
Suppose that we have a cloned liquid of $m$ replicas which are subjected 
different shear-strains $\gamma^{a}$ ($a=1,2,\ldots,m$): the replicas are put into different containers which can be deformed independently.  This is just an example of the
static perturbations on the cloned liquid discussed in sec. \ref{subsubsec-disentangle}.
Then we may formally expand the free-energy of the cloned liquid as,
\beq
mF_{m}(\gamma)=mF_{m}(0)+ N \sum_{a=1}^{m} \langle \sigma^{a} \rangle \gamma_a
+\frac{N}{2} \sum_{a,b=1}^{m} \mu_{ab} \gamma^{a} \gamma^{b} +
O(\gamma^{d}) \ .
\label{eq-F-cloned-1st-expansion}
\eeq
The argument for the decomposition of the response \eq{eq-chi-decomposition-clone} implies that the generalized rigidity $\mu_{ab}$ may also be parametrized as,
\beq
\mu_{ab} =\hat{\mu} \delta_{a,b} + \tilde{\mu}
\label{eq-generalized-rigidity}
\eeq
with the {\it intra-state rigidity} or {\it effective rigidity},
\beq
\hat{\mu}=\left [\!\left[ \left. \frac{\partial^{2}f_{\alpha}(T,\gamma)}{\partial \gamma^{2}} \right |_{\gamma=0} \right]\!\right]_{T,m}
\eeq
and the {\it negative} correction term due to inter-state fluctuations,
\beq
\tilde{\mu}= - N  \beta \left( [\![ \sigma_{\alpha}^{2} ]\!]_{T,m}-
[\![\sigma_{\alpha}]\!]_{T,m}^{2} \right) 
\label{eq-tilde-mu}
\eeq
where $\sigma_{\alpha}$ is the 'internal stress' of state $\alpha$,
\beq
\qquad \sigma_{\alpha}=\left. \frac{\partial f_{\alpha}(T,\gamma)}{\partial \gamma} \right |_{\gamma=0}.
\label{eq-sigma-alpha}
\eeq
As we discuss below $\tilde{\mu}$, which is negative, can be interpreted as
{\it inter-state modulus}.

The trace over the replica index of $\mu_{ab} $ yields the rigidity of a
molecular liquid. This must be zero
since cloned liquid as a whole is just a liquid (!) (see the discussion 
in the end of sec. \ref{subsubsec-static-fluctuation-formula})
\beq
\sum_{b=1}^{m}\mu_{ab}=\hat{\mu}+m \tilde{\mu}=0.
\label{eq-sum-rule}
\eeq

Following the discussion in sec. \ref{subsubsec-disentangle},
we find the total rigidity of the glass from \eq{eq-sum-rule} as,
\beq
\mu=\lim_{m \to 1^{-}}\sum_{b=1}^{m}\mu_{ab}=\hat{\mu}+m^{*} \tilde{\mu}=0.
\label{eq-mu-total-glass}
\eeq
where $m^{*}=m^{*}(T)$ is the glass transition line. Thus we find the inter-state correction term
$\hat{\mu}$ is simply related to the intra-state rigidity $\hat{\mu}$ as,
\beq
\tilde{\mu}=-\frac{1}{m^{*}}\hat{\mu}.
\label{eq-tilde-mu-hat-mu}
\eeq

\subsection{Static analogue of yielding or shocks} 
\label{subsec-intermittent}

Let us discuss a physical interpretation of the above result \eq{eq-tilde-mu-hat-mu}.
Suppose that internal stress
$\sigma_{\alpha}$ defined in \eq{eq-sigma-alpha} 
has a Gaussian distribution with zero mean and variance
$\Delta^{2}/N$. Then \eq{eq-tilde-mu} and \eq{eq-tilde-mu-hat-mu} implies
that the intra-state rigidity is related to $\Delta$ as\cite{YM2010},
\beq
\hat{\mu}=m^{*} \beta \Delta^{2}.
\label{eq-mu-delta}
\eeq
Note that this corresponds to \eq{eq-mu-p-slow-mu-state}. (We mentioned 
similar relations for the instantaneous rigidity
\eq{eq-b-sigma-variance} 
and the rigidity of inherent structures \eq{eq-mu-p-fast-mu-IS}).
\blue{A very recent numerical study \cite{harrowell-2012} explicitly demonstrated that
the internal stresses of inherent structures indeed follow Gaussian distributions.}

\begin{figure}[t]
\includegraphics[width=0.4\textwidth]{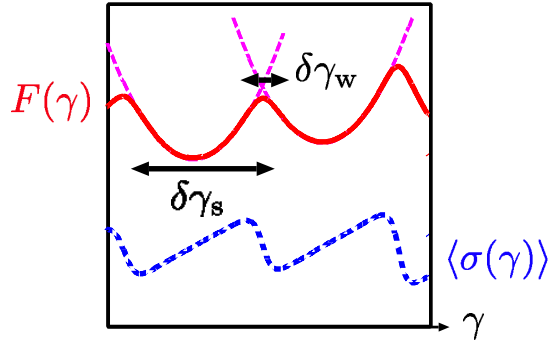} 
\caption{Schematic mean-field picture of free-energy landscape and
 static stress-strain curve below $T_{\rm K}$. The metastable states (metabasins) with
 rigidity $\hat{\mu}$ are represented by parabolas}
\label{fig-intermittent}
\end{figure}

The {\it static} response of a system with many metastable states
against variation of the shear-strain $\gamma$ at low temperatures 
$T < T_{\rm K}$ may be viewed pictorially as shown in Fig.~\ref{fig-intermittent}.
Variation of the free-energy $f_{\alpha}(\gamma)$ 
associated with a given metastable state $\alpha$ along the
$\gamma$-axis may be described as,
\beq
f_{\alpha}(\gamma)=\frac{\hat{\mu}}{2}(\gamma-\gamma^{\rm
min}_{\alpha})^{2}+f^{\rm min}_{\alpha}
\label{eq-f-alpha-landscape}
\eeq
where $\hat{\mu}$ is the intra-state rigidity. The free-energy takes
a minimum value $f^{\rm min}_{\alpha}$ at $\gamma=\gamma^{\rm min}_{\alpha}$.
Then the 
internal stress $\sigma_{\alpha}(\gamma)$ of the metastable state at a
given $\gamma$ is given by,
\beq
\sigma_{\alpha}(\gamma)=\hat{\mu}(\gamma-\gamma^{\rm min}_{\alpha})
\label{eq-sigma-alpha-landscape}
\eeq
Of course we should expect that the metastable state itself disappears 
at some distance from $\gamma^{\rm min}_{\alpha}$ approaching its
metastability limit where a saddle-node bifurcation takes place\cite{maloney-lemaitre-2004b}.

Let us consider a set of metastable states which lives
at around a given point $\gamma$ along the $\gamma$-axis.
The locations of their minima $\gamma^{\rm min}_{\alpha}$ will be distributed 
randomly around the point $\gamma$.
Then the internal stress $\sigma_{\alpha}(\gamma)$ 
defined by \eq{eq-sigma-alpha-landscape} at a given $\gamma$
will take both positive and negative random values depending on the sign
of $\gamma-\gamma_{\alpha}^{\rm min}$. Then the relation
\eq{eq-mu-delta} implies the variance of the distribution 
of the distances $\gamma-\gamma^{\rm min}_{\alpha}$ is $(1/m^{*} \beta \hat{\mu})N^{-1}$.

At any point along the $\gamma$-axis, 
there will be exponentially large number of metastable
states $\propto e^{N\Sigma(f)}$ at a given free-energy 
level (per particle) $f$ where $\Sigma(f)$ is the complexity or structural entropy \eq{eq-complexity-def}.
However, at low temperatures $T< T_{\rm K}$, 
the ground state and at most only a few other low lying states 
would have significant statistical weights \cite{REM}. 
Under variation of $\gamma$, there will be level crossings
among such low lying states as shown in Fig.~\ref{fig-intermittent}.
Just around the points where there are level crossings, the
rigidity should become locally {\it negative}. Correspondingly 
the ground state energy exhibits a cusp singularity where a level crossing takes place.
This would be the physical interpretation of the
{\it negative} inter-state modulus $\tilde{\mu}=-(1/m^{*})\hat{\mu}$
(See \eq{eq-tilde-mu-hat-mu}). 
We can consider this feature as a static analogue of {\it yielding}.

Typical spacing between such level crossings along the $\gamma$-axis will be of order,
\beq
\delta \gamma_{\rm s} =\frac{k_{\rm B}T_{\rm K}}{\Delta\sqrt{N}}.
\label{eq-delta-gamma-s}
\eeq
This is because typical free-energy gap between low lying metastable states will be of order $k_{\rm B}T_{\rm K}$
while free-energy change of each of them for a small change of the strain $\gamma$ is proportional to their internal stress $\sigma_{\alpha}$.
Another important scale is typical width of the thermal rounding of the level crossings. This would be simply give by,
\beq
\delta \gamma_{\rm w} =\frac{k_{\rm B}T}{\Delta\sqrt{N}}.
\label{eq-delta-gamma-w}
\eeq
In the zero temperature limit $T \to 0$, the free-energy landscape
should exhibit cusp singularity. Correspondingly we find that the
intra-state modulus $\tilde{\mu}$ defined in \eq{eq-tilde-mu-hat-mu} diverges negatively because $\lim_{T \to 0}m^{*}(T) \to 0$. 

Based on the above observations the probability $p_{\rm crossing}$ to meet a level crossing
at a given $\gamma$ may be estimated as \cite{yoshino-rizzo},
\beq
p_{\rm crossing} \sim \delta \gamma_{\rm w}/\delta \gamma_{\rm s} \sim T/T_{\rm K}
\label{eq-p-crossing}
\eeq
It is interesting to recall that the factor $m^{*}(T)$ which appears in front of the inter-state modulus $\hat{\mu}$
in \eq{eq-mu-total-glass} can be regarded as the probability to make an inter-state transitions 
within the standard interpretation of the replica symmetry breaking ansatz \cite{Parisi-1983,Cavagna-review-sg}.
(See sec. \ref{subsubsec-disentangle} and Appendix \ref{appendix-p-spin} for related discussions)
Typically it is found that $m^{*}(T) \sim T/T_{\rm K}$ \cite{mezard-parisi-1999,coluzzi-mezard-parisi-verrocchio-1999}.
Thus \eq{eq-p-crossing} is consistent with the standard interpretation of the RSB ansatz.

Note that in the thermodynamic limit $N \to \infty$, the spacing between the cusps vanishes \cite{rizzo-yoshino,yoshino-rizzo}. It means that 
if we define a coarse-grained rigidity for any small but {\it non-zero} interval along the $\gamma$-axis, we obtain $0$ rigidity in the $N \to \infty$ limit.
Thus the static intermittency of the stress-strain curve sketched in Fig.~\ref{fig-intermittent} explains how we can solve the apparent paradox 
discussed in sec. \ref{subsubsec-paradox} in the static mean-field theoretical context.
Similar static intermittencies have been found in disordered systems including spin-glasses \cite{KY,YBM,Krzakala-Martin-2002,rizzo-yoshino,yoshino-rizzo,doussal-muller-wiese-2010},
and elastic manifolds in random media \cite{balentz-bouchaud-mezard,Bouchaud-Mezard-1997,doussal-wiese-2007}, which are actually related to the problem of intermittency 
in turbulent flows due to shocks \cite{Bouchaud-Mezard-Parisi-1995}.

The above observation implies the inter-state fluctuations can be considered as Goldstone modes characteristic in glasses which try to restore the broken translational symmetry.
On the other hand, the elastic fluctuation within a metastable state is more usual Goldstone mode which is also present in crystals.

Although we have assumed $T < T_{\rm K}$ in the above discussions, the transitions between different metastable states
under external strain can be relevant even at higher temperatures in the following sense.
Within the so called mosaic picture of the RFOT theory  \cite{RFOT,bouchaud-biroli-2004,biroli-bouchaud-cavagna-grigera-verrocchio-2008,RFOT-review}, 
the configuration of the system in the intermediate temperature range $T_{\rm K} < T < T_{\rm c}$
consists of mosaics or patches of local regions such that the
subsystem in each of the patch stays in a metastable state.
The size of the local regions is called as the mosaic length $\xi_{\rm
mosaic}$. It means that the subsystem within a mosaic is as if at a temperature below $T_{\rm K}$.

Then an interesting possibility is that the intermittent stress-strain
curves shown schematically in Fig.~\ref{fig-intermittent} may be 
observable by some rheological experiment in a confined geometry.
Suppose to exert some local shear-strain $\gamma$ in a quasi-static
manner on a system confined in a narrow region of size $L$. 
If the size of the region is small enough such that $L < \xi_{\rm mosaic}$, then 
the resultant stress-strain curve may exhibit the intermittent feature with typical spacing between the shocks  $\delta \gamma_{\rm s} \propto 1/\sqrt{L^{d}}$.

Finally also note for clarity that yielding at mesoscopic scales can occur by the saddle-node bifurcation mechanism at metastability limits
\cite{maloney-lemaitre-2004a,maloney-lemaitre-2006,procaccia-2011}. If the energy level
at the metastability limit is order $O(1)$, we find again that $\delta \gamma_{\rm s} \propto 1/\sqrt{L^{d}}$.

\subsection{Cage expansion of the rigidity}
\label{sec-Cage expansion-of-the-rigidity}

Now let us examine the rigidity $\mu_{ab}$ defined in
\eq{eq-F-cloned-1st-expansion}. 
Its microscopic expression 
can be obtained generalizing the fomulae 
Eqs. \eq{eq-def-shearmodulus},\eq{eq-def-born} and \eq{eq-thermal-average},
\beq
\mu_{ab}=\frac{\partial^{2} m F_{m}}{\partial \gamma^{a}\partial \gamma^{b}}
=\langle b_{[\{{\bf r}^{a}_{i}\}]} \rangle \delta_{ab}-\beta [\langle \sigma_{[\{{\bf r}^{a}_{i}\}]} \sigma_{[\{{\bf r}^{b}_{i}\}]} \rangle -
\langle \sigma_{[\{{\bf r}^{a}_{i}\}]} \rangle \langle \sigma_{[\{{\bf r}^{b}_{i}\}]} \rangle].
\label{eq-generalized-rigidity-expression}
\eeq
where the subscript, for instance, ${[\{{\bf r}^{a}_{i}\}]}$ indicates that coordinates
${\bf r}_{i}$ ($i=1,2,\ldots,N$) of particles in replica $a$. 
The 1st term $b$ on the r.h.s of the above equation is
 the born term defined in \eq{eq-def-born} which reads,
\beq
 b_{[\{{\bf r}_{i}\}]}  =\frac{1}{N}\sum_{i <
j} b({\bf r}_{ij}) \qquad b({\bf r}_{ij})=\hat{z}_{ij}^{2} \left [
 r^{2}v^{(2)}(r)
\hat{x}_{ij}^{2}
+  r v^{(1)}(r)
(1- \hat{x}_{ij}^{2}) \right]_{r=r_{ij}}.
\label{eq-born-expression}
\eeq
The 2nd term is the correlation function of the fluctuation of the shear-stress,
\beq
\sigma_{[\{{\bf r}_{i}\}]}= \frac{1}{N} 
\sum_{i < j}  \sigma({\bf r}_{ij}) \qquad \sigma({\bf r}_{ij})=  \left. r v^{(1)}(r) \right |_{r=r_{ij}}
\hat{x}_{ij}\hat{z}_{ij}. 
\label{eq-stress-expression}
\eeq

Our task is to evaluate the rigidity $\mu_{ab}$ by the cage expansion as
follows. It is essentially the same as the analysis of the generic susceptibility $\chi_{a b}$ 
we discussed in sec. \ref{sec-recipe-disentangle}. First we explicitly decompose the coordinates of the particles as \eq{eq-molecular-coordinate},
\bmat
{\bf r}^{a}_{i}={\bf R}_{i}+{\bf u}_{i}^{a}.
\emat
Then we expand the thermal averages which appear in the fluctuation formula
of the rigidity \eq{eq-generalized-rigidity-expression}
in power series of the molecular coordinates ${\bf u}^{a}_{i}$ 
($a=1,2,\ldots,m$, $i=1,2,\ldots,N$) using the formula \eq{eq-O-cage-expansion}.
Then we plug-in the expansions in the microscopic expression of the
the rigidity $\mu_{ab}$ given by \eq{eq-generalized-rigidity-expression}.
Finally we evaluate the thermal averages using \eq{eq-pdf-molecular} 
(\eq{eq-variance-molecular}) together with \eq{eq-variance-molecular-1st} for the
molecular coordinates ${\bf u}_{i}^{a}$ and \eq{eq-pdf-CM} for the
center of mass coordinates ${\bf R}_{i}$ of the molecules. The result
may be put into a  power series of the parameter $\alpha$ (or the cage size $A$),
\beq
\mu_{ab}=(\mu_{0})_{ab}+(\mu_{1})_{ab}\/\/\alpha+O(\alpha^{2})
\label{eq-mu-alpha-expansion}
\eeq
or
\beq
\mu_{ab}=\hat{\mu}\delta_{ab}+\tilde{\mu} \qquad 
\hat{\mu}=\hat{\mu}_{0}+ \hat{\mu}_{1}\/\/\alpha+O(\alpha^{2}) \qquad
\tilde{\mu}=\tilde{\mu}_{0}+ \tilde{\mu}_{1}\/\/\alpha+O(\alpha^{2}) 
\label{eq-mu-alpha-expansion-2}
\eeq
which must satisfy the sum rule \eq{eq-sum-rule},
\beq
\sum_{b}(\mu_{n})_{ab}=\hat{\mu}_{n}+m\tilde{\mu}_{n}=0
\label{eq-sum-rule-expansion}
\eeq
at each other in the cage expansion.

\subsubsection{Zero-th order}

At zero-th order of the cage expansion, all particles are bound to 
the CM positions of the  molecules. We find,
\beq
(\mu_{0})_{ab}= \langle b_{[\{{\bf R}_{i}\}]} \rangle _{*}\delta_{ab}-\beta[\langle \sigma_{[\{{\bf R}_{i}\}]}^{2} \rangle_{*} -\langle \sigma_{[\{{\bf R}_{i}\}]} \rangle_{*}^{2} ]
\eeq
which obviously satisfies the sum rule \eq{eq-sum-rule-expansion},
\beq
\sum_{a=1}^{m} (\mu_{0})_{ab}=
\langle b_{[\{{\bf R}_{i}\}]} \rangle _{*}-\beta m [\langle \sigma_{[\{{\bf R}_{i}\}]}^{2} \rangle_{*} -\langle \sigma_{[\{{\bf R}_{i}\}]} \rangle_{*}^{2} ]=0,
\eeq
since this is just the rigidity of a simple liquid at $T=T/m^{*}$.
Thus at the zero-th order we find the intra-state and inter-state rigidities as,
\beqa
\hat{\mu}_{0}=\langle b_{[\{{\bf R}_{i}\}]} \rangle _{*} \qquad 
\tilde{\mu}_{0}=-\frac{\hat{\mu}_{0}}{m}.
\label{eq-mu0}
\eeqa

\subsubsection{First order}

At the 1st order of the cage expansion we find the following result,
\beq
(\mu_{1})_{ab}\/\/\alpha =J_{1}(1-m\delta_{ab})+(1-m)[(J_{2}+J_{3})\delta_{ab}+J_{4}+J_{5}]
\label{eq-mu1}
\eeq
where
\begin{eqnarray}
&& J_{1}=c\frac{1}{N}\sum_{i}\sum_{j_{1} (\neq i)}\sum_{j_{2} (\neq i)} 
\beta
\left \langle 
\nabla \sigma({\bf r}_{i j_{1}}) \cdot \nabla \sigma({\bf r}_{i j_{2}})
\right \rangle_{*}  \nonumber \\
&& J_{2}=-c\frac{1}{N} \sum_{i < j} 
\left \langle 
\nabla^{2} b({\bf r}_{ij})
\right \rangle_{*}  \nonumber \\
&& J_{3}=c \frac{1}{N}\sum_{i < j}\sum_{k < l}
\beta^{*} [
\langle 
 b({\bf r}_{ij}) \nabla^{2} v(r_{kl})
\rangle_{*} -\langle 
 b({\bf r}_{ij}) \rangle_{*} \langle \nabla^{2} v(r_{kl})
\rangle_{*}] \nonumber \\
&& J_{4}=2c\frac{1}{N}\sum_{i_{1} < j_{1}}\sum_{i_{2} < j_{2}}\beta
 [\langle \nabla^{2} \sigma({\bf r}_{i_{1}j_{1}}) \sigma({\bf r}_{i_{2}j_{2}})
 \rangle_{*}-\langle \nabla^{2} \sigma({\bf r}_{i_{1}j_{1}})\rangle_{*}
 \langle \sigma({\bf r}_{i_{2}j_{2}}) \rangle_{*}] \nonumber \\
&& J_{5}=-c \frac{1}{N}\sum_{i_{1} < j_{1}}\sum_{i_{2} < j_{2}}\sum_{k <
 l}\beta\beta^{*} [\langle  \sigma({\bf r}_{i_{1}j_{1}}) \sigma({\bf r }_{i_{2}j_{2}}) \nabla^{2}v(r_{kl}) \rangle_{*}-\langle
 \sigma({\bf r}_{i_{1}j_{1}})\sigma({\bf r}_{i_{2}j_{2}})\rangle_{*}
 \langle \nabla^{2}v(r_{kl}) \rangle_{*}]
\label{eq-I1-I2-I3-I4-I5}
\end{eqnarray}
Here we introduced a parameter,
\beq
c \equiv \langle u^{2} \rangle_{\rm cage}=
\frac{1}{\beta^{*}\kappa_{\rm eff}}
\label{eq-c-A}
\eeq
with $\kappa_{\rm eff}$ defined in \eq{eq-hookian-const}.
In the above equations \eq{eq-I1-I2-I3-I4-I5} and in the following 
quantities which appear inside the brackets $\langle \ldots \rangle_{*}$ are
functions of only the CM coordinates ${\bf R}_{i}$ ($i=1,2,\dots,N$).

As discussed in Appendix \ref{sec-identity}, the group of terms $J_{2}$,$J_{3}$,$J_{4}$,$J_{5}$ can be related to the rigidity of {\it renormalized} liquid interacting with the effective potential $v_{\rm eff}(r)$ given in \eq{eq-veff}.  Then the following identity,
\beq
J_{2}+J_{3}+m(J_{4}+J_{5})=0,
\label{eq-J-identity}
\eeq
holds. Using this identity, the result \eq{eq-mu1} can be simplified as,
\beq
\mu_{1}\alpha=(\hat{\mu}_{1}\delta_{ab}+\tilde{\mu})\alpha=\left[J_{1}-\frac{1-m}{m}(J_{2}+J_{3})\right](1-m\delta_{ab}).
\eeq
so that
\beq
\hat{\mu}_{1}\alpha=-m \left[J_{1}-\frac{1-m}{m}(J_{2}+J_{3})\right] \qquad 
\tilde{\mu}_{1}\alpha=J_{1}-\frac{1-m}{m}(J_{2}+J_{3})
\label{eq-mu1-rev}
\eeq
The result satisfies the sum rule \eq{eq-sum-rule-expansion},
\beq
\sum_{b} (\mu_{1})_{ab}=0
\eeq
as it should.  To summarize, up to 1st order in the cage expansion we find,
\begin{eqnarray}
\hat{\mu}=\langle b_{[\{{\bf R}_{i}\}]} \rangle_{*} -mJ_{1}+(J_{2}+J_{3})(1-m) \qquad 
\tilde{\mu}=-\frac{\hat{\mu}}{m}.
\label{eq-rigidity-result}
\end{eqnarray}

\subsubsection{Summary and discussions}
\label{sec-summary-shearmodulus-clonedliquid}

\begin{figure}[t]
\includegraphics[width=0.2\textwidth]{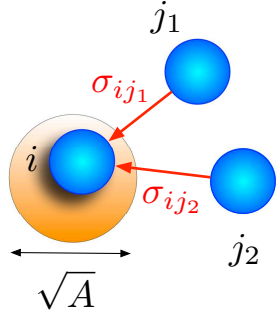}
\caption{Fluctuation of stresses due to fluctuations of particles inside a cage
}
\label{fig-stress-fluctuation-in-cage}
\end{figure}

To summarize, we find the rigidity of the cloned liquid 
\eq{eq-generalized-rigidity-expression} as,
\beqa
&& \mu_{ab}=\hat{\mu} (\delta_{ab}-\frac{1}{m}) \nonumber \\
&& \hat{\mu}=\langle b^{\rm eff} \rangle_{T/m,[v_{\rm eff}(r)]}-
c\beta^{*} \frac{1}{N} \sum_{i} \left \langle |{\bf \Xi}_{i}|^{2}\right
\rangle_{*}  
\label{eq-mu-clonedliquid}
\eeqa
In practice, it is useful to express various terms in \eq{eq-rigidity-result}
in terms of particle distributions so that standard techniques of the 
density functional theories of liquids \cite{hansen-mcdonald} can be incorporated.
The details of the technical aspects are presented in Appendix \ref{appendix-Representation-in-terms-of-particle distribution-functions}, \ref{sec-J12} and \ref{appendix-Formulations-for-the-binary-mixture}.

Let us discuss the two terms on the r. h. s. of \eq{eq-mu-clonedliquid} below,
\begin{itemize}
\item Born term of the renormalized liquid

The terms $J_{2}$ and $J_{3}$ are corrections to the Born term due to
the renormalization of the potential \eq{eq-veff}.
It is easy to realize that the sum of the three terms, i.~e. the Born term, $J_{2}$ and $J_{3}$
is equivalent to the Born term of the renormalized liquid with the effective potential \eq{eq-veff},
\begin{eqnarray}
&&\langle b^{\rm eff}_{[\{{\bf R}_{i}\}]}\rangle_{T/m,[v_{\rm eff}(r)]}=
\langle b_{[\{{\bf R}_{i}\}]}\rangle_{*}+(1-m)(J_{2}+J_{3})+ O(A^{2})
\label{eq-def-born-renormalized}
\end{eqnarray}
with $\langle \ldots \rangle_{T/m,[v_{\rm eff}(r)]}$ defined in \eq{eq-averaging-effective}.
Here $b^{\rm eff}_{[\{{\bf R}_{i}\}]}$ is the Born term
of the renormalized liquid which can be obtained by substituting 
$v_{\rm eff}(r)$ into $v(r)$ of \eq{eq-def-born}. 

\item Non-affine correction

The term $J_{1}$ is clearly distinct from others. While the contributions of the 
terms $J_{2}$ and $J_{3}$ vanish above $T_{\rm K}$ where $m^{*}(T)=1$,
the contribution of the term $J_{1}$ does not. Presumably it continues to
exist up to $T_{\rm c}$ where the metastable solid states disappear.
Note that it has the same structure as the generic intra-state susceptibility
$\hat{\chi}$ (See \eq{eq-intra-chi-cage}) discussed in sec. \ref{sec-recipe-disentangle}.
Let us rewrite the $J_{1}$ term as,
\beq
mJ_{1}= \frac{1}{\kappa_{\rm eff}} \frac{1}{N} \sum_{i} \left \langle |{\bf \Xi}_{i}|^{2}\right \rangle_{*}  
\label{eq-j1-cage-expansion}
\eeq
where $\kappa_{\rm eff}$ is defined in \eq{eq-hookian-const} and  ${\bf \Xi}$ is the derivative of the shear-stress introduced in \eq{eq-def-xi}.
In \cite{maloney-lemaitre-2004b} it is
pointed out that the field ${\bf \Xi}_{i}$ defined in \eq{eq-def-xi} 
is a random field with  very short-ranged correlation in space.

The physical interpretation of this term is straightforward and quite suggestive.
Consider time scales shorter than the $\alpha$-relaxation time, where the 
stress-field does not evolve much so that, as a first approximation, we can assume it is a frozen-in vector field associated with a metastable state.
As shown schematically in Fig.~\ref{fig-stress-fluctuation-in-cage}, the fluctuation
of a particle, say particle $i$, inside its cage induces fluctuations of the 
stresses between the particle and the surrounding particles, say $j_{1}$ and $j_{2}$.
The point is that the resultant fluctuations of the stresses $\sigma_{i j_{1}}$
and $\sigma_{i j_{2}}$ are {\it correlated} since they originate from the common source
of fluctuation, the fluctuation of particle $i$.

\end{itemize}

Although we have only performed the cage expansion up to 1st order,
we expect that the structure \eq{eq-mu-clonedliquid} is generic.
Higher order cage expansion will yield more accurate effective 
potential $v_{\rm eff}(r)$ \cite{parisi-zamponi}. This will amount to
refine the renormalized Born term \eq{eq-def-born-renormalized} and the $J_{1}$ term. Similarly, the $\Xi$ field \eq{eq-def-xi} will be replaced by that of
renormalized potential at higher orders in the cage expansion.

It is instructive to compare the above result with that of
the rigidity $\mu_{\rm IS}$ of inherent structures (IS) discussed in sec. \ref{sec-IS-shearmodulus}.
The comparison is essentially the same as the comparison of the susceptibility associated with inherent structures $\chi_{\rm IS}$ and the intra-state
susceptibility $\hat{\chi}$ we discussed in \ref{sec-intra-state-responses}. 
Comparing the full expression of the rigidity of inherent structures $\mu_{\rm IS}$  \eq{eq-mu-IS-AQS}
with our results of the intra-state rigidity $\hat{\mu}$ \eq{eq-j1-cage-expansion} together with \eq{eq-c-A},
we find that the our scheme approximate the full Hessian matrix by that of an effective Einstein model.
The other, more important difference is that the reference state
itself used  in the evaluation of the intra-state susceptibility
$\hat{\mu}$ is {\it not} completely frozen-in one of energy minima, at marked variance with that for $\mu_{\rm IS}$.
The reference state is just took from a thermalized liquid configuration in a metastable state.
Then it is natural to expect that the amplitude of the $\Xi$ field
$\left \langle |{\bf \Xi}_{i}|^{2}\right \rangle_{*}$, which appears in
\eq{eq-j1-cage-expansion},  is {\it larger} than that at inherent structures (See \eq{eq-xi-IS}).
Thus we consider that the intra-state rigidity $\hat{\mu}$ obtained above is
most likely {\it smaller} than the rigidity of inherent structure $\mu_{\rm IS}$
for the same reason as the generic intra-state susceptibility
$\hat{\chi}$ is larger than the susceptibility $\chi_{\rm IS}$
associated with inherent structures as we discussed in sec. \ref{sec-intra-state-responses}.
We consider that the underlying physical processes are thermal
excitations between different inherent structures within a common metabasin, which are probably localized plastic deformations as we discussed in
sec. \ref{subsubsec-Rigidity-of-metabasin-or-intra-state-rigidity}. 


Finally let us recall the discussion in sec \ref{subsec-melting-orderparameter}
where we pointed out that mean-field theories generically imply that
the rigidity of solids is closely related to the order parameter,
more precisely the rigidity is simply proportional to the square of the
order parameter (see also Appendix. ~\ref{appendix-spin-wave-vectorSG} for a
corresponding problem in a certain vectorial spin-glasses) . Indeed the computation of the intra-state rigidity $\hat{\mu}$
of glasses performed in the present section is consistent to some extent
with this generic picture. We found the intra-state rigidity $\hat{\mu}$
can be expanded in power series of the order parameter $A$ as
\eq{eq-mu-alpha-expansion-2} combined with \eq{eq-alpha-expansion}),
which reads,
\beq
\hat{\mu}(A)=c_{0}-c_{1} A + O(A^{2}).
\label{eq-mu-A}
\eeq
with $c_{0}$ and $c_{1}$ being positive constants related to
$\hat{\mu}_{0}$ and $-\hat{\mu}_{1}$ respectively (See \eq{eq-mu0} and
\eq{eq-mu1-rev}). 
The above expression \eq{eq-mu-A} implies 
larger cage size $A$ yields smaller rigidity $\hat{\mu}$.
Here it is instructive to note 
that the equivalent of the cage size $A$ in the spin-glass case is
$1-q$ with $q$ being the Edwards-Anderson order parameter as we point
out in Appendix \ref{appendix-p-spin} (see also Appendix \ref{appendix-spin-wave-vectorSG}).


\subsection{Fluctuation of the shape of the container : shear-strain fluctuations}
\label{sec-Integration-over-the-shear-strain-fluctuations}

So far we have implicitly assumed that the shape of the container,
parametrized by the shear-strain $\gamma$, is rigid and fixed. What will happen if we allow it to fluctuate? 
As long as the rigidity $\hat{\mu}$ is positive, we can safely 
integrate over the shear-strain $\gamma$ to obtain an elastic contribution 
to the free-energy of the cloned liquid. Using \eq{eq-mu-clonedliquid}
we find,
\beq
-\beta m \frac{F_{m}^{\rm elastic}(\alpha)}{N}
=\log \int \prod_{a}d \gamma^{a}
e^{-\beta \sum_{ab}\frac{\mu_{ab}}{2}\gamma^{a}\gamma^{b}}
=\log \sqrt{\frac{(2\pi)^{m-1}}{m^{m-2}\left(\beta \hat{\mu}(\alpha)/m\right)^{m-1}}}.
\eeq
with
\beq
\mu_{ab}=\hat{\mu}(\alpha)\left(\delta_{ab}-\frac{1}{m}\right).
\eeq
Here the intra-state rigidity $\hat{\mu}$ is given by 
\eq{eq-mu-alpha-expansion-2},
\beq
\hat{\mu}(\alpha)=\hat{\mu}_{0}+\hat{\mu}_{1}\alpha+O(\alpha^{2}).
\label{eq-mu-clonedliquid-again}
\eeq

\subsubsection{First order}
\label{sec-First-order-rigidity}

Let us first discuss the consequence of the fluctuation of the shear-strain $\gamma$
at the 1st order in the cage expansion. 
By adding $F_{m}^{\rm elastic}(\alpha)$ to the free-energy $F_{m}(\alpha)$
\eq{eq-free-energy-1st} and performing the Legendre transform \eq{eq-legendre-transform} we find
\beq
\alpha(A)=mA + \left( \frac{2m\beta}{d}\frac{1}{N}\sum_{i < j} \langle
\nabla^{2}  v(r_{ij}) \rangle_{*}+ \frac{m^{2}}{d}
\frac{\hat{\mu}_{1}}{\hat{\mu}_{0}}\right) A^{2}+O(A^{3})
\label{eq-alpha-A-mod}
\eeq
Then the free-energy $G_{m}(A)$ \eq{eq-G-1storder} is modified as,
\beq
\frac{\beta}{N} G_{m}(A)=
\frac{d(1-m)}{2m}\log(2\pi
A/\Lambda^{2})- A \frac{1-m}{m}\frac{1}{N}\sum_{i < j} \beta \langle \nabla^{2}
v(r_{ij}) \rangle_{*}
-\frac{1-m}{2m}
\log (\beta \hat{\mu}(\alpha(A))) + \ldots,
\label{eq-G-1stoder-corrected}
\eeq
where we omitted terms independent of the cage size $A$.
Now the equation $0 = \partial G_{m}(A)/\partial A$ by which the cage
size is determined \eq{eq-sp-A} is replaced  by,
\beq
0=\frac{1}{A}-\frac{2\beta}{d}  \frac{1}{N}\sum_{ i < j}
\langle \nabla^{2}v(r_{ij})\rangle_{*} 
-\frac{\hat{\mu}_{1}}{\hat{\mu}_{0}}\frac{m}{d} + O(A)
\label{eq-cage-size-1st-with-elasticity}
\eeq
which yields,
\beq
A=\frac{d}{2\beta \frac{1}{N} \sum_{ i < j}
\langle \nabla^{2}v(r_{ij})\rangle_{*}+ m \hat{\mu}_{1}/\hat{\mu}_{0}} .
\label{eq-cage-size-1st-corrected}
\eeq
Since $\hat{\mu}_{1}$ is negative (see \eq{eq-mu1-rev}) the result \eq{eq-cage-size-1st-corrected}
means that the fluctuation of the container tends to increase the cage size and thus weakens the glassy order.

In turn, the increase of the cage size will make the rigidity smaller.
Using \eq{eq-cage-size-1st-corrected} back 
in \eq{eq-alpha-A-mod},
we find again $\alpha=2mA$, which is the same as \eq{eq-variance-molecular-1st}. Thus
the intra-state rigidity of the system which is contained in the deformable container becomes,
\beq
\hat{\mu}=\hat{\mu}_{0}- 2m A \hat{\mu}_{1}.
\eeq
This is formally the same as the previous one \eq{eq-mu-clonedliquid} but it is now smaller than
the latter because the cage size $A$ is increased as found above \eq{eq-cage-size-1st-corrected}.

\subsubsection{Coupling of the cage size and the shear-strain fluctuations : a possible scenario}
\label{sec-coupling-scenario}

An interesting feature found above is that fluctuation of the shear-strain is {\it coupled}
to the cage size $A$, which is the order parameter of the glassy state, and makes it larger.
We suspect that the actual effect of the shear-strain fluctuation would
be much stronger than the behaviour found at the 1st order level.
Let us discuss a possible consequence of the coupling below.

By fully taking into account the elastic free-energy beyond the 1st order cage expansion,
the saddle point equation for the cage size $A$ \eq{eq-cage-size-1st-with-elasticity} may be cast into the following simple form,
\beqa
0=\frac{1}{A}-\frac{1}{A_{0}}+\frac{r}{1-r\frac{A}{A_{0}}}\frac{1}{A_{0}},
\label{eq-reduced-cage-size-sce0}
\eeqa
where $A_{0}$ is the cage size in the absence of the shear-strain fluctuation.
We also introduced a dimensionless parameter $r$ which represents the strength of the non-affine correction,
\beq
r \equiv -\frac{\hat{\mu}_{1}}{\hat{\mu}_{0}}  m A_{0}.
\label{eq-non-affinity}
\eeq
Note that $\hat{\mu}_{0} > 0$ and  $\hat{\mu}_{1} < 0$ (see \eq{eq-mu0}
and \eq{eq-mu1-rev}) so that $r$ is a positive parameter. 
We anticipate that it is an increasing function of the temperature $T$.
Although we have not performed higher order cage expansions explicitly, we expect that the 
structure of \eq{eq-reduced-cage-size-sce0} is generic. 
It can be simplified as,
\beqa
a=\frac{1}{1-r a}
\label{eq-reduced-cage-size-sce}.
\eeqa
where
\beq
a \equiv \frac{A}{A_{0}}.
\label{eq-reduced-cage-size}
\eeq
The solution of \eq{eq-reduced-cage-size-sce} is easily found as,
\beq
a=\frac{1-\sqrt{1-4r}}{2r}= 1+r+O(r^{2})  
\eeq
where we have took the physical solution which becomes $1$ in the limit $r \to 0$.
Most interesting feature is that the cage size exhibits a square-root singularity approaching 
a critical point,
\beq
a \simeq a_{c} \left (1- \sqrt{\frac{r}{r_{c}}-1}\right) \qquad (r \to r_{c}^{-}),
\eeq 
with $r_{c}=1/4$ and $a_{c}=2$ and disappears at $r_{c}$. 
Correspondingly this is reflected on the intra-state rigidity as,
\beq
\hat{\mu} \simeq \hat{\mu}_{c} \left(1 + \sqrt{\frac{r}{r_{c}}-1}\right) \qquad (r \to r_{c}^{-}),
\label{eq-mu-coupling}
\eeq
where $\hat{\mu}_{c}=\hat{\mu}_{0}/2$. Thus the rigidity exhibits a discontinuous jump 
preceded by a square-root singularity.

The above observations imply the consequence of the coupling between the
order parameter and the strain fluctuation can be significant leading to collapse of
metastable solid states when the strength of non-affinity $r$ become strong enough.
\blue{Interestingly enough the the discontinuous vanishment of the rigidity preceded by 
the square-root behaviour \eq{eq-mu-coupling} is apparently similar
to the expectation in the pure mean-field limit discussed in 
sec.~\ref{subsec-melting-orderparameter}. However here the rigidity is not merely
subjected to the order parameter at variance to the pure mean-field scenario 
but plays an essential role for the melting, partly supporting the intuition of Born's 
rigidity crisis scenario \cite{Born}.}


\subsection{Discussion: how the rigidity disappears?}
\label{subsec-discussion-melting}

An interesting general question is how the rigidity of the metastable solids 
disappear by raising the temperature. Of course glasses melt at the 
glass transition temperature $T_{\rm g}$.  But the effective rigidity $\hat{\mu}$ 
remains finite in the supercooled liquid state above $T_{\rm g}$
so that a supercooled liquid can be viewed as a solid with {\it finite} life time.
Our interest is how the rigidity of this temporally solid disappear by
increasing temperature.

\subsubsection{Mean-field theories}

In sec.~\ref{subsec-melting-orderparameter} we pointed out that mean-field
theories generically imply that the rigidify of solid states are simply 
proportional to the square of the
order parameters. Then it immediately follows that the  rigidity 
vanish discontinuously signaled by preceding
square-root behaviours approaching spinodal like melting
temperatures from below, simply reflecting the same behaviour of the order parameters.
We showed that this scenario holds in a Ginzburg-Landau theory for
superheated ferromagnet (sec.~\ref{subsubsec-rigidity-superheated-ferro}),
the mode coupling theory for supercooled liquids (sec.~\ref{subsubsec-MCT}) and a replica field
theory for vectorial spin-glasses (Appendix.~\ref{appendix-spin-wave-vectorSG}).

In sec \ref{sec-Cage expansion-of-the-rigidity} we performed a
microscopic computation of the rigidity of supercooled liquids and
glasses based on the cloned liquid theory.
Indeed the result also suggests that the rigidity is intimately related to the
order parameter $A$ as \eq{eq-mu-A}, which reads as,
\beq
\hat{\mu}(A)=c_{0}-c_{1} A + O(A^{2}).
\label{eq-mu-A-2}
\eeq
with $c_{0}$ and $c_{1}$ being certain  positive constants.
Since both the mode coupling theory \cite{MCT,MCT2} and the cloned liquid
theory \cite{parisi-zamponi} predict that 
the cage size $A$ exhibits a discontinuous jump 
preceded by a square-root behaviour 
$A(T_{\rm c})-A(T) \propto \sqrt{T_{\rm c}-T}$ as $T \to T_{\rm c}^{-}$, 
the rigidity $\hat{\mu}$ as given by \eq{eq-mu-A-2}
should also  exhibit a discontinuous jump at $T_{\rm c}$ preceded by a square
root singularity, at least at the mean-field level.
Let us note that a recent formulation of a replica approach to the rigidity of glasses by
Szamel and Flenner  \cite{szamel-flenner} is consistent with this scenario.

To summarize, at the {\it pure} mean-field level, 
the rigidity does behave as the order parameter. 
Said differently, at this level the {\it transverse} 
fluctuations such as shear-strains, spin-waves, e.t.c.
do {\it not} play any significant roles compared with the {\it longitudinal} fluctuations
such as the fluctuation of
the density, amplitude of vectorial order parameters, e.t.c. 
because the rigidity is totally subjected to the order parameter which parametrize
the strength of the longitudinal fluctuations.
The situation may change if we consider explicitly influences of the
transverse fluctuations going beyond the pure mean-field level as we discuss below.

\subsubsection{Influence of transverse fluctuations}

In sec \ref{sec-Integration-over-the-shear-strain-fluctuations}, we
discussed an alternative melting mechanism due to coupling between the order parameter, i.~e. the cage size $A$ and fluctuations of the shape of the container, i.~e. shear-strain.
Interestingly enough it also predicts discontinuous jumps of the cage
size $A$ and the rigidity $\hat{\mu}$ with square-root singularities.
However the driving mechanism of the melting is different from the one discussed above.
In this case the control parameter is the {\it non-affinity} $r$ defined
in \eq{eq-non-affinity} which is presumably an increasing function of
the temperature $T$.

At the pure mean-field level, fluctuation of the shear-strain can be
forgotten by just considering a container with a fixed shape. 
However we may view a finite dimensional system 
as an assembly of {\it deformable cells} for each of which we can define an effective
mean-field free-energy. Such a local mean-field free energy on a cell around a point ${\bf r}$ may be expressed in terms of the local order parameter $A({\bf r})$ 
and the shape of the cell parametrized by local shear-strain
$\gamma({\bf r})$. In this description the shear-strain $\gamma({\bf
r})$ naturally becomes a fluctuating continuous field (elastic strain
field) which we must integrate out. 

The equivalent of the shear-strain field $\gamma({\bf r})$ 
in spin systems are the derivative of the angular field
$\nabla \theta({\bf r})$ which parametrizes the spin-waves discussed in sec
\ref{subsubsec-rigidity-superheated-ferro} (ferromagnets) 
and Appendix \ref{appendix-spin-wave-vectorSG} (spin-glasses).
Apparently the the spin-wave $\nabla \theta({\bf r})$ must be integrated out in
order to take into account fluctuations around the mean-field theories.
As the result we will obtain certain {\it renormalized} effective potential
for the longitudinal fluctuations, which should be analogous to the one we discussed in sec. \ref{sec-Integration-over-the-shear-strain-fluctuations}.

However, as long as  the effects of the transverse fluctuations
can be took into account perturbatively, the overall basic characters of the spinodal 
like criticality themselves would remain intact. It will be very interesting to examine
possible non-perturbative effects of the transverse fluctuations in future.

\subsubsection{Comparison with related problems}

The spinodal like criticality is very different from the case of usual 2nd order phase
transitions (e.g. $\phi^4$ theory) where the order parameter behaves
continuously at the transition point so that the rigidity also exhibits a continuous behaviour across the transition temperature as we noted in the end of
sec. \ref{subsubsec-rigidity-superheated-ferro}. In the latter case the vanishing of the
rigidity is related to the divergence of static correlation length at
the transition through the so called Josephson's scaling relation \cite{Chaikin-Lubenski}.
Such continuous behavior of the order parameter and the rigidity have been
predicted for the cases of spin-glass transitions \cite{Kotliar-Sompolinski-Zippelius} and vulcanization transitions \cite{goldpart-goldenfeld-1989,zippelius-group,castillo-goldbart}.

Jamming transition is quite intriguing in this respect because it
cannot be understood simply along the line of thoughts discussed above: some simplified model systems for granular matters exhibit discontinuous jump of
the contact number preceded by a square-root singularity but continuous
behaviour of the rigidity at the (un) jamming transition
\cite{Nagel-group,hecke-2010,Zaccone-Romano-2011}.


\subsection{A model computation - binary soft sphere system}
\label{subsec-softsphere}

We now perform an explicit computation of the rigidity for the case of a binary mixture of soft-spheres in three dimensions \cite{binary-soft-sphere}, which is a standard model system to study the supercooled liquid state and the glassy state. We compare our results with the rigidity of the same system obtained by Barrat et. al. \cite{barrat-roux-hansen-klein-1988} using molecular dynamic (MD) simulations. 
Thermodynamic properties of the same system has been analyzed using the cloned liquid theory by Coluzzi~et.~al. \cite{coluzzi-mezard-parisi-verrocchio-1999}.

\subsubsection{Model}

The system consists of two kinds of particles $\epsilon=+,-$ 
which have different radius $\sigma_{+}$ and $\sigma_{-}$  and interact
with each other by a repulsive potential,
\beq
v^{\epsilon_{i}\epsilon_{j}}(r_{ij})=u_{0} \left(\frac{{\sigma}_{\epsilon_{i}}+\sigma_{\epsilon_{j}}}{r_{ij}} \right)^{12}.
\eeq
We introduce a smooth long-range cut-off of the potential following \cite{grigera-2002} which endures continuity up to the 2nd derivative of the potential.  
We choose the number fractions of the two types of particles as $x_{+}=x_{-}=1/2$ and the choose the ratio of the diameters as $\sigma_{+}/\sigma_{-}=1.2$ \cite{barrat-roux-hansen-klein-1988,roux-barrat-hansen-1989,barrat-roux-hansen-1990}. 
The radii $\sigma_{+}$ and $\sigma_{-}$ are fixed by requiring that the effective diameter to be equal to the unit length scale $l_{0}$, i.~e. $(\sum_{\epsilon=\pm,\epsilon'=\pm}x_{\epsilon}x_{\epsilon'}(\sigma_{\epsilon}+\sigma_{\epsilon'})^{3})^{1/3}=l_{0}$. Then the thermodynamic properties are determined by a single parameter $\Gamma=\rho l_{0}^{3}(k_{\rm B}T/u_{0})^{-1/4}$ . Here $\rho=N/V$ is
the number density where $V$  is the volume of the system measured in the unit of $l^{d}_{0}$. 
In the following we choose $\rho=1$ and study the temperature range $0.05 < T < 0.3$ where the temperature is measured in the unit of
$u_{0}/k_{\rm B}$. The mode coupling critical temperature $T_{\rm c}$ and Kauzmann transition temperature $T_{\rm K}$ of the system has been found to be 
$T_{\rm c} \sim 0.19-0.22$ by molecular dynamics simulations \cite{roux-barrat-hansen-1989}\cite{berthier-private-communication-2012} and $T_{\rm K} \simeq 0.125$ by the cloned liquid theory \cite{coluzzi-mezard-parisi-verrocchio-1999}. 

\subsubsection{Analysis}

Our analysis proceeds as follows. First we obtain some basic thermodynamic parameters.
To this end we analyzed the thermodynamics of the system following the scheme by Coluzzi~et.~al. in \cite{coluzzi-mezard-parisi-verrocchio-1999}. 
More precisely we used the 1st order cage expansion for the binary system which is a straight forward extension of the prescription \cite{mezard-parisi-1999} summarized in sec. \ref{subsubsec-molecular-liquid}.
We employed the binary Hyper Netted Chain (HNC) approximation
\cite{coluzzi-mezard-parisi-verrocchio-1999}, which is the HNC approximation \cite{morita-hiroike} adapted to the binary mixture, 
to obtain the radial distribution functions $g^{\epsilon \epsilon'}(r)$ between particles of type $\epsilon=\pm$ and
$\epsilon'=\pm$ of the liquid state of the binary system, and the free-energy of the liquid state \eq{eq-free-ene-liquid} which are needed in the thermodynamic analysis.
As the main outcome of the thermodynamic analysis, we obtain radial distribution functions $g^{\epsilon \epsilon'}(r)$ , the the glass-transition line $m=m^{*}(T)$ and the cage sizes $A_{\epsilon}(T)$ of the two kinds of particles $\epsilon=\pm$
which are indispensable in the following analysis.

Next we compute the intra-state rigidity $\hat{\mu}$ of the system at the level of 1st order cage expansion. To this end we need to evaluate various terms in \eq{eq-rigidity-result}
by representing them using particle distribution functions as explained in Appendix \ref{appendix-Representation-in-terms-of-particle distribution-functions}, \ref{sec-J12} and \ref{appendix-Formulations-for-the-binary-mixture}.
The values of the parameters $c_{\epsilon}$ \eq{eq-c-epsilon}, which controls the strength of the non-affine corrections, 
are obtained using the cage sizes $A_{\epsilon}$ and the parameter $m^{*}(T)$ determined in the thermodynamic analysis. We use the radial distribution functions $g^{\epsilon \epsilon'}(r)$ 
in the evaluation of the Born term \eq{eq-born-av-binary} and the non-affine correction term $J_{2}$ \eq{eq-j2-cage-expansion-binary}. 
All integrals $\int dr \ldots$ are done numerically. 
To evaluate the non-affine correction term $J_{1}$, which involves three-point particle distribution function (see \eq{eq-J1-decomposition}), we use the Kirkwood superposition approximation \eq{eq-kirkwood} which approximates
the three-point function as a product of two-point functions $g^{\epsilon \epsilon'}(r)$ (See  \eq{eq-J12-legendre-binary} and \eq{eq-legendre-kirkwood-binary}). 
To evaluate the non-affine correction term $J_{3}$ \eq{eq-j3-cage-expansion-binary}, we need to evaluate the derivatives $\left. \partial g^{\epsilon' \epsilon''}_{*}(r;\delta_{+},\delta_{-})/\partial \delta_{\epsilon} \right |_{\delta=0}$. 
In practice, we obtain them as follows: first we evaluate the radial distribution functions $g^{\epsilon' \epsilon''}(r)$ using the binary HNC \cite{coluzzi-mezard-parisi-verrocchio-1999}
for the modified potential \eq{eq-modified-v-binary} with some sufficiently small values of the perturbation $\delta_{\epsilon}$ and then we take numerical derivatives of the results with respect to $\delta_{\epsilon}$.

\begin{figure}[t]
\includegraphics[width=0.6\textwidth]{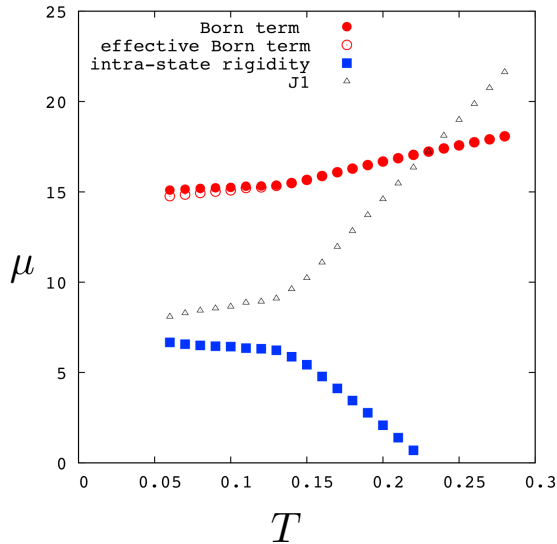}
\caption{Rigidity of a binary soft-sphere system, computed by
the cloned liquid approach. The Kauzmann temperature of
the system is $T_{\rm K} \sim 0.12$
 \cite{coluzzi-mezard-parisi-verrocchio-1999}
and the mode coupling critical temperature is around
$T_{\rm c} \sim 0.19-0.22$ \cite{roux-barrat-hansen-1989}\cite{berthier-private-communication-2012}.
The intra-state rigidity $\hat{\mu}$ (squares) 
is nearly independent of temperature below $T_{\rm K}$
while it sharply decreases with increasing temperature above $T_{\rm K}$.
On the other hand, the instantaneous rigidity, i.~e. the Born term (filled circles) 
\blue{increases with the temperature.} 
The renormalization of the Born term is very weak (open circles).
The significant part of the non-affine correction is due to 
the $J_{1}$ term (triangles), whose magnitude sharply increases with
 temperature above $T_{\rm K}$.
}
\label{fig-shearmodulus}
\end{figure}

\begin{figure}[t]
\includegraphics[width=0.7\textwidth]{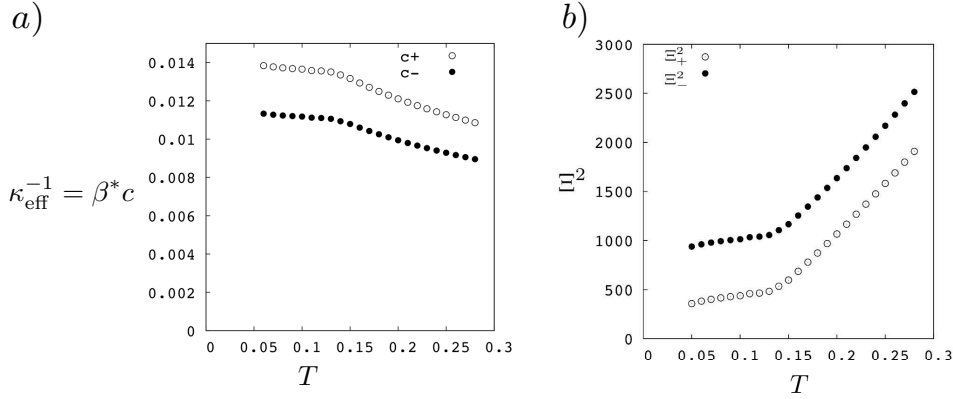}
\caption{Two temperature dependent factors in the non-affine correction term $J_{1}$.
In the binary system the $J_{1}$ term \eq{eq-j1-cage-expansion} 
can be written as (see \eq{eq-j1-cage-expansion-binary}),
$
mJ_{1}= \sum_{\epsilon=\pm} c_{\epsilon} \beta^{*} \Xi^{2}_{\epsilon}
$.
The panel a) shows the inverse of the effective Hookian spring constants $\kappa_{\epsilon}^{-1}=\beta^{*}c_{\epsilon}$ with $c_{\epsilon}$
as defined in \eq{eq-c-epsilon}, multiplied by $\beta^{*}=m^{*}(T) \beta$.
The panel b) shows the squared amplitudes of the ${\bf \Xi}$ fields,
 $\Xi^{2}_{\epsilon}$ as defined in \eq{eq-xi-binary}.
}
\label{fig-c-xi}
\end{figure}

\subsubsection{Results}

As shown in Fig.~\ref{fig-shearmodulus}, the intra-state
shear-modulus $\hat{\mu}$ exhibits the following characteristic temperature dependence.
On one hand it is almost independent of the temperature below the Kauzmann transition temperature $T_{\rm K} \sim 0.12$
and takes a value around $\hat{\mu} \sim 6$. On the other hand it becomes strongly dependent on the temperature at $T > T_{\rm K}$.
Evidently there is a significant difference between the Born term, which represents the affine part of the response, 
and the intra-state rigidity $\hat{\mu}$ {\it at all temperatures}. Even at very low temperatures the reduction of the rigidity due to non-affine corrections is significant, 
in agreement with numerical studies of static response to shear of energy minima (inherent structures)  \cite{Tanguy2002,maloney-lemaitre-2004b}. 
Moreover the sharp decrease of $\hat{\mu}$ above $T_{\rm K}$ is clearly due to the non-affine correction.
The fact that the Born term $b$ increases with increasing temperature
would appear somewhat puzzling but it simply means that particles, 
which are interacting with each other by purely repulsive interactions,
are colliding each other more often at higher temperatures
so that {\it instantaneous} increase of the energy (without any relaxation) with a given shear-strain $\gamma$ becomes larger at higher temperatures.

Let us examine the non-affine corrections in more detail. 
As discussed in sec. \ref{sec-summary-shearmodulus-clonedliquid}, the non-affine corrections can be divided into two groups: 1) the group of terms ($J_{2}$ and $J_{3}$) 
which yields the renormalized born term $b^{\rm eff}$ \eq{eq-def-born-renormalized} together with the original born term $b$ and 2) the $J_{1}$ term \eq{eq-j1-cage-expansion} which is distinct from others.
In Fig.~\ref{fig-shearmodulus} we included the curves of both the renormalized born term $b^{\rm eff}$ and the $J_{1}$ term. 
Apparently the major contribution to the non-affine correction is the $J_{1}$ term. 
As can be seen in Fig.~\ref{fig-shearmodulus}, it only weakly depends on the temperature 
at $T < T_{\rm K}$ but sharply increases with the temperature above $T_{\rm K}$. 

From the definition \eq{eq-j1-cage-expansion} (more precisely
\eq{eq-j1-cage-expansion-binary} which is adapted for the binary case),
it can be seen that there are two different sources of the temperature
dependence of the $J_{1}$ term.  
One is the factors $\kappa_{\epsilon}^{-1}=c_{\epsilon} \beta^{*}$  which are the inverse of the effective Hookian spring constants (see \eq{eq-c-A} and \eq{eq-hookian-const})
and the other is the amplitudes of the $\Xi$ fields.
As shown in Fig.~\ref{fig-c-xi}, the main source of the strong temperature dependence of the non-affine correction term above $T_{\rm K}$ is clearly the amplitude of the $\Xi$ field.
We have argued in sec. \ref{subsubsec-Rigidity-of-metabasin-or-intra-state-rigidity} and sec. \ref{sec-summary-shearmodulus-clonedliquid} that 
the non-affine correction term can explicitly depend on the temperature
because it reflects thermal fluctuations within metastable states (metabasins). 

\subsubsection{Discussions}

\begin{figure}[t]
\includegraphics[width=0.4\textwidth]{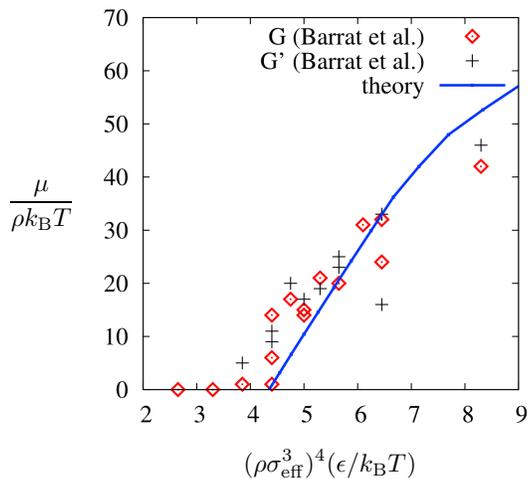}
\caption{Comparison of the rigidity of a binary soft-sphere system
 computed by Barrat et. al. (1988) \cite{barrat-roux-hansen-klein-1988}
 with $N=3456$ and the theoretical 
one. Here $G$ and $G'$ represent respectively the rhombohedral and tetragonal shear-modulus which become identical
in isotropic elastic systems.} 
\label{fig-shearmodulus-comparison}
\end{figure}

We found the intra-state shear-modulus $\hat{\mu}$ becomes almost
independent of the temperature below $T_{\rm K}$ suggesting that thermal fluctuations are significantly suppressed  at $T < T_{\rm K}$.
Then the value of $\hat{\mu} \sim 6$ at $T < T_{\rm K}$ may be compared
with the rigidity of inherent structures $\mu_{\rm IS}$  (see \eq{eq-T-0-mu-mu}).
Quite interestingly an analysis of $\mu_{\rm IS}$ \eq{eq-mu-IS-AQS} of
the same system indicates $\mu_{\rm IS} \sim 5$ \cite{Yoshino-Lemaitre} independently of the temperatures.
Similarly it may also be compared with the onset value of the plateau $G_{\infty}$ in the stress relaxation (see Fig.~\ref{fig-two-step-func-model}).
Indeed a molecular dynamic (MD) simulation of the same system indicates $G_{\infty} \sim 5$ \cite{Okamura-Yoshino-Yukawa} independently of the temperatures.

On the other hand, we found the intra-state shear-modulus $\hat{\mu}$ becomes smaller at higher temperatures at  $T > T_{\rm K}$ (See Fig.~\ref{fig-shearmodulus-comparison}).
This is very different from both $\mu_{\rm IS}$ \cite{Yoshino-Lemaitre} and $G_{\infty}$ \cite{furukawa,harrowell-2012,Okamura-Yoshino-Yukawa} 
which appear to depend little on the temperature up to surprisingly much higher temperatures. 
Thus we expect $G_{\rm p} \sim \hat{\mu}$ lies below $G_{\infty}$ at $T >T_{\rm K}$.
However, somewhat puzzlingly,  in numerical observations of stress-stress auto-correlation function $C_{\sigma}(\tau)$
\cite{furukawa,harrowell-2012,Okamura-Yoshino-Yukawa} there are no indications of the existence of such an rigidity $G_{\rm p}$ lying below $G_{\infty}$.
Probably this  means that the distinction between the $\beta$-relaxation
and the $\alpha$-relaxation is just a smooth crossover in the real dynamics at variance with the idealized picture shown in Fig.~\ref{fig-two-step-func-model}).

In Fig.~\ref{fig-shearmodulus-comparison} we compare the intra-state
rigidity $\hat{\mu}$ with the {\it static} rigidity obtained by a previous MD simulation on exactly the same system done by Barrat et. al. (1988) \cite{barrat-roux-hansen-klein-1988}. 
In the MD simulation \cite{barrat-roux-hansen-klein-1988} 
the {\it static} fluctuation formula for the elastic modulus (see sec. \ref{subsubsec-static-fluctuation-formula}) was used to compute the
elastic moduli. 
It appears that the theoretical curve, which is obtained with no adjustable parameters, compares reasonably well with the data points of the MD simulation.
However we should note that this comparison is delicate since static
rigidity of a truly equilibrated system of large enough system sizes
should become  $0$ at least down to $T_{\rm K} \sim 0.12$ (corresponding to $(\rho \sigma_{\rm eff}^{3})^{4}(\epsilon/k_{\rm B}T) \sim 8.3$ in Fig.~\ref{fig-shearmodulus-comparison}).

It is interesting to note that the intra-state shear-modulus $\hat{\mu}$, as well as the rigidity observed by the MD simulation \cite{barrat-roux-hansen-klein-1988}, crosses $0$ at around $T \sim 0.23$ (corresponding to $(\rho \sigma_{\rm eff}^{3})^{4}(\epsilon/k_{\rm B}T) \sim 4.3$ in Fig.~\ref{fig-shearmodulus-comparison}) ,  
which is rather close to the putative mode coupling critical temperature $T_{\rm c} \sim 0.19-0.22$ 
estimated by MD simulations
\cite{roux-barrat-hansen-1989}\cite{berthier-private-communication-2012}. This
feature may be interpreted in favor of the Born's original criteria for
melting \cite{Born}, i.~e. continuous vanishing of the rigidity approaching the spinodal temperature is valid for the present system. The situation is similar to the numerical observations of the rigidity by some MD simulations on superheated crystals \cite{wang,sorkin}.
However we note that our computation is limited to the 1st order cage expansion and does not tell us the location of the putative spinodal temperature
where the glassy metastable states actually disappear. 

The discussion in sec \ref{subsec-discussion-melting} suggests rather
that the system would actually exhibit discontinuous melting but
preceded by a continuous, square-root behavior
of both the cage size $A$ and the intra-state rigidity $\hat{\mu}$ at a somewhat lower temperature. 
Then the seemingly continuous vanishing of the shear-modulus $\hat{\mu}$ around $T_{\rm c}$ is a pure coincidence and
a small discontinuity may be hidden in the numerical data of \cite{barrat-roux-hansen-klein-1988} shown in  Fig.~\ref{fig-shearmodulus-comparison}.
We have discussed two qualitatively different driving mechanisms for such a behavior.
The strong temperature dependence of the non-affine correction term
$J_{1}$ shown in Fig.~\ref{fig-shearmodulus} and the fairly good
comparison of $\hat{\mu}$ with the data of the MD simulation shown in  Fig.~\ref{fig-shearmodulus-comparison} 
suggest that the coupling between the
cage size and the shear-strain fluctuation may be
important in the present system. In this respect it is interesting to note that the conventional mode coupling theory which does not take into account the coupling
of the shear-strain fluctuation and the density predicts a too large critical temperature $T_{\rm c} \sim 0.33$ \cite{barrat-latz-1990}.
However we must remind ourselves once more that melting of amorphous
solids can be at most just a smooth crossover in finite dimensional systems.

\section{Discussions}
\label{sec-discussions}

Before concluding this paper, let us present discussions on some problems.
\begin{itemize}
\item In the present paper we analyzed {\it static} rigidity of the amorphous metastable states of structural glasses. 
We discussed a possible interpretation of the hierarchy of the static
      rigidities in the context of rheology in
sec. \ref{sec-landscape} - sec. \ref{subsubsec-Rigidity-of-metabasin-or-intra-state-rigidity}.
In particular, we argued that the rigidity of the inherent structures $\mu_{\rm IS}$
and the intra-state rigidity $\hat{\mu}$ correspond respectively to the
      rigidity at the {\it onset} of the plateau regime $G_{\infty}$
and the {\it large time  limit} of the plateau regime $G_{\rm p}$ (see Fig.~\ref{fig-two-step-func-model}).
However the plateaus become obscure at higher temperatures and it becomes difficult to separate the $\beta$ and $\alpha$-relaxations.
A possible way out of this difficulty is to perform the analysis of fluctuation-dissipation relation (FDR)
discussed in Appendix \ref{sec-Out-of-equilibrium}
      \cite{Okamura-Yoshino-Yukawa} which allows one to ``eliminate
      times'' and focus on the breaking point of the FDR, which is
      expected to correspond to the crossover point between the $\beta$ and
      $\alpha$ relaxations, i.~e. $G_{\rm p}$.

\item From a broader perspective, the intra-state rigidity $\hat{\mu}$ is
just one of many intra-state susceptibilities which characterize {\it quasi-equilibrium} material properties of amorphous solids,
which are well accessible to experimental observations and important in practical use of amorphous solids. 
It is interesting to note that properties of the underlying ``ground state'' of a supercooled liquid can show up in the intra-state susceptibilities
even if the system as a whole is out of equilibrium.
As we discussed in sec. \ref{subsubsec-disentangle} and in sec. \ref{sec-recipe-disentangle} 
the cloned liquid approach realises decomposition of a generic susceptibility into
intra-state ($\beta$-relaxation) and inter-state ($\alpha$-relaxation).
This suggests possibilities to  make further progress in understanding material properties of 
amorphous solids out of equilibrium starting from microscopic Hamiltonians.

\item A natural step to go beyond the mean-field theoretical level is to consider spatial fluctuations of the order parameter i.~e. the cage size $A({\bf r})$, shear-strain $\gamma({\bf r})$ and e.t.c.
In particular we expect that the coupling of the shear-strain $\gamma({\bf r})$ and the cage size $A({\bf r})$ becomes important in finite dimensional systems
as we discussed in sec. \ref{sec-Integration-over-the-shear-strain-fluctuations} and sec. \ref{subsec-discussion-melting}.
Following the discussion in sec. \ref{subsec-discussion-melting}, we may write a phenomenological Ginzburg-Landau like free-energy functional of the cloned liquid as, 
\beq
F[A({\bf r}),\gamma({\bf r})]=\int d^{d}r \left[\sum_{ab}\frac{\rho\hat{\mu}(A({\bf r}))}{2}\left(\delta_{ab}-\frac{1}{m}\right)\gamma^{a}({\bf r})\gamma^{b}({\bf r})+\frac{c}{2} \left( \nabla A({\bf r})\right)^{2}+ g_{m}(A({\bf r})) \right]
\label{eq-GL}
\eeq
where $\rho=N/V$ is the number density, $g_{m}(A)=\rho (G_{m}(A)/N)$ is
      the local 'mean-field' free-energy with $G_{m}(A)$ given in
      \eq{eq-G-1stoder-corrected} and $\hat{\mu}(A)$ is the intra-state
      rigidity. \blue{This can be compared with the free-energy functional 
\eq{eq-GL-vectorSG-3} of the replica field theory of vectorial
      spin-glasses discussed in Appendix \ref{appendix-spin-wave-vectorSG}
where the equivalent of the shear-strain field $\gamma({\bf r})$ is the
      derivative of the angular field $\nabla \theta({\bf r})$ which
      parametrizes the spin-waves.}
The free-energy functional defines an ``effective Debye model'' for amorphous solids. To be completed the elastic free-energy associated with normal strain (compression) must also be added.
In principle all key parameters which
parametrizes the free-energy functional can be determined by microscopic
      computations using the mean-field theory (cloned liquid theory).
Especially computation of the bulk modulus which parametrizes the elastic free-energy
      associated with the normal strains can be done in the same way as done
      for the shear-modulus $\hat{\mu}$ in the present paper. 
However, we must note that such a coarse-grained
description cannot describe important {\it non}-Debye features 
in amorphous solids such as the boson peak and anomalous soft-modes around jamming point \cite{Wyart-2005}.

We found the rigidity depends on the cage size $A$ 
as \eq{eq-mu-A} which reads as,
\beq
\hat{\mu}(A)=c_{0}-c_{1} A + O(A^{2}).
\eeq
This implies the local rigidity $\hat{\mu}({\bf r})$ is smaller in the
      region where the local cage size $A({\bf r})$ is larger.
It might be related to the numerical observation of the elastic heterogeneity \cite{Tanguy2010}. 
It will be very interesting to study more in detail the consequences of
      the coupling between the transverse fluctuation (shear-strain) and
      longitudinal fluctuation (cage size) in the static and dynamic
      properties of amorphous solids and supercooled liquids, such as
      the dynamic heterogeneity
      \cite{yamamoto-onuki-1997,Widmer-Cooper-2006,Widmer-Cooper-2009,DH-book,furukawa}.

\item We argued that inter-state part of the response to quasi-static
      shear brings about jerky, intermittent stress-strain curves at
      mesoscopic scales (see sec. \ref{subsec-intermittent}). They are
      signatures of the inter-state fluctuations which are
      Goldstone modes in glasses which try to restore the broken the translational symmetry.  
It will be very interesting to perform some micro-rheological experiment
      \cite{Habdas2004} 
of slow strain rate in supercooled liquids to look for some signatures of the putative quasi-static intermittency.
A possible gedanken experiment may be to slowly drag a single particle 
immersed in a deeply supercooled liquid through a Hookian spring attached it.
Such an experiment can be realized for instance by applying magnetic
field to a magnetic particle immersed in a colloidal glass system \cite{Habdas2004}. 
In order to suppress melting, the system may be confined in a narrow
region between two walls with separation $L$ comparable or smaller than the mosaic 
length $\xi_{\rm mosaic}$ \cite{RFOT,RFOT-review}.
Then the interior of the local region (mosaic) around the particle will
      behave as a bulk system in the glass phase $T < T_{\rm K}$. 
By dragging the end of the spring slowly over some a small distance $x$, this region 
will be subjected to a local shear-strain of order $\delta \gamma \propto x/\xi_{\rm mosaic}$. 
Then the force curve $F(x)$ of the spring will exhibit intermittent
profiles either due to level crossings among different metastable states within the mosaic or yielding due to saddle-node bifurcation at metastability limits \cite{maloney-lemaitre-2004a,maloney-lemaitre-2006}.
A useful observable may be the force-force correlation function $C_F(x) \equiv \overline{F(0)F(x)}$ (where $\overline{\cdots}$ means averaging over the space or trajectories). 
Presumably the correlation function $C_F(x)$ exhibit a cusp like singularity at $x=0$ (but rounded over some width 
$\delta x_{\rm w} \propto T$ at finite temperatures) and decays rapidly
      as function of $x/\delta x_{\rm s}$ where $\delta x_{\rm s}$ is
      the typical spacing between the shocks. 
The conjecture \eq{eq-delta-gamma-s} implies $\delta x_{\rm s} \sim \xi_{\rm mosaic} \delta \gamma_{\rm s} \sim \xi_{\rm mosaic}^{1-d/2}$. 

\end{itemize}

\section{Conclusions}
\label{sec-conclusions}
In this paper we developed a first principle scheme to evaluate the rigidity of supercooled liquids and glasses by analyzing the rigidity of a cloned liquid, 
which might appear absurd because the rigidity of a liquid is just zero. We showed that the replica technique allows one to switch-on or off contributions
of the inter-state fluctuations of the shear-stresses to find non-zero rigidity of amorphous metastable solids and inter-state responses with intermittent nature.  
We discussed physical interpretation of the result within the context of
linear rheology and the phenomenological free-energy landscape picture.
We also presented a model computation of the rigidity of the binary
soft-spheres and found that the result compares well with the result of
a previous molecular dynamic simulation. A strong advantage of the cloned liquid
approach is that it fully makes use of conventional liquid theories such that one can choose the best one for a given problem at hand.
Since the rigidity is a basic property which is very important both to understand the physics of glassy systems and to use amorphous materials in practice,  
it will be certainly interesting to develop the method further and apply it in various glasses and jammed matters.

{\bf Acknowledgment} 
The author is grateful to Marc M\'ezard for the collaboration at the early stage of the present work and for many helpful discussions in the course of the present work.
He thanks Giulio Biroli, Jean-Philippe Bouchaud, Peter Harrowell, Ana\"el Lema\^{\i}tre, Satoshi Okamura and Grzegorz Szamel for constructive comments and discussions.
He thanks Jean-Louis Barrat, Ludovic Berthier, Andrea Cavagna, Jeppe C. Dyre, Silvio Franz, Akira Furukawa, Hikaru Kawamura, 
Jorge Kurchan, Florent Krzakala, Kunimasa Miyazaki, Tommaso Rizzo, Anne Tanguy, Eric Vincent and Francesco Zamponi for useful discussions. 
The author thanks LPTMS University of Paris Sud and SPEC CEA Saclay where some part of the present work has been done and for kind hospitality.
This work is supported by a {\it Triangle de la physique} grant number 117 "Intermittent response of glassy systems at mesoscopic scales"
and JPS Core-to-Core Program ``International research network for non-equilibrium dynamics of soft matter''.

\appendix

\section{Fluctuation formula for stress relaxation}
\label{appendix-ft}

\subsection{In equilibrium}

Here we outline a simple derivation of the generic fluctuation formula for the stress relaxation \eq{eq-linear-response-1} \cite{Williams-Evans-2009}.
To be specific we consider the time evolution of the system
in terms of Liouville equation of motion,
\beq
\frac{d}{dt} |P(t) \rangle  = {\cal L} |P(t) \rangle\!\rangle
\eeq
where vector $|P(t) \rangle\!\rangle$ represents probability distribution
of the positions ${\bf r}_{i}$ and  momentum ${\bf p}_{i}$ of particles
($i=1,2,\ldots,N$) and ${\cal L}$ is the Liouville operator;
\beq
{\cal L}=\prod_{i=1}^{N} \left(-\frac{{\bf p}_{i}}{m}\cdot \frac{\partial}{\partial {\bf r}_{i}}+\frac{d U}{d {\bf r}_{i}}\cdot \frac{\partial}{\partial {\bf p}_{i}}
\right).
\eeq
The equilibrium distribution, which is an eigen vector of the Liouville operator with zero eigen value ${\cal L}|P_{\rm eq} \rangle\!\rangle=0$,
is given by,
\beq
|P_{\rm eq} \rangle\!\rangle =Z^{-1}
e^{-\beta\sum_{i=1}^{N}\frac{ |{\bf p}_{i}|^{2}}{2m}} 
e^{-\beta U},
\label{eq-p-eq}
\eeq
where $Z$ is the partition function \eq{eq-def-free-energy}. 

The expectation value of the stress $\sigma$ observed at time $t$ can be written formally as,
\beq
\langle \sigma(t)  \rangle= \langle\!\langle \sigma e^{{\cal L}t} | P(0)\rangle\!\rangle
\eeq
where $\langle\!\langle$ on the r. h. s  of the above equation represents 
traces over the positions ${\bf r}_{i}$ and momentum ${\bf p}_{i}$ 
of the particles.
The stress is defined in \eq{eq-def-stress} which reads as,
\beq
\sigma= \left. \frac{1}{N}\frac{d U}{d\gamma} \right |_{\gamma=0}
\eeq
where $U$ is the potential part of the Hamiltonian of the system defined in \eq{eq-hamiltonian}. We assume that $U$ depends only on the positions ${\bf r}_{i}$ of the particles.

The linear response of the stress with respect to infinitesimal changes of
the strain $\delta \gamma(t)$ can be obtained formally as,
\begin{eqnarray}
\langle \delta \sigma(t) \rangle
&=& \langle\!\langle \frac{d \sigma}{d \gamma}  e^{{\cal L}t} |
P(0) \rangle\!\rangle \delta \gamma(t) +\int_{0}^{t} dt' \langle\!\langle \sigma e^{{\cal L}(t-t')} \frac{d {\cal
L}}{d\gamma}  e^{{\cal L}t'} | P(0) \rangle\!\rangle \delta \gamma(t') \\
&=& \langle b(t) \rangle \delta \gamma(t)
+\int_{0}^{t} dt' \langle\!\langle \sigma e^{{\cal L}(t-t')} \frac{d {\cal
L}}{d\gamma}  e^{{\cal L}t'} | P(0) \rangle\!\rangle \delta \gamma(t') 
\label{eq-linear-response-to-strain}
\end{eqnarray}
where 
\beq
b=\frac{d \sigma}{d\gamma}=\left. \frac{1}{N}\frac{d^{2} U}{d\gamma^{2}} \right |_{\gamma=0}
\eeq 
is the Born term defined in \eq{eq-born-expression}
which represents the instantaneous, affine response to shear. The 2nd term on the r.h.s  of
\eq{eq-linear-response-to-strain} represents non-affine responses to shear due to stress relaxations.

If the system is equilibrated at the beginning, i.~e.
$|P(0)\rangle\!\rangle=|P_{\rm eq}\rangle\!\rangle$ we find the fluctuation formula \eq{eq-linear-response-1},
\begin{eqnarray}
\langle \delta \sigma(t) \rangle
&=& \langle b \rangle \delta \gamma(t)-\beta\int_{0}^{t} dt'
\frac{\partial C_{\sigma}(t,t')}{\partial t'} 
\delta \gamma(t') 
\label{eq-fluctuation-formula-appendix}
\end{eqnarray}
where $C_{\sigma}(t,t')$ is the auto-correlation function of the stress, 
\beq
C_{\sigma}(t,t')=\langle \sigma(t) \sigma(t') \rangle
=\langle\!\langle \sigma e^{{\cal L}(t-t')} \sigma e^{{\cal L}t'}| P(0)\rangle\!\rangle
\eeq
Derivation of \eq{eq-fluctuation-formula-appendix} from
\eq{eq-linear-response-to-strain} is a standard one: one just need to notice,
\beq
\langle\!\langle \ldots  e^{{\cal L}(t-t')}\frac{d {\cal L}}{d\gamma} e^{{\cal L}t'}  | P_{\rm eq} \rangle\!\rangle 
= \beta \langle\!\langle \ldots  e^{{\cal L}(t-t')}\left( {\cal L} \frac{d U}{d \gamma}-\frac{d U}{d \gamma}{\cal L}  \right) e^{{\cal L}t'} | P_{\rm eq} \rangle\!\rangle 
= -\beta \frac{\partial}{\partial t'} \langle\!\langle \ldots  e^{{\cal L}(t-t')}\frac{d U}{d \gamma} e^{{\cal L}t'} | P_{\rm eq} \rangle\!\rangle.
\eeq
Here the 2nd equation follows from the fact that $|P_{\rm eq}\rangle\!\rangle$ is the equilibrium distribution given by \eq{eq-p-eq}. 

\subsection{Out-of-equilibrium}
\label{sec-Out-of-equilibrium}

In out-of equilibrium glasses, i.~e. glasses under aging or driven by external forces,
the equilibrium fluctuation formula \eq{eq-fluctuation-formula-appendix} cannot be expected. 
However the linear response can still be written formally as,
\begin{eqnarray}
\langle \delta \sigma(t) \rangle
&=& \langle b \rangle \delta \gamma(t)-\beta\int_{0}^{t} dt'
X(t,t')\frac{\partial C_{\sigma}(t,t')}{\partial t'} 
\delta \gamma(t')
\label{eq-fluctuation-formula-aging-appendix}
\end{eqnarray}
by introducing an unknown factor $X(t,t')$ which is $1$ in the case of equilibrium response. 

As suggested by the studies of the dynamical mean-field theory of
spin-glasses \cite{Cuku}, a natural way to investigate large time behaviors of out-of-equilibrium
glassy systems is to consider large time limits
$t, t' \to \infty$  such that the value of the correlation function is fixed
to a certain value $C_{\sigma}(t,t')=C$.  In such a limit the $X$ factor 
becomes parametrized solely by the value of the correlation function $C$,
\beq
\lim_{t' \to \infty, C_{\sigma}(t,t')=C}X(t,t')=X(C).
\eeq
Then in the case of the typical stress relaxation protocol discussed in
sec. \ref{subsubsec-two-step-relaxation},
the linear response in the large time limits $t, t' \to \infty$ with 
fixed $C=C_{{\sigma}}(t,t')$ can be written simply as,
\begin{eqnarray}
\lim_{t \to \infty, C_{\sigma}(t,t')=C} \frac{\langle \delta \sigma(t) \rangle}{\gamma}
&=& \lim_{t \to \infty}\langle b(t) \rangle -\beta\int_{C}^{\lim_{t \to \infty} C_{\sigma}(t,t)} dC' X(C')
\end{eqnarray}

The simplest ansatz for the function $X(C)$ would be to assume that it consists of two pieces,
\begin{eqnarray}
X(C) = \
\left \{
\begin{aligned}
 1 \hspace*{2cm} (\beta C > \mu_{\rm p}/x) \\
 x  \hspace*{2cm} (\beta C <  \mu_{\rm p}/x)
\end{aligned}
\right.
\label{eq-1step-x}
\end{eqnarray}
This is exactly the case for the class of mean-field spin-glass (MFSG) models which exhibit 1 step replica symmetry breaking (RSB)
including the $p$-spin spherical MFSG model \cite{Cuku,Cavagna-review-sg}. Because of the intimate analogy between the MFSG models which exhibit 1RSB
and glass phenomenology
\cite{Kirkpatrick-Thirumalai-1987,Kirkpatrick-Wolynes-1987a,RFOT,RFOT-review},  it is often assumed that $X(C)$ function of structural glasses
also has the form \eq{eq-1step-x}. This ansatz is supported by numerical
simulations \cite{Parisi-1999,Barrat-Kob-1999,Berthier-Barrat-2002}.

\section{Cloned liquid computation of the linear response of $p$-spin spherical mean-field spin-glass model}
\label{appendix-p-spin}

Here let us analyze static magnetic linear response of the $p$-spin spherical mean-field spin-glass (MFSG) model using the cloned liquid approach. 
The validity of the cloned liquid approach for this system was demonstrated 
in \cite{mezard-physicaA-1999}.
Here we demonstrate that the decomposition of the intra-state and inter
state linear magnetic susceptibility can be done exactly as discussed in
sec. \ref{subsubsec-disentangle} and also by using the cage expansion approach
discussed in sec. \ref{sec-recipe-disentangle}.

The Hamiltonian of the $p$-spin spherical MFSG model \cite{crisanti-sommers-1992}  is given by,
\beq
H= -\sum_{1 \leq i_{1} \leq i_{2} \leq \ldots i_{p} \leq N}
J_{i_{1}i_{2}\ldots i_{p}}s_{i_{1}}s_{i_{2}} \ldots s_{i_{p}}
-h \sum_{i=1}^{N} s_{i} 
\label{eq-p-spin-model}
\eeq
where the scalar spin variables $s_{i}$ at sites $i=1,2,\ldots,N$
Here we have introduced a uniform probing external magnetic field $h$.
are forced to satisfy the spherical constraint $\sum_{i=1}^{N}s_{i}^{2}=N$.
The random couplings $J_{i_{1}i_{2}\ldots i_{p}}$ are statistically independent
from each other and drawn from a Gaussian distribution
with zero mean and variance $J^{2}p!/(2N^{p-1})$.
In the case of $p \to \infty$ limit it reduced to the random energy model \cite{REM}. 

Let us consider a cloned system with $m$ replicas $a=1,2,\ldots,m$ 
with fixed overlap,
\beq
\frac{1}{N}\sum_{i=1}^{N}s^{a}_{i}s^{b}_{i}=q \qquad a \neq b.
\eeq
Static properties can be computed from the partition function of the 
cloned system, which is obtained as,
\beq
Z_{m}(T,{h^{a}})
=\int \prod_{a, b} D \lambda_{ab}
e^{-N(G(Q_{a,b},\lambda_{ab})+\delta G(\lambda_{ab},{h^{a}}))}
\label{eq-zm-pspin}
\eeq
where
\beq
G(Q_{ab},\lambda_{ab})=-\frac{\beta^{2} J^{2}}{4}\sum_{a  b} Q_{a b}^{p}
-\sum_{a b} \lambda_{ab} Q_{ab}+\frac{1}{2}\log ({\rm det} (2\lambda_{ab})/\pi^{m})
\eeq
with
\beq
Q_{ab}=(1-q)\delta_{ab}+q.
\eeq
and
\beq
\delta G(\lambda_{ab}, {h^{a}})= - \log
\frac{1}{\sqrt{\pi^{m} / \rm{det} (2\lambda_{ab})}}
\int \prod_{a} D s^{a}
e^{ -\sum_{ab} \lambda_{ab} s^{a}s^{b}
-\sum_{a} h^{a}s^{a} }=-\frac{1}{4}\sum_{ab}  h^{a}(\lambda^{-1})_{ab} h^{b}
\label{eq-dG}
\eeq
In \eq{eq-zm-pspin} we have simply took average of the partition function 
over disorder of the random couplings.
This is the so called as annealed average which is valid only in the
high temperature phase where replica symmetry is not broken. 
We analyze the low temperature region choosing small enough $m$ so that
we remain in the high temperature phase (See
Fig.~\ref{fig-m-T-diagram-complexity} a)).

In \eq{eq-zm-pspin} we have introduced different probing field $h^{a}$ on different replicas. They are assumed to be infinitesimally small. 
Performing the integration over $\lambda_{ab}$ by the saddle point method (
at $h^{a}=0)$ we find,
\beq
\lim_{N \to \infty} -\beta m \frac{F_{m}(T,{h^{a}})}{N}
= -G(Q_{ab},\lambda^{*}_{ab})+\frac{1}{2}\sum_{ab}h^{a}Q_{ab}h^{b}
\qquad \lambda^{*}_{ab}=\frac{1}{2}Q^{-1}_{ab}
\eeq
This is formally the same as the 1 step replica symmetry breaking (RSB)
solution \cite{crisanti-sommers-1992}. 
Thus by performing extremization with respect to the parameters $q$ and $m$
one recovers the exact 1RSB free-energy of the system \cite{mezard-physicaA-1999}.

Using the above results the linear-susceptibility matrix $\chi_{ab}$
\eq{eq-chi-decomposition-clone} is readily obtained as
\beq
\chi_{ab}=\beta Q_{ab} =\beta(1-q)\delta_{ab}+\beta q.
\label{eq-sus-matrix-pspin}
\eeq
Comparing with \eq{eq-chi-decomposition-clone} we can easily
identify the intra-state linear susceptibility
$\hat{\chi}=\beta(1-q)$ and inter-state susceptibility$\tilde{\chi}=m\beta  q$.
The total linear magnetic susceptibility becomes
\beq
\chi=\sum_{b=1}^{m} \chi_{ab}= \beta (1-q)+m \beta q 
\label{eq-1RSB}
\eeq
As expected these results agree with those of the usual 1 step RSB
solution \cite{crisanti-sommers-1992}.
In the standard interpretation of the RSB solution \cite{Parisi-1983}
for the spin-glass phase below the static spin-glass transition temperature $T_{\rm SG}$,
the factor $m$ (written rather as $x$ in the context of RSB in quenched
disordered systems) in front of the 2nd term on the r.~h.~s is understood as the probability 
that two replicas stay at different metastable states. More physically
the factor $m$ roughly corresponds to the probability that the system stay at an low lying 
'excited' metastable state rather than 'ground' metastable state in the glass phase. 
The parameter $m$ is found to behave typically behaves as $m \sim T/T_{\rm SG}$ \cite{REM}.

In the temperature range $T_{\rm SG} < T < T_{\rm d}$ (corresponding to
$T_{\rm K} < T < T_{\rm c}$), the system in the liquid state 
in the thermodynamic sense. However the RSB solution \eq{eq-1RSB} 
is still very useful if we consider $m \to 1$ limit in the following sense. 
First note that the total
susceptibility becomes $\lim_{m \to 1^{-}}\chi= \beta$ so that the correct
paramagnetic susceptibility, which could be obtained by the replica symmetric (RS) ansatz
which yields $q=0$, is recovered {\it independently of the value of} $q$.
The value $q$ should be interpreted as self-overlap of metastable states
which can be obtained as $\lim_{m \to 1^{-}} q$ using the RSB ansatz 
or equivalently the cloning liquid like methods \cite{monasson-1995,Franz-Parisi-1997,mezard-physicaA-1999}. Indeed the latter procedure correctly reproduce the self-overlap value $q$ of
metastable states which dominates the thermodynamics of the ``liquid state'' at
$T_{\rm SG} < T < T_{\rm d}$. Thus we can still decompose the total susceptibility 
into the intra-state susceptibility $\hat{\chi}=\beta(1 - \lim_{m \to 1^{-}}q)$
and the inter-state susceptibility $\tilde{\chi}=\beta \lim_{m \to 1^{-}}q$
in this intermediate temperature range.

Now let us try the cage expansion approach to evaluate the linear
susceptibilities as discussed in sec. \ref{sec-recipe-disentangle}.
To this end we decompose the spin variables into two parts as,
\beq
s_{i}^{a}=S_{i}+\delta s^{a}_{i}
\eeq
where $S_{i}$  is the 'center of mass' of the spin
\beq
S_{i}=\frac{1}{m}\sum_{a=1}^{m}s_{i}^{a}
\label{eq-decomposition-spin}
\eeq
and $\delta s^{a}$ is the molecular coordinate which describes fluctuations
around the 'center of mass'. 
As far as the spherical model is concerned,
the effective action is already quadratic in terms of the spin variables (see 
\eq{eq-dG}) so that the decomposition \eq{eq-decomposition-spin}
is not particularly needed in order to compute the free-energy.
Nonetheless, examination of the linear susceptibility by this approach
may be instructive as we outline below.

The Hamiltonian of the cloned system can be written in terms of the new variables as,
\beq
H_{m}=-\sum_{1 \leq i_{1} \leq i_{2} \leq \ldots i_{p} \leq N}
J_{i_{1}i_{2}\ldots i_{p}} \sum_{a=1}^{m}(S_{i_{1}}
+ \delta s^{a}_{i_{1}})(S_{i_{2}}+\delta s^{a}_{i_{2}}) \ldots (S_{i_{p}}+\delta s^{a}_{i_{p}})
-h \sum_{i}S_{i} - \sum_{a}  \delta h^{a} \sum_{i} \delta s^{a}_{i}
\eeq
where we have introduced probing fields conjugated to the 'center of
mass coordinate'  $S_{i}$ and 'molecular  coordinate' $\delta s^{a}_{i}$.
Then we only need to replace \eq{eq-dG}) with,
\beqa
\delta G(\lambda_{ab}, h, {\delta h^{a}}) &=& - \log 
\frac{1}{\sqrt{\pi^{m} / \rm{det} (2\lambda_{ab})}}
\int DS \prod_{a}D \delta s^{a} m \delta(\sum_{a}\delta s^{a})
e^{-\sum_{ab} \lambda_{ab} (S+\delta s^{a})(S+\delta s^{b})
-h S- \sum_{a} \delta h^{a}\delta s^{a}} \nonumber \\
&& = - \frac{1+(m-1)q}{m}\frac{h^{2}}{2}
+\frac{1}{2}\sum_{ab} \delta h^{a}  (1-m \delta_{ab})\frac{1-q}{m}\delta h^{b}
\eeqa
Now we can readily find,
\beq
\langle \delta s^{a} \delta s^{b} \rangle = -(1-m \delta_{ab})\frac{1-q}{m}.
\label{eq-1-q}
\eeq
This result is reminiscent of \eq{eq-variance-molecular}
combined with \eq{eq-alpha-expansion}). We can notice indeed that
$1-q$ plays precisely the role of cage size $A$.
Using the above results the linear-susceptibility matrix $\chi_{ab}$ can be obtained as,
\beq
\chi_{ab}=\beta \langle s^{a} s^{b} \rangle = \beta \langle (s+\delta
s^{a})(s+\delta s^{a}) \rangle 
= \beta  \langle S^{2}\rangle  + \beta  \langle \delta s^{a}\delta s^{b}\rangle 
=\beta[(1-q)\delta_{ab}+ q].
\eeq
The last result agrees with \eq{eq-sus-matrix-pspin} obtained without
using the cage expansion.


\section{Rigidity of spin-waves in vectorial spin-glass systems}
\label{appendix-spin-wave-vectorSG}

Here we present a simple mean-field theoretical analysis of the rigidity of spin-waves in a class of vectorial spin-glass systems.
Our purpose is to extend the analysis of the rigidity of spin-waves in
the superheated ferromagnets discussed in sec. \ref{subsubsec-rigidity-superheated-ferro} to the case of 
some spin-glass systems approaching the dynamical transition temperature
$T_{d}$. To this end we consider a replica field theory with rotational
invariance following Ref. \cite{Kotliar-Sompolinski-Zippelius} which analyzed
the case of the Sherrington-Kirkpatrick model with vectorial spins. 

Although the following discussion does not depend on the details of specific
choices of microscopic models but we may consider
as a specific example the $p$-spin spherical MFSG model \eq{eq-p-spin-model} 
extended to a model with vectorial spins,
\beq
H= -\sum_{1 \leq i_{1} \leq i_{2} \leq \ldots i_{p} \leq N}
J_{i_{1}i_{2}\ldots i_{p}} (\vec{s}_{i_{1}} \cdot \vec{s}_{i_{2}}) 
 (\vec{s}_{i_{3}} \cdot \vec{s}_{i_{4}}) 
\ldots  (\vec{s}_{i_{p-1}} \cdot \vec{s}_{i_{p}}). 
\label{eq-vector-p-spin-model}
\eeq
The spin variables are two-component vectors $\vec{s}_{i}=(s_{i,1},s_{i,2})$
and the symbol $\cdot$ represents the inner product,
i.~e. $\vec{s}_{i}\cdot \vec{s}_{j} \equiv
\sum_{\mu=1,2}s_{i,\mu}s_{j,\mu}$. 
Note that the model has the $O(2)$ symmetry just as the ferromagnetic
system considered in sec. \ref{subsubsec-rigidity-superheated-ferro}.
The vectorial mean-field model can be solved by introducing a tensorial order parameter,
\beq
Q^{\mu \nu}_{ab}=\frac{1}{N}\sum_{i=1}^{N} \langle S^{a}_{i,\mu}S^{b}_{i,\nu} \rangle,
\eeq
where $\langle \ldots \rangle$ stands for thermal averages.

This class of systems with $p$ being an even integer greater than $2$,
i.~.e $p=4,6,8,\ldots$ exhibit the dynamical transition at $T_{\rm d}$ and
static spin-glass transition at $T_{\rm SG}$ 
at a lower temperature just as the original scalar $p$-spin MFSG model 
\cite{crisanti-sommers-1992} discussed in appendix \ref{appendix-p-spin}. The case of $p=2$ is the
vectorial version of the Sherrington-Kirkpatrick model studied in
\cite{Kotliar-Sompolinski-Zippelius} which exhibits the usual spin-glass
transition at a single critical temperature.

To analyze generic properties of spin-waves in system as the one given above, 
let us consider a replica field theory of $m$ replicas ($a=1,2,\ldots,m$)
with a schematic free-energy functional,
\beq
F\left[Q_{ab}^{\mu \nu}\right]=\int d^{d}r \left[
\frac{c}{4} \sum_{a <b}\sum_{\mu \nu}\left(\nabla Q^{\mu \nu}_{ab}\right)^{2}
+w\left(
\sum_{\mu \nu}\left(Q^{\mu \nu}_{ab}\right)^{2}
\right)
\right].
\label{eq-GL-vectorSG}
\eeq
Here the order parameter $Q^{\mu \nu}_{ab}({\bf r})$ is considered
as a continuous field in a $d$-dimensional space.
Just as considered in appendix \ref{appendix-p-spin} we assume that this is
a cloned system: replicas are forced to belong to a common equilibrium
state. The local potential $w\left(\sum_{\mu \nu}\left(Q^{\mu \nu}_{ab}\right)^{2}\right)$ in the integrand  of \eq{eq-GL-vectorSG}
can be computed as the mean-field free-energy of the underlying
microscopic model such as the one given by \eq{eq-vector-p-spin-model}
but the details are not important in the following discussion.

As noted in \cite{Kotliar-Sompolinski-Zippelius}, an 
important symmetry property of the free-energy functional \eq{eq-GL-vectorSG}
is that it is invariant under global rotations of any one (or more) of the replicas:
it is invariant under a generic transformation of the form,
\beq
Q^{\mu \nu}_{ab} \to \sum_{\mu' \nu'}
(R^{t})^{\mu \mu'}(\theta_{a})Q^{\mu' \nu'}_{ab} R^{\nu' \nu}(\theta_{b})
\eeq
where $\theta_{a}$ and $\theta_{b}$ are the angles of the global rotations of the replica $a$ and $b$ respectively.
The rotations are represented by the rotational matrix $R^{t}(\theta)$,
\beq
R^{t}(\theta) \equiv
\left(
\begin{array}{cc}
\cos(\theta) & -\sin(\theta) \\
\sin(\theta) & \cos(\theta) \\
\end{array}
\right).
\eeq
Then by choosing a specific reference coordinate system for the
$2$-component spin space in which
the matrix $Q^{\mu \nu}_{ab}$ becomes diagonal, i.~.e. $Q^{\mu \nu}_{ab}=q_{ab}\delta_{ab}$,
a generic replica field $Q^{\mu \nu}_{ab}$ can be parametrized as,
\beq
Q^{\mu \nu}_{ab}=
q_{ab}\sum_{\gamma}
(R^{t})^{\mu \gamma}(\theta_{a}) R^{\gamma \nu}(\theta_{b})
=q_{ab}(R^t)^{\mu \nu}(\theta_{a}-\theta_{b}).
\eeq
Using the latter parametrization the
free-energy functional \eq{eq-GL-vectorSG} becomes,
\beq
F\left[Q_{ab}^{\mu \nu}\right]=\int d^{d}r \left[
\frac{c}{2} \sum_{a<b}q^{2}_{ab} \left(\nabla (\theta_{a}- \theta_{b})\right)^{2}
+\frac{c}{2} \sum_{a< b}\left(\nabla q_{ab}\right)^{2}
+w\left( q_{ab}^{2}\right)
\right].
\label{eq-GL-vectorSG-2}
\eeq
Note that the latter two terms of the integrand appear also in the usual
scalar replica field theories. On the other hand, the 1st term can be naturally interpreted as the spin-wave term
just as in the case of the ferromagnetic model (See \eq{eq-GL-model-2}).

As discussed in appendix \ref{appendix-p-spin}, the matrix $q_{ab}$ of
the cloned system can be parametrized as,
\beq
q_{ab}=(1-q)\delta_{ab}+q
\eeq
with $q$ being the self-overlap, i.~e. the Edwards-Anderson order
parameter.  Then the free-energy functional can be simplified as,
\beq
F\left[Q_{ab}^{\mu \nu}\right]=\int d^{d}r \left[
\frac{1}{2} \sum_{a,b} \mu_{ab} \nabla \theta_{a}\nabla \theta_{b}
+\frac{c}{2} \sum_{a< b}\left(\nabla q\right)^{2}
+w(q^{2})
\right]
\label{eq-GL-vectorSG-3}
\eeq
Here $\mu_{ab}$ is the rigidity matrix of the spin-waves which is
obtained as,
\beq
\mu_{ab}= c mq^{2}\left(\delta_{ab}-\frac{1}{m}\right).
\label{eq-rigiditymatrix-spin-wave}
\eeq
Remarkably the rigidity matrix \eq{eq-rigiditymatrix-spin-wave} has precisely the same structure as that for the shear-modulus of glasses (see sec. \ref{subsec-shear-on-the-cloned-liquid}):
$\mu=\hat{\mu}\delta_{ab}+\tilde{\mu}$ with
$\tilde{\mu}=-(1/m)\hat{\mu}$ so that the sum rule $\sum_{b=1}^{m}\mu_{ab}=0$
is satisfied. Now the 'intra-state spin-wave rigidity' can be easily read off as,
\beq
\hat{\mu}=cm q^{2}.
\label{eq-rigidity-spin-wave-vectorSG}
\eeq
Evidently this is quite similar to the case of ferromagnet discussed in
sec. \ref{subsubsec-rigidity-superheated-ferro} in the sense that the spin-wave
rigidity is proportional to the square of the order parameter (here the
Edwards-Anderson order parameter $q$). 

We readily know that the Edwards-Anderson order parameter $q$ jumps
discontinuously to $0$ approaching the dynamical transition temperature
$T_{\rm d}$ from below, preceded by a square-root singularity,
\beq
q(T)-q(T_{\rm d}) \propto \sqrt{T_{\rm d}-T}.
\eeq
Using the latter in \eq{eq-rigidity-spin-wave-vectorSG}, we immediately
find that the 'intra-state spin-wave rigidity' should discontinuously 
vanish approaching $T_{\rm d}$ from below preceded by a square-root
singularity,
\beq
\hat{\mu}(T)-\hat{\mu}(T_{\rm d}) \propto c m \sqrt{T_{\rm d}-T},
\eeq
where $m=1$ at $T_{\rm SG} < T < T_{\rm d}$.
Apparently these features are the same as 
the superheated ferromagnet discussed in
sec. \ref{subsubsec-rigidity-superheated-ferro}.


\section{Proof of the relation \eq{eq-J-identity} }
\label{sec-identity}

Suppose that the original 2-body interaction potential $v(r)$ 
(See \eq{eq-hamiltonian}) is slightly deformed as
$v(r) \to v(r)+\delta w(r)$ where $w(r)$ is an additional potential
and $\delta$ is a small parameter.
Notice a trivial fact that the rigidity \eq{eq-def-shearmodulus} must be zero
irrespective of the deformation of the potential as long as the system remains in
the liquid phase. Then the identity,
\beq
\langle b \rangle -\beta [\langle \sigma^{2} \rangle -\langle \sigma \rangle^{2} ]=0,
\label{eq-rigidity-of-liquid-is-zero}
\eeq
must continue to hold under variation of the parameter $\delta$. 
Especially the form of the renormalized potential $v_{\rm eff}(r)$ given 
in \eq{eq-veff} suggests us to consider the case $w(r)=\nabla^{2}v(r)$.

Let us examine changes of the thermal averages on the l.~h.~s. of 
\eq{eq-rigidity-of-liquid-is-zero} under variation of the parameter $\delta$.
From  the microscopic definitions of the shear-stress $\sigma$
and the Born term $b$ given in \eq{eq-def-stress} and \eq{eq-def-born}, we
find that they are actually derivatives of the potential $v(r)$ so that
they change under variation of $\delta$.
Thus the thermal averages can expressed formally as,
\beq
\langle O_{[v(r)+\delta w(r)]} \rangle = Z^{-1} \int_{\cal V} \prod_{i=1}^{N}
\frac{d^{d}r_{i}}{\Lambda^{d}} e^{-\beta
\sum_{i<j}v(r)+\delta w(r)} O_{[v(r)+\delta w(r)]} 
\qquad Z=\int_{\cal V} \prod_{i=1}^{N}
\frac{d^{d}r_{i}}{\Lambda^{d}} e^{-\beta
\sum_{i<j}v(r)+\delta w(r)}
\label{eq-thermal-average-inflation}
\eeq
with $O$ being the Born term $b$, the shear-stress $\sigma$ or $\sigma^{2}$.
Expanding \eq{eq-thermal-average-inflation} in power series of $\delta$
and inserting the results into \eq{eq-rigidity-of-liquid-is-zero},
we can find an identity at each order of $\delta$.

Choosing the additional potential in particular as $w(r)=\nabla^{2}v(r)$, 
we find at $O(\delta)$ the following identity,
\begin{eqnarray}
0 = &&\frac{1}{N}\sum_{i < j}  \langle  \nabla^{2} b({\bf r}_{ij}) \rangle
-\frac{1}{N}\sum_{i < j}\sum_{k < l} \beta [ \langle  b({\bf r}_{ij}) \nabla^{2} v(r_{kl})
\rangle 
-\langle  b({\bf r}_{ij}) \rangle \langle \nabla^{2} v(r_{kl})\rangle]
\nonumber \\
&&-2\frac{1}{N}\sum_{i_{1} < j_{1}}\sum_{i_{2} < j_{2}}\beta [\langle \nabla^{2}
\sigma({\bf r}_{i_{1}j_{1}}) \sigma({\bf r}_{i_{2}j_{2}}) \rangle-\langle \nabla^{2}
\sigma({\bf r}_{i_{1}j_{1}})\rangle \langle \sigma({\bf r}_{i_{2}j_{2}}) \rangle]
\nonumber \\
&&
+\frac{1}{N} \sum_{i_{1} < j_{1}}\sum_{i_{2} < j_{2}}\sum_{k <
l}\beta\beta [\langle  \sigma({\bf r}_{i_{1}j_{1}}) \sigma({\bf
r}_{i_{2}j_{2}}) \nabla^{2}v(r_{kl}) \rangle-\langle \sigma({\bf
r}_{i_{1}j_{1}})\sigma({\bf r}_{i_{2}j_{2}})\rangle \langle \nabla^{2}v(r_{kl}) \rangle]
\label{eq-identity}
\end{eqnarray}

Using the above result for the liquid at temperature $T/m^{*}$, we find
the relation \eq{eq-J-identity},
\beq
J_{2}+J_{3}+m^{*}(J_{4}+J_{5})=0.
\eeq

\section{Representation in terms of particle distribution functions}
\label{appendix-Representation-in-terms-of-particle distribution-functions}

In order to compute each terms in \eq{eq-rigidity-result}, we need to 
evaluate thermal averages $\langle \ldots \rangle_{*}$ over the thermal fluctuations
of several quantities which are multi-point functions of the CM coordinates of the molecules. To this end it is convenient to represent each terms in \eq{eq-rigidity-result} and related quantities in terms of particle distribution functions,
\begin{eqnarray}
&& \rho^{2} g_{*}^{(2)}({\bf r},{\bf r}')=\sum_{i \neq j} \langle \delta^{d}({\bf r}_{i}-{\bf r})\delta^{d}({\bf r}_{j}-{\bf r}') \rangle_{*} \qquad 
\rho^{3}g_{*}^{(3)}({\bf r},{\bf r}',{\bf r}'')=\sum_{i \neq j \neq k}
\langle \delta^{d}({\bf r}_{i}-{\bf r})\delta^{d}({\bf r}_{j}-{\bf r}')\delta^{d}({\bf r}_{k}-{\bf r}'')\rangle_{*}, \qquad \ldots  \nonumber \\
\end{eqnarray}
where $\rho=N/V$ is the number density. Here we introduced a short-hand notations 
$\sum_{i \neq j} = \sum_{i} \sum_{j} (1-\delta_{ij})$, $\sum_{i \neq j \neq k} = \sum_{i} \sum_{j}\sum_{k} (1-\delta_{ij}\delta_{jk})$.  
Since we will consider homogeneous and isotropic systems, we can write,  
\begin{eqnarray}
&& \rho^{2} g(r)=\frac{\rho}{N}\sum_{i \neq j} 
\langle \delta^{d}({\bf r}_{ij}-{\bf r}) \rangle   \\
&& \rho^{3}g^{(3)}({\bf r},{\bf r}')=\frac{\rho}{N} \sum_{i \neq j \neq k}
\langle \delta^{d}({\bf r}_{ij}-{\bf r})
\delta^{d}({\bf r}_{ik}-{\bf r}')\rangle=\sum_{l=0}^{\infty} g^{(3)}_{l}(r,r')P_{l}(\cos(\theta))
\qquad \cos(\theta)=\frac{{\bf r}\cdot{\bf r}'}{rr'}
\label{eq-g3-legendre}
\end{eqnarray}
In the first equation $g(r)=g^{(2)}(r)$ is the usual radial distribution function.
In the last equation $P_{l}(x)$ ($l=0,1,2,\ldots$) are the Legendre polynomials ($P_{0}(x)=1$, $P_{1}(x)=x$,$P_{2}(x)=\frac{1}{2}(3x^{2}-1)$,...) and $\theta$ is the angle
between vector ${\bf r}$ and ${\bf r'}$. The coefficients $g^{(3)}_{l}(r,r')$ are defined as,
\beq
g^{(3)}_{l}(r,r') \equiv \frac{2}{2l+1}\int_{0}^{\pi} d\theta \sin(\theta)
P_{l}(\cos(\theta))g^{(3)}({\bf r},{\bf r}').
\eeq

The radial distribution function $g(r)$ can be computed using
the standard closures of the liquid theory \cite{hansen-mcdonald} such
as the hyper-netted-chain approximation (HNC) \cite{morita-hiroike} and e.t.c. A simple way to evaluate the three-point correlation function $g^{(3)}({\bf r},{\bf r}')$ is the Kirkwood superposition approximation,
\beq
g^{(3)}({\bf r},{\bf r}')= g(r)g(r')g(|{\bf r}-{\bf r}'|).
\label{eq-kirkwood}
\eeq

\begin{enumerate}

\item Cage size

The cage size $A$ at the 1st order cage expansion \eq{eq-cage-size-1st} of $d=3$ dimensional system can be expressed as,
\beq
A=\frac{3}{\beta \rho \int_{0}^{\infty} dr 4\pi r^{2}  g_{*}(r) \nabla^{2} v(r)}
\label{eq-cage-size-av}
\eeq
with the radial distribution function $g_{*}(r)$ at temperature $T^{*}=T/m^{*}(T)$ and,
\beq
\nabla^{2}v(r)=v^{(2)}(r)+2\frac{v^{(1)}}{r}.
\eeq

\item Affine response

Similarly the Born term  can also be expressed just by the radial distribution function $g(r)$ as,
\begin{eqnarray}
 \langle b_{[\{{\bf R}_{i}\}]}\rangle_{*} && =
\frac{1}{N}\sum_{i < j} \langle b({\bf r}_{ij}) \rangle_{*} 
 = \frac{\rho}{2}\int_{0}^{\infty} dr 4\pi r^{2} g_{*}(r)
\left( (r^{2}v^{(2)}-rv^{(1)})\langle \hat{x}^{2}\hat{z}^{2} \rangle_{\rm angle}
+rv^{(1)}\langle \hat{z}^{2}\rangle_{\rm angle} \right) \nonumber \\
&& =\frac{\rho}{2} \int_{0}^{\infty} dr 4\pi r^{2} g_{*}(r) 
\left ( \frac{1}{15} r^{2}v^{(2)} +\frac{4}{15}rv^{(1)} \right)
\label{eq-born-av}
\end{eqnarray}
where we used $b({\bf r})=r^{2}v^{(2)}\hat{x}^{2}\hat{z}^{2}+rv^{(1)}(1-\hat{x}^{2})\hat{z}^{2}$ given in \eq{eq-def-born}.
Here $\langle \ldots \rangle_{\rm angle}$ denotes an ``angular average'' 
over isotropic fluctuations of the orientations of the unit vectors
 $(\hat{x},\hat{y},\hat{z})=(x/r,y/r,z/r)$ with
      $r=\sqrt{x^{2}+y^{2}+z^{2}}$.

The terms $J_{2}$ and $J_{3}$ are corrections to the Born term due to
the renormalization of the potential \eq{eq-veff}.
The term $J_{2}$ just involves 2-particle distribution function and we find,
\begin{eqnarray}
 J_{2}=-c\frac{1}{N} \sum_{i < j} 
\left \langle 
\nabla^{2} b({\bf r}_{ij})
\right \rangle_{*}  
= -c\frac{\rho}{2}\int_{0}^{\infty} dr 4\pi r^{2} g_{*}(r)  \left[
\frac{1}{15} r^{2}v^{(4)}(r)
+\frac{2}{3}rv^{(3)}(r)+\frac{22}{15}v^{(2)}(r)+\frac{8}{15}\frac{v^{(1)}(r)}{r}
\right]
\label{eq-J2-av}
\end{eqnarray}
On the other hand $J_{3}$ is more complicated since
it involves connected $4$-particle distribution function.
This term can be evaluated by considering a liquid with a modified potential,
\beq
v_{\rm eff}(r;\delta)=v(r)+\delta \nabla^{2} v (r),
\eeq
at temperature $T^{*}=T/m$. Let us denote its radial distribution function as $g_{*}(r;\delta)$.
Then
\begin{eqnarray}
&& J_{3}=c \frac{1}{N}\sum_{i < j}\sum_{k < l}
 [
\langle 
 b({\bf r}_{ij}) \beta^{*}\nabla^{2} v(r_{kl})
\rangle_{*} -\langle 
 b({\bf r}_{ij}) \rangle_{*} \langle \beta^{*}\nabla^{2} v(r_{kl})
\rangle_{*}]  \nonumber \\
&& =c \frac{\rho}{2} \int_{0}^{\infty} dr 4 \pi r^{2} \left. \frac{\partial g_{*}(r;\delta)}{\partial \delta} \right |_{\delta=0} b({\bf r}_{ij})
=c \frac{\rho}{2} \int_{0}^{\infty} dr 4\pi r^{2} \left. \frac{\partial g_{*}(r;\delta)}{\partial \delta} \right |_{\delta=0}
\left ( \frac{1}{15} r^{2}v^{(2)} +\frac{4}{15}rv^{(1)} \right).
\label{eq-J3-av}
\end{eqnarray}

\item  Non-affine response

Next let us examine the $m J_{1}$ term defined in \eq{eq-I1-I2-I3-I4-I5}.
By noting $\sum_{i} \sum_{j_{1} (\neq i)}\sum_{j_{2} (\neq i)}=\sum_{i \neq j}+\sum_{i \neq j_{1} \neq j_{2}}$, the term $m J_{1}$ can be decomposed into two parts as,
\begin{eqnarray}
 m J_{1} &=& c\frac{1}{N}\sum_{i}\sum_{j_{1} (\neq i)}\sum_{j_{2} (\neq i)} 
\beta^{*}
\left \langle 
\nabla \sigma({\bf r}_{i　j_{1}}) \cdot \nabla \sigma({\bf r}_{i　j_{2}})
\right \rangle_{*}  
 = c\frac{1}{N}\sum_{i \neq j}
\beta^{*}
\left \langle 
|\nabla \sigma_{i　j}|^{2}
\right \rangle_{*}  
 + c\frac{1}{N}\sum_{i \neq j_{1} \neq j_{2}}
\beta^{*}
\left \langle 
\nabla \sigma({\bf r}_{i　j_{1}}) \cdot \nabla \sigma({\bf r}_{i　j_{2}})
\right \rangle_{*}  \nonumber \\
&=& \underbrace{c \beta^{*} \rho \int d^{d}r g_{*}(r) |\Xi({\bf r})|^{2}}_{mJ_{11}}+
\underbrace{c \beta^{*} \rho^{2} \int d^{d}r_{1}d^{d}r_{2} g^{(3)}_{*}({\bf r}_{1},{\bf r}_{2}) 
\Xi({\bf r}_{1}) \cdot \Xi({\bf r}_{2})}_{m J_{12}}
\label{eq-J1-decomposition}
\end{eqnarray}

Using \eq{eq-def-stress} we find the explicit expression of the vector 
$\Xi=(\Xi_{x},\Xi_{y},\Xi_{z})$,
\begin{eqnarray}
&& {\bf \Xi}_{x}({\bf r})=\frac{\partial \sigma({\bf r})}{\partial x} =
[r v^{(2)}(r)\hat{x}^{2}+v^{(1)}(1-\hat{x}^{2})]\hat{z} \nonumber \\
&& {\bf \Xi}_{y}({\bf r})=\frac{\partial \sigma({\bf r})}{\partial y} =
[r v^{(2)}(r)-v^{(1)}(r)]\hat{x}\hat{y}\hat{z} \nonumber \\
&& {\bf \Xi}_{z}({\bf r})=\frac{\partial \sigma({\bf r})}{\partial z} =
[r v^{(2)}(r)\hat{z}^{2}+v^{(1)}(1-\hat{z}^{2})]\hat{x}.
\label{eq-def-Xi-expression}
\end{eqnarray}
The 1st term in the r.h.s of the last equation of \eq{eq-J1-decomposition}, i.~.e $mJ_{11}$
can be easily evaluated using the radial distribution function $g(r)$,
\beq
mJ_{11}=c \beta^{*} \rho \int d^{d}r g_{*}(r) |\Xi({\bf r})|^{2}
=c \beta^{*}\rho \int dr 4\pi r^{2} g_{*}(r) \left[ 
 \frac{1}{15}r v^{(2)}(r))^{2} + \frac{2}{15}r v^{(2)}(r)v^{(1)}(r)) + \frac{7}{15}v^{(1)}(r))^{2} 
 \right].
\label{eq-J11}
\eeq
The 2nd term in the r.h.s of the last equation of \eq{eq-J1-decomposition}, i.~.e $mJ_{12}$
involves the 3-particle distribution function $g^{(3)}({\bf r},{\bf r'})$. As shown in 
Appendix \ref{sec-J12} it can be cast into the following form,
\begin{eqnarray}
&& mJ_{12}= c \beta^{*} \rho^{2} \int d^{d}r_{1}d^{d}r_{2}
 r^{2}_{1}r^{2}_{2} g^{(3)}_{*}({\bf r}_{1},{\bf r}_{2}) 
\Xi({\bf r}_{1}) \cdot \Xi({\bf r}_{2}) \nonumber \\
&& = c \beta^{*} \rho^{2}
\int dr_{1}dr_{2} r_{2} r^{2}_{1}r^{2}_{2}\left[ (g^{(3}_{1})_{*}(r_{1},r_{2})
\frac{32\pi^{2}}{9} \left( \frac{v^{(1)}(r_{1})-r_{1}v^{(2)}(r_{1})}{5}-v^{(1)}(r_{1}) \right)
\left( \frac{v^{(1)}(r_{2})-r_{2}v^{(2)}(r_{2})}{5}-v^{(1)}(r_{2}) \right) \right. \nonumber \\
&&  \left. +(g^{(3}_{2})_{*}(r_{1},r_{2})\frac{16\pi^{2}}{175} (r_{1}v^{(2)}(r_{1})-v^{(1)}(r_{1}))(r_{2}v^{(2)}(r_{2})-v^{(1)}(r_{2})) \right]
\label{eq-J12-legendre}
\end{eqnarray}
where $g^{(3)}_{1}(r_{1},r_{2})$ and $g^{(3)}_{2}(r_{1},r_{2})$ are the coefficients of 
of the expansion of the three particle correlation function
$g^{(3)}({\bf r}_{1},{\bf r}_{2})$  by the Legendre polynomials (see \eq{eq-g3-legendre}).

\end{enumerate}

\section{Evaluation of $mJ_{12}$}
\label{sec-J12}

Here we show derivation of the 2nd equation of \eq{eq-J12-legendre}. The term $mJ_{12}$ is defined as,
\beq
mJ_{12} \equiv c \beta^{*} \rho^{2}\int d^{d}r_{1}d^{d}r_{2} g^{(3)}_{*}({\bf r}_{1},{\bf r}_{2}) {\bf \Xi}({\bf r}_{1}) \cdot {\bf \Xi}({\bf r}_{2}).
\label{eq-def-mJ12}
\eeq
By using the formal expansion of the three particle distribution function 
by the Legendre polynomials,
\beq
g^{(3)}({\bf r}_{1},{\bf r}_{2})=\sum_{l=0}^{\infty}g^{(3)}_{l}(r_{1},r_{2})P_{l}(\cos \theta)
\qquad \cos \theta=\frac{{\bf r}\cdot{\bf r}'}{rr'}
\eeq
and a formal expansion of the field ${\bf \Xi}({\bf r})$ by spherical harmonics $Y^{m}_{n}(\theta,\phi)$,
\beq
{\bf \Xi}({\bf r})=\sum_{n}\sum_{m} {\bf C}^{m}_{n}({\bf r}) Y^{m}_{n}(\theta,\phi)
\eeq
where $\theta$ and $\phi$ are the angular variables of the 3-dimensional polar coordinate
${\bf r}=r(\sin \theta \cos \phi, \sin \theta \sin \phi, \cos \theta )$, 
\eq{eq-def-mJ12} can be rewritten as,
\beq
mJ_{12}=c\beta^{*}\int dr_{1} \int dr_{2} r_{1}^{2} r_{2}^{2}
\sum_{l}(g^{(3)}_{l})_{*}(r_{1},r_{2})\frac{4\pi}{2l+1}\sum_{m=-l}^{l} {\bf C}^{m}_{l}({\bf r})
\cdot [ {\bf C}^{m}_{l}(r_{2})]^{*}
\label{eq-mJ12-expansion}
\eeq
where $[\ldots]^{*}$ means the complex conjugate.

Explicit expressions of the components of the vector field ${\bf \Xi}({\bf})$ is given in
\eq{eq-def-Xi-expression}. From the latter we find, the coefficients ${\bf C}^{m}_{n}$ of their expansions
by the spherical harmonics as,
\begin{eqnarray}
&& ({\bf C}^{m}_{l}({\bf r}))_{x}=({\bf C}^{m}_{l}({\bf r}))_{z}=\delta_{l,1}\delta_{m,0}
\sqrt{\frac{4\pi}{3}} \left  \{
\frac{1}{5}\left(-rv^{(2)}+v^{(1)} \right) -v^{(1)} \right\}
+\delta_{l,3}\delta_{m,0}
\sqrt{\frac{4\pi}{7}} \left  \{-\frac{1}{5}\left(-rv^{(2)}+v^{(1)} \right)   \right\} \nonumber \\
&& +\delta_{l,3}(\delta_{m,2}-\delta_{m,-2})
\sqrt{\frac{4\pi}{7}} \left  \{\frac{\sqrt{5!}}{60}\left(-rv^{(2)}+v^{(1)} \right)  \right\} \nonumber \\
&& ({\bf C}^{m}_{l}({\bf r}))_{y}=\sqrt{-1}
\delta_{l,3}(\delta_{m,2}+\delta_{m,-2})
\sqrt{\frac{4\pi}{7}} \left  \{\frac{\sqrt{5!}}{60}\left(-rv^{(2)}+v^{(1)} \right)  \right\}.
\end{eqnarray}
Using the above result in \eq{eq-mJ12-expansion} we obtain \eq{eq-J12-legendre} which reads,
\begin{eqnarray}
&& mJ_{12}= c \beta^{*} \rho^{2}
\int dr_{1}dr_{2} r_{1}^{2}r_{2}^{2}\left[ (g^{(3}_{1})_{*}(r_{1},r_{2})
\frac{32\pi^{2}}{9} \left( \frac{v^{(1)}(r_{1})-r_{1}v^{(2)}(r_{1})}{5}-v^{(1)}(r_{1}) \right)
\left( \frac{v^{(1)}(r_{2})-r_{2}v^{(2)}(r_{2})}{5}-v^{(1)}(r_{2}) \right) \right. \nonumber \\
&&  \left. +(g^{(3}_{2})_{*}(r_{1},r_{2})\frac{16\pi^{2}}{175} (r_{1}v^{(2)}(r_{1})-v^{(1)}(r_{1}))(r_{2}v^{(2)}(r_{2})-v^{(1)}(r_{2})) \right].
\end{eqnarray}

\section{Formulations for the binary mixture}
\label{appendix-Formulations-for-the-binary-mixture}

Following Coluzzi~et.~al. in  \cite{coluzzi-mezard-parisi-verrocchio-1999} we denote the radial distribution function between particles of types $\epsilon=+,-$ and $\epsilon'=+,-$ as $g^{\epsilon \epsilon'}(r)$. At the order of 1st order cage expansion, the cage sizes of the two types of the particles are found to be,
\beq
A_{\epsilon}=\frac{3}{\beta \rho 
\int_{0}^{\infty} dr 4\pi r^{2} \sum_{\epsilon'}
x_{\epsilon'} g^{\epsilon \epsilon'}_{*}(r) \nabla^{2} v^{\epsilon \epsilon'}(r)
}.
\label{eq-cage-size-av-binary}
\eeq
which is a generalization of \eq{eq-cage-size-av}.

The Born term \eq{eq-born-av} becomes in the binary case,
\begin{eqnarray}
 \langle b_{[\{{\bf R}_{i}\}]}\rangle_{*} 
=\frac{\rho}{2} \sum_{\epsilon \epsilon'} \int_{0}^{\infty} dr 4\pi r^{2} x_{\epsilon} x_{\epsilon'} g^{\epsilon \epsilon'}_{*}(r) 
\left ( \frac{1}{15} r^{2}(v^{\epsilon \epsilon'})^{(2)} 
+\frac{4}{15}r(v^{\epsilon,\epsilon'})^{(1)} \right).
\label{eq-born-av-binary}
\end{eqnarray}
Similarly the $J_{2}$ term \eq{eq-J2-av} becomes,
\begin{eqnarray}
 J_{2}
= -\sum_{\epsilon \epsilon'}c_{\epsilon}\frac{\rho}{2} \int_{0}^{\infty} dr 4\pi r^{2} x_{\epsilon} x_{\epsilon'}g^{\epsilon \epsilon'}_{*}(r)  \left[
\frac{1}{15} r^{2}(v^{\epsilon \epsilon'})^{(4)}(r)
+\frac{2}{3}r(v^{\epsilon \epsilon'})^{(3)}(r)+\frac{22}{15}(v^{\epsilon \epsilon'})^{(2)}(r)+\frac{8}{15}\frac{(v^{\epsilon \epsilon'})^{(1)}(r)}{r}
\right]
\label{eq-j2-cage-expansion-binary}
\end{eqnarray}
where the parameters $c_{\epsilon}$ are related to the cages sizes $A_{\epsilon}$
as,
\beq
c_{\epsilon} = 2 \frac{A_{\epsilon}}{m},
\label{eq-c-epsilon}
\eeq
which is a generalization of \eq{eq-c-A}. 
Similarly the $J_{3}$ term \eq{eq-J3-av} becomes,
\begin{eqnarray}
J_{3}=\sum_{\epsilon} c_{\epsilon} \frac{\rho}{2} \int_{0}^{\infty} dr 4\pi r^{2} \left. 
\sum_{\epsilon' \epsilon''} x_{\epsilon'} x_{\epsilon''}
\frac{\partial g^{\epsilon' \epsilon''}_{*}(r;\delta_{+},\delta_{-})}{\partial \delta_{\epsilon}} \right |_{\delta=0}
\left ( \frac{1}{15} r^{2}(v^{\epsilon' \epsilon''})^{(2)} 
  +\frac{4}{15}r(v^{\epsilon' \epsilon''})^{(1)} \right),
\label{eq-j3-cage-expansion-binary}
\end{eqnarray}
where
$g^{\epsilon' \epsilon''}(r; \delta_{+}, \delta_{-})$ 
is the radial distribution function of a system with modified potential,
\beq
v^{\epsilon_{1} \epsilon_{2}}_{\rm eff}(r;\delta_{+},\delta_{-})=v^{\epsilon_{1} \epsilon_{2}}(r)+\sum_{\epsilon_{3}=+,-}\delta_{\epsilon_{3}} \nabla^{2} v^{\epsilon_{3} \epsilon_{2}} (r).
\label{eq-modified-v-binary}
\eeq

The non-affine correction term $J_{1}$ \eq{eq-j1-cage-expansion} becomes,
\beq
mJ_{1}= \sum_{\epsilon=\pm} c_{\epsilon} \beta^{*} \Xi^{2}_{\epsilon}
\label{eq-j1-cage-expansion-binary}
\eeq
where 
\beq
\Xi^{2}_{\epsilon} \equiv \frac{1}{N_{\epsilon}} \sum_{i \in \epsilon}
\left \langle | \sum_{j (\neq i)}\nabla \sigma({\bf r}_{ij})  |^{2}
\right \rangle_{*}
\label{eq-xi-binary}
\eeq
with $N_{\epsilon}=N x_{\epsilon}$.  Using the particle distribution functions we find $J_{11}$ \eq{eq-J11} and $J_{12}$ \eq{eq-J12-legendre} become,
\begin{eqnarray}
&& mJ_{11} 
=\sum_{\epsilon}
c_{\epsilon} \beta^{*}\rho \int dr 4\pi r^{2} 
\sum_{\epsilon'}
x_{\epsilon}x_{\epsilon'}g^{\epsilon \epsilon'}_{*}(r) \left[ 
 \frac{1}{15}r (v^{\epsilon \epsilon'})^{(2)}(r))^{2} + \frac{2}{15}r (v^{\epsilon \epsilon'})^{(2)}(r)v^{(1)}(r)) + \frac{7}{15}(v^{\epsilon \epsilon'})^{(1)}(r))^{2} 
 \right].\\
&&  mJ_{12}
 = \sum_{\epsilon} c_{\epsilon} \beta^{*} \rho^{2}
\int dr_{1}dr_{2} r_{1}^{2} r_{2}^{2} \sum_{\epsilon' \epsilon''}
\left[ \right. \nonumber \\
&& \left.
((g^{\epsilon \epsilon' \epsilon''})^{(3)}_{1})_{*}(r_{1},r_{2})
\frac{32\pi^{2}}{9} \left( \frac{(v^{\epsilon \epsilon'})^{(1)}(r_{1})-r_{1}(v^{\epsilon \epsilon'})^{(2)}(r_{1})}{5}-(v^{\epsilon \epsilon'})^{(1)}(r_{1}) \right) 
\left( \frac{(v^{\epsilon \epsilon''})^{(1)}(r_{2})-r_{2}(v^{\epsilon \epsilon''})^{(2)}(r_{2})}{5}-(v^{\epsilon \epsilon''})^{(1)}(r_{2}) \right) \right. \nonumber \\
&&  \left. +((g^{\epsilon \epsilon' \epsilon''})^{(3)}_{2})_{*}(r_{1},r_{2})\frac{16\pi^{2}}{175} (r_{1}(v^{\epsilon \epsilon'})^{(2)}(r_{1})-(v^{\epsilon \epsilon'})^{(1)}(r_{1}))(r_{2}(v^{\epsilon \epsilon''})^{(2)}(r_{2})-(v^{\epsilon \epsilon''})^{(1)}(r_{2})) \right]
\label{eq-J12-legendre-binary}
\end{eqnarray}
where $(g^{\epsilon \epsilon' \epsilon''})^{(3)}_{l}(r_{1},r_{2})$ with $l=1,2,\ldots$ are the coefficients of the expansion of the three-particle distribution function by the Legendre polynomials (See \eq{eq-g3-legendre}). 
Using the  Kirkwood superposition approximation \eq{eq-kirkwood} we evaluate them as,
\beq
(g^{\epsilon \epsilon' \epsilon''})^{(3)}_{l}(r,r') = \frac{2}{2l+1}\int_{0}^{\pi} d\theta \sin(\theta)
P_{l}(\cos(\theta))
g^{\epsilon \epsilon'}(r)g^{\epsilon' \epsilon ''}(r')g^{\epsilon'' \epsilon}(|{\bf r}-{\bf r}'|)
\qquad \cos(\theta)=\frac{{\bf r}\cdot{\bf r}'}{rr'}.
\label{eq-legendre-kirkwood-binary}
\eeq

\end{document}